\documentclass[review]{elsarticle}

\usepackage{amssymb,amsmath,array,adjustbox,xcolor}
\usepackage{etoc,longtable,caption}
\usepackage[hidelinks]{hyperref}
\newcolumntype{C}[1]{>{\arraybackslash}p{#1}}

\journal{Annual Reviews in Control}

\begin{document}

\begin{frontmatter}

\title{\qquad Neural Network-based Flight Control Systems: \newline Present and Future}

\author[1]{Seyyed Ali Emami}

\author[2]{Paolo Castaldi}

\author[1]{Afshin Banazadeh \corref{mycorrespondingauthor}}
\cortext[mycorrespondingauthor]{Corresponding author}
\ead{banazadeh@sharif.edu}

\address[1]{Department of Aerospace Engineering, Sharif University of Technology, Tehran, Iran}
\address[2]{Department of Electrical, Electronic and Information Engineering "Guglielmo Marconi", University of Bologna, Via Dell’Universit`a 50, Cesena, Italy}

\begin{abstract}
As the first review in this field, this paper presents an in-depth mathematical view of Intelligent Flight Control Systems (IFCSs), particularly those based on artificial neural networks. The rapid evolution of IFCSs in the last two decades in both the methodological and technical aspects necessitates a comprehensive view of them to better demonstrate the current stage and the crucial remaining steps towards developing a truly intelligent flight management unit. To this end, in this paper, we will provide a detailed mathematical view of Neural Network (NN)-based flight control systems and the challenging problems that still remain. The paper will cover both the model-based and model-free IFCSs. The model-based methods consist of the basic feedback error learning scheme, the pseudocontrol strategy, and the neural backstepping method.
Besides, different approaches to analyze the closed-loop stability in IFCSs, their requirements, and their limitations will be discussed in detail.
Various supplementary features, which can be integrated with a basic IFCS such as the fault-tolerance capability, the consideration of system constraints, and the combination of NNs with other robust and adaptive elements like disturbance observers, would be covered, as well. On the other hand, concerning model-free flight controllers, both the indirect and direct adaptive control systems including indirect adaptive control using NN-based system identification, the approximate dynamic programming using NN, and the reinforcement learning-based adaptive optimal control will be carefully addressed. Finally, by demonstrating a well-organized view of the current stage in the development of IFCSs, the challenging issues, which are critical to be addressed in the future, are thoroughly identified. As a result, this paper can be considered as a comprehensive road map for all researchers interested in the design and development of intelligent control systems, particularly in the field of aerospace applications.
\end{abstract}


\end{frontmatter}


\tableofcontents

\section{Introduction}
\subsection{Intelligent control systems}
Although the words \emph{Intelligence} and \emph{Autonomy} have been widely employed interchangeably, there is an essential conceptual difference between them \cite{Clough.2002}. Different definitions have been given for both concepts in the literature \cite{Gottfredson.1997,Long.2009}. However, in a general view, the intelligence may be defined as \emph{a very general mental capability that involves the ability to reason, plan, solve problems, think abstractly, comprehend complex ideas, learn quickly and learn from experience} \cite{Long.2009}. On the other hand, \emph{the ability to generate one's own purposes without any instruction from outside} can be interpreted as the autonomy of a system \cite{Clough.2002}.
Nevertheless, concerning these two general definitions, in some cases, distinguishing the high level of autonomy from the low level of intelligence is not trivial at all. Within the framework of the control theory, typically, the final purpose is to develop an autonomous system (rather than an intelligent system), which can fulfill a set of predefined missions in a satisfactory manner. However, like the common literature, one may interpret the high level of autonomy of some Unmanned Aerial Vehicles (UAVs) as intelligence.
Accordingly, despite the conceptual difference of these two words, in the current study, (like existing literature,) we will use the words \emph{Intelligence} and \emph{Autonomy}, interchangeably.

Different metrics have been provided in literature to specify the Level of Autonomy (LoA) of a system \cite{Eze.2012}, particularly a UAV. Despite the lack of a unique definition categorization \cite{Vagia.2016}, a beneficial division has been given in \cite{Clough.2002,Protti.2007} in the case of UAVs. Considering it, the highest LoA for a single UAV (level $4$) indicates the self-accomplishment of an assigned tactical plan, where it is capable of on-board trajectory replanning, event-driven self resource management, and compensating for most faults and disturbances in different flight conditions. This, in turn, requires different self-adaptive mechanisms in the entire control system including the entire range from low-level control such as attitude control to high-level control such as path planning. Such a scheme will be known as an intelligent control system in the rest of the paper.

Nowadays, there are different types of Intelligent Flight Control Systems (IFCSs) in the literature that have been designed by employing neural networks, fuzzy systems \cite{Sarabakha.2018b}, behavior tree \cite{Scheper.2016,Croon.2016}, reinforcement learning \cite{Hwangbo.2017}, different data-driven approaches \cite{Cheng.2011b,Ju.2013}, evolutionary algorithms \cite{Floreano.2005,Silva.2016}, etc.
Such a widespread and scattered use of IFCSs in the literature necessitates a comprehensive survey, which can clearly demonstrate the evolution of IFCSs in both theoretical and practical aspects in recent years.
As the first review in this field, authors in \cite{Santoso.2018}, have addressed various technical and practical aspects of IFCSs, where different approaches including fuzzy inference systems, Neural networks, genetic algorithms, swarm intelligence, and hybrid evolutionary systems have been discussed in the paper. However, due to the breadth of the subject, it could not provide an in-depth theoretical view of IFCSs. To deal with such an issue, in the current survey, we will mainly focus on a specific type of IFCSs, namely the Neural Network (NN)-based flight control systems as the most commonly used approach in the literature within recent years.
NNs have been satisfactorily employed in both the dynamic model identification and controller design process. Due to their universal approximation property, they can be employed to estimate different nonlinearities in dynamic systems. In addition, unlike the basic fuzzy control schemes, which highly depend on expert knowledge and pure experiments to construct the fuzzy rule base \cite{Lee.1990}, NNs can effectively learn the system or controller dynamics with no prior information about the system. They can also be integrated with different learning-based methodologies and have been successfully utilized in both direct and indirect control structures, which are discussed in the following. Further, due to their inherent property of parallel processing, neural networks can be suitably employed in real-time implementations \cite{Campa.2002,Kacprzyk.2010b}.

Historically, the concept of IFCS was introduced in the 1990s by adopting NNs in the structure of flight control systems as a learning element to adapt to unexpected fault and flight conditions \cite{Tomayko.2003}.
However, although the beginning of using NNs in flight control systems dates back to early 1990s, due to both technological and methodological limitations, dynamic NNs were not employed in practical flight control problems until 2001 \cite{Steinberg.2005b}.
As the largest project in this field, the \emph{IFCS program} has been conducted in a collaboration between NASA and Boeing between 1999 to 2009 \cite{Schumann.2010,Kacprzyk.2010b,Kaminski.2017}. This program consists of two main phases. The first phase of the program focused on the development of an indirect adaptive flight control system. The first set of flights using a highly modified F-15B prototype occurred in 1999, where the stability and control derivatives of the air vehicle have been estimated using pre-trained NNs. Subsequently, dynamic cell structure NNs have been adopted in the control scheme for online modification of the estimated derivatives. The second set of flights using such an online identification scheme have been performed in 2003 \cite{Williams.2004b}. Although the obtained results in the closed-loop simulation were reported as a \emph{promising} achievement, due to the fact that the online identified model was not utilized in the control structure in real flights, the control scheme was not yet really adaptive \cite{Hageman.2003}.
On the other hand, the second phase of the program dealt with developing a direct adaptive control architecture in which a dynamic inversion block was augmented by an online NN \cite{Smith.2010}. Flight tests of the second phase began in 2006 and continued into 2008.
Flight tests consisted of performance evaluation, with and without dynamic NN augmentation, in the presence of structural damages and control surface faults. The evaluation was based on performance measurements and pilot ratings. As reported in \cite{Bosworth.2007,Burken.2009}, for structural damages, NN augmentation was generally found to provide significant improvements in overall pitch performance. However, control surface faults led to mixed results from slight improvements in pitch rate response to a propensity for roll pilot-induced oscillation. A modification was also introduced in the designed control scheme employed in the second phase of the program. The utilized modifications included the use of alternate NN inputs in the designed framework which can satisfactorily tackle high-correlation and high-gain problems in the basic design, the adoption of a weight decay term (in updating the NN weights) to avoid the overfitting problem, and using \emph{scalar} dead-zones in adaptation laws for simplicity. The results obtained by the modified control scheme in 2009 indicated a significant improvement over the basic design \cite{Burken.2009}. With the retirement of the F-15B air vehicle in January 2009, the IFCS program was finally finished \cite{Kaminski.2017}.

The lesson learned from the IFCS program demonstrated that the high complexity of the control design, as well as the unpredictable behavior of the control scheme in the presence of unexpected flight and fault conditions, could be serious concerns in utilizing adaptive control approaches in real applications, particularly manned aircraft \cite{Hanson.2012,Hanson.2014}. Another program has been accordingly launched by NASA in 2009, namely the Integrated Resilient Aircraft Control (IRAC) project, where one of its main objectives was to investigate simple, yet effective, adaptive control methods to address the issue of verification and validation of adaptive flight controls to a safety-critical level. Addressing this project in more detail is beyond the scope of this paper.
Motivated by the above discussion, in this survey, we will address both indirect and direct NN-based adaptive control schemes and their evolution towards more reliable approaches with less computational complexity, in detail. However, as will be discussed in the following, although the indirect and direct adaptive control approaches arise from two different points of view, they can be formulated within a similar mathematical framework with the same updating rules for dynamic NNs in the case of model-based approaches as in the IFCS project of NASA. Accordingly, we will deal with IFCSs in a different primary categorization, i.e. the model-based and model-free NN-based control methods. In addition, in the current research, we will address a variety of model-free flight control systems that have been built upon some other machine learning approaches such as Reinforcement Learning (RL).

\subsection{Direct versus indirect adaptive control}
There are a variety of flight control systems in the literature in which neural networks have been utilized to solve an online optimization problem within the control block \cite{Dalamagkidis.2011,Li.2016} or to mimic the behavior of a classical controller \cite{Efe.2011b}. In this paper, we will focus on control methods where neural networks have been directly adopted in the control design procedure as \emph{intelligent elements} to bring a degree of intelligence into the closed-loop behavior.
This can be performed in different manners:
Various studies have attempted to employ NNs to estimate model uncertainties, which are subsequently utilized in designing the control command. Such a framework is known as \emph{indirect} neural control.
The training process of NNs can be performed online using well-known learning algorithms such as feedback error learning or offline to provide a pre-trained dynamic model of the system. On the other hand, in the \emph{direct} neural controller, NNs have been utilized to directly construct the control command.

To be more specific, in the case of model-based control approaches, if the system dynamics can be formulated as $\dot{x}=f(x)+g(x)u$, where $f(x)$ and $g(x)$ denote unknown nonlinear functions of system states, we have two general choices to design the control command $u$. In the first approach, we can estimate both the unknown functions $f$ and $g$ using NNs and then employ them in the designed control command. In the second approach, however, we attempt to directly design the control command by estimating $g^{-1}(\dot{x}_d-f)$, which is required in the control command, using a NN ($x_d$ represents the reference trajectory).
In the literature, the first method is known as an indirect adaptive control, while the second approach corresponds to a direct adaptive control scheme \cite{Xu.2015,Xu.2018}. However, as will be discussed in Section \ref{Sec:Model_based_IFCS}, concerning the mathematical formulation and the structure of updating rules of NNs, there is no fundamental difference between these two model-based control approaches.

On the other hand, in the case of model-free adaptive control methods, it is not easy to provide a general view of different types of indirect and direct adaptive control schemes employed in the literature. As one of the most commonly used model-free indirect IFCSs, (different types of) NNs are used to identify the entire system dynamics, and subsequently an adaptive control scheme is designed based on the online identified model. In addition, the direct model-free control could be originated from various design methodologies, while in the current survey, two more common schemes, i.e. the adaptive optimal and the RL-based control methods will be discussed in detail.

Although due to the simpler structure and less computational complexity \cite{KarlJohanAstrom.2008}, direct adaptive control has been widely employed in different applications, there are various concerns regarding its applicability in serious missions. Briefly, increasing the learning rate in direct adaptive control, known as aggressive learning, is a typical approach to rapid reduction of the dynamic inversion error \cite{Gu.2019}. In this regard, high-gain control due to aggressive learning in direct adaptive control is a problematic issue which can lead to actuator saturation, the excitation of unmodeled dynamics, and other well-known problems of high learning rates \cite{Ioannou.2014}.
Besides, in the case of a damaged aircraft, the system dynamics can significantly change, and the lack of reliable knowledge about the current system dynamics may result in inefficient control commands, particularly, when the control system consists of a \emph{nominal} controller augmented by an adaptive NN-based control command \cite{Nguyen.2008}.

\subsection{Model-based versus Model-free control}
In this section, we present a more detailed classification of NN-based flight control systems to be used in the remainder of the paper.
Neural networks have been extensively employed in the structure of flight control systems for the past three decades \cite{Calise.1996}.
They can be generally studied in two fundamentally different categories, i.e. the model-base and the model-free control approaches.

\subsubsection{Model-based approach}
The model-based neural control, which utilizes a nominal model of the system in the control design process, has significantly evolved during the last two decades. Feedback Error Learning (FEL) is the most popular learning scheme, which has been widely incorporated in intelligent control systems. By employing the tracking error of the system, the prediction error of the model, the output of a baseline controller, or a combination of them, an unsupervised learning approach is developed in such a way that both the tracking error of the system and the estimation error of the neural networks remain bounded (unsupervised learning occurs when the NN is trained to respond to a certain pattern in the absence of output examples \cite{Hagan.2016}). Several variations of FEL-based IFCSs have been proposed by researchers in recent years. In this method, the neural network attempts either to estimate the model uncertainties (or/and external disturbances) or to determine the control command. The first approach leads to an indirect control structure, while the second one results in a direct adaptive control scheme.
In addition, different types of feedforward or recurrent neural networks including Radial Basis Function (RBF) neural networks, multilayer perceptron, High-Order NNs (HONNs) \cite{Xu.2015c}, Extreme Learning Machine (ELM) \cite{Xu.2014}, Elman NN, etc have been employed in FEL-based control methods. The FEL scheme, its characteristics, and different variants within the flight control framework will be intently studied in Section \ref{Sec:Model_based_IFCS}.

One of the main drawbacks of FEL-based control systems is that all the uncertain terms in the model are typically estimated as a single term using NNs. This may result in poor training performance, particularly in the lack of Persistent Exciting (PE) signals. In addition, in most of the FEL-based neural controllers, a baseline control, which is designed based on a nominal model of the system, is employed where the NN acts as an aid to the controller. This may cause large control actions under severe structural damages or dynamic changes. Such concerns and other design considerations result in incorporating several features in the basic FEL-based control methods, which will be addressed in Section \ref{Sec:SumplementaryFeatures}.

\subsubsection{Model-free approach}
On the other hand, the model-free scheme does not require any prior information about the system dynamics to be used in the control design procedure.
As a traditional model-free approach, the \emph{entire} controller was modeled by a single NN \cite{Rinaldi.2013,Artale.2016}, where the error back-propagation technique was typically utilized to train the NN. Although such a control scheme, in some cases, could provide an acceptable response even under severe external disturbances \cite{Juang.2006}, the stability of the closed-loop system could not be mathematically analyzed \cite{Kurnaz.2010}. In addition, such a training method may occasionally converge to local minima. Thus, this control approach could not be safely utilized in important missions.

In recent years, a class of model-free intelligent control systems has been proposed in the literature using the concept of Approximate Dynamic Programming (ADP) and Reinforcement Learning (RL). Indeed, although the introduced schemes have originated from different scientific points of view (from the control theory to machine learning and information theory), the principal methodology employed by them are fundamentally similar. More specifically, in many of such control structures, an actor-critic framework is defined (where NNs may be used to estimate both the actor and critic functions).
In this regard, the critic corresponds to the cost-to-go function (or the value function), while the actor determines the control input applied to the system. With a focus on the RL framework, the entire control design process is typically transformed to a Markov Decision Process (MDP). Accordingly, the value function demonstrates an accumulative discounted reward function obtained by the system (from the present time) using the current policy. The control objective is then to endeavor to maximize the value function by changing the policy function. This can be performed using the conventional policy-gradient method. Owing to the advancement of the computational power of processors, different RL-based IFCSs have been proposed in few recent years, where the control design process has been completely performed in the simulation environment and subsequently, the designed controller is satisfactorily applied to a real application. Such a control methodology will be discussed in detail in Sections \ref{Sec:Optimal} and \ref{Sec:RL}.

There is also a variety of model-free indirect IFCSs in literature in which a separate identification process has been defined in the control design procedure. Different types of neural networks including nonlinear autoregressive with exogenous inputs (NARX) network \cite{Rinaldi.2013}, Elman networks (as recurrent NN), Convolutional NN, Wavelet NN, ELMs \cite{Emami.2018b}, and Fuzzy NNs \cite{Kayacan.2016b} have been utilized in the identification step to identify different unmodeled dynamics. The identified models can also be updated online in order to adapt to dynamic changes in the system. Subsequently, the control system is designed based on the identified NN. Although the analysis of the closed-loop stability in such a multi-step control design process may be more challenging, this can result in a more efficient control system in comparison with FEL-based control methods, especially in the presence of severe dynamic changes. We will address this type of indirect IFCS in Section \ref{Sec:Model-free-Indirect}.

Finally, a set of concluding remarks and possible future directions for NN-based IFCSs will be provided in Section \ref{Sec:Conclusions}, which indeed attempts to illustrate the main existing gaps in the framework of IFCSs to be employed in serious missions as a reliable, effective, and really intelligent control scheme with acceptable computational cost.


\section{Foundations of model-based intelligent control} \label{Sec:Model_based_IFCS}

In this section, we deal with model-based NN-based flight control systems. Regarding the dynamic model of air vehicles, they can be categorized in different manners. As a general classification, an air vehicle can be modeled as a nonlinear affine or nonaffine dynamic model. Concerning the affine model, we will pay more attention to two types of more popular dynamic models: dynamic models with $dim(x)=dim(u)$ (Section \ref{Sec:FEL}) and the dynamic models in the strict feedback form (Section \ref{Sec:Backstepping}).
Most of the current approaches in the literature to control the aerial vehicles attempt to transform the system dynamics into one of the above-mentioned variants by defining intermediate control variables, designing multi-loop control systems, etc, where such techniques will also be briefly discussed.
It is notable that, both the continuous-time and discrete-time models will be covered, as well.
In addition, the control of nonaffine systems, mainly using the pseudocontrol strategy or similar methods, will be discussed in Section \ref{Sec:PseudoControl}.

Besides, regarding the consideration of model uncertainties, internal faults, and external disturbances, it should be noted that they can be modeled using either additive or multiplicative uncertain terms. Although both schemes have been utilized in the literature, the employment of additive uncertain terms is more general than the multiplicative case \cite{Baldi.2013,Baldi.2016}. Indeed, multiplicative uncertainties can also be modeled using additive terms, though the unknown terms may become a function of both the system inputs and states \cite{Verhaegen.2010}.
Accordingly, in the following, we mainly focus on the control of dynamic systems with additive uncertain terms and will transform the possible multiplicative uncertainties into additive terms.
In addition, here, we will consider uncertain terms in the dynamic model as a lumped disturbance, which will be estimated by a NN.
However, dealing with different types of uncertain dynamics such as model uncertainties, atmospheric disturbances, and operational faults may necessitate different learning strategies with their own requirements, which are addressed in the following section. More specifically, combined approaches that utilize a combination of NNs, disturbance observers, and/or state estimators to tackle different uncertain terms in the system dynamics will be discussed in Sections \ref{Sec:NNDO} and \ref{Sec:FTC}.
Further, as the consideration of the multiplicative representation of uncertain terms can be more beneficial in case of identification of unknown gains corresponding to \emph{actuator faults}, we will also address this type of model uncertainties in the framework of NN-based Fault-Tolerant Control (FTC) systems in Section \ref{Sec:FTC}.

\subsection{Feedback Error Learning} \label{Sec:FEL}
Here, we will introduce the fundamental theoretical basis for the most commonly used approach to incorporate NNs within the adaptive control design process, i.e. the FEL method, while the application of such an algorithm in flight control systems would be addressed in the following subsections.
FEL can effectively integrate the control design procedure and the online updating law for the parameters of the NN, which is utilized to compensate for model uncertainties and disturbances.
Accordingly, in a general view, the control block can consist of a conventional controller in the inner loop to stabilize the system dynamics, and the neural controller acts as an aid to the controller to compensate for model uncertainties. Thus, employing a composite Lyapunov function including both the tracking error and the estimation error of NN parameters, the closed-loop system can satisfy the Bounded-Input–Bounded-Output (BIBO) stability requirement in the presence of model uncertainties and external disturbances.
To be more precise, consider the dynamic model of an aircraft (in the affine form) as follows:
\begin{equation}\label{Model_1}
\dot{x}=F(x)+B(x) u+ \Delta,
\end{equation}
where $x,u \in \mathbb{R}^n$, and $\Delta$ stands for model uncertainties and external disturbances. Defining the desired trajectory as $x_d$, the tracking error is obtained as $e=x-x_d$. Now, the control command can be defined as
\begin{equation}\label{Ctrl_0}
u=B^{-1}\left(-F(x)-\Delta+\dot{x}_d-k_1 e\right),
\end{equation}
where $k_1$ is a positive-definite matrix. However, the vector $\Delta$ is unknown. Thus, it is approximated by a NN (such as RBFNN or multilayer perceptron) as $\hat{\Delta}=\hat{W}^T \mu (x)$, where $\mu$ represents the vector of basis functions (corresponding to hidden layers of the NN) and $W$ indicates the matrix of unknown weights which should be identified.
Such a formulation can be used to represent different feedforward and recurrent NNs. Here, we will use such a general formulation, and for brevity, do not address different possible network structures and their advantages and disadvantages in flight control systems (for more details, see \cite{Gu.2020}).
Accordingly, due to the universal approximations property of NNs, we have:
\begin{equation}\label{NN_1}
\Delta = {W^*}^T \mu (x) + \varepsilon,
\end{equation}
where $W^*$ denotes the (unknown) optimal weight and $\varepsilon$ indicates the \emph{bounded} approximation error ($\|\varepsilon\|\leq \varepsilon_M$).
The control command can now be constructed as follows:
\begin{equation}\label{Ctrl_1}
u=B^{-1}\left(-F(x)-\hat{\Delta}(x)+\dot{x}_d-k_1 e\right).
\end{equation}
Now, consider a Lyapunov function as
\begin{equation}\label{Lyap_0}
  V=\frac{1}{2}e^T e + \frac{1}{2} tr \left(\tilde{W}^T \Gamma^{-1} \tilde{W}\right),
\end{equation}
where $\tilde{W}=\hat{W} - W^*$ and $\Gamma$ is a positive definite matrix. Next, we have
\begin{align}
\begin{split}\label{Lyap_1}
   \dot{V} & = e^T \dot{e}+ tr \left(\tilde{W}^T \Gamma^{-1} \dot{\tilde{W}}\right)  \\
     & = - e^T \left(\tilde{\Delta}+ k_1 e\right)+ tr \left(\tilde{W}^T \Gamma^{-1} \dot{\tilde{W}}\right),
\end{split}
\end{align}
where, $\tilde{\Delta}=\hat{\Delta} - \Delta$. Accordingly, defining
\begin{equation}\label{Learn_1}
 \dot{\hat{W}}=\Gamma \mu e^T,
\end{equation}
and considering $\dot{\tilde{W}}\approx  \dot{\hat{W}}$ (as a consequence of assuming a \emph{constant} optimal weight $W^*$, while such an assumption is reasonable even in the case of a time-dependent uncertain term $\Delta={W^*}^T(t) \mu (x) + \varepsilon$ with $\dot{W}^* \ll \dot{\hat{W}}$), we have
\begin{align}
\begin{split}\label{Lyap_2}
\dot{V} & = - e^T \left(\tilde{\Delta}+ k_1 e\right)+ tr \left(\tilde{W}^T \Gamma^{-1} \dot{\tilde{W}}\right) \\
 & = - e^T \left(\tilde{\Delta} + k_1 e\right) + tr \left(\tilde{W}^T \mu e^T \right) \\
 & = - e^T \left(\tilde{\Delta} + k_1 e\right) + e^T \tilde{W}^T \mu \\
 & = - k_1 e^T e + e^T \varepsilon = - e^T (k_1 e - \varepsilon),
\end{split}
\end{align}
which leads to $\dot{V}<0$ for $\|k_1 e\| > \|\varepsilon\|$ thereby guaranteeing the bounded tracking error. The determination of optimal design parameters such as $k_1$ and $\Gamma$ is not an easy task, where it is typically done by trial and error. It is also possible to define an optimization problem in terms of these parameters and solve it using well-known optimization methods (such as evolutionary algorithms) to determine their optimal values according to predefined criteria \cite{MohdBasri.2015}.
In cases where the matrix $B$ is also uncertain, we have:
\begin{equation}\label{Model_2}
\dot{x}=F(x)+(B+\Delta B) u+ \Delta,
\end{equation}
where $B$ represents the nominal part. Thus, it is possible to define
\begin{equation}
  \bar{\Delta}:=\Delta B u+ \Delta
\end{equation}
and estimate it using an NN as $\hat{\Delta}(x,u)=\hat{W}^T \mu (x,u)$. Consequently, the control command can be calculated as follows:
\begin{equation}\label{Ctrl_2}
u=B^{-1}\left(-F(x)-\hat{\Delta}(x,u)+\dot{x}_d-k_1 e\right).
\end{equation}
As seen, the control command results in an equation as $u=h(.,u)$. The existence and uniqueness of a solution for $u$ require a contraction assumption \cite{Calise.2001b}. Sufficient conditions for satisfying this assumption are given in \cite{NakwanKim.2003}. Notably, this assumption implicitly requires the sign of the control gain function to be known \cite{Chowdhary.2014}.
Note that, it is also possible to update the weights of the hidden layer to provide more effective learning. This can be performed using a similar Lyapunov stability analysis by taking advantage of the Taylor expansion of the hidden layer output ($\mu(x)$) \cite{Lee.2001}. However, due to the more complicated formulation and excessive computational burden, in this paper, we will only update the output weights, $\hat{W}$, and the other parts of the NN remain unchanged.

Besides, one can replace the tracking error $e$ in (\ref{Lyap_0}) and (\ref{Learn_1}) by a filtered tracking error as $s=e+\lambda \int e \enspace dt$ (with $\lambda$ denotes a positive constant or a positive definite matrix) to compensate for the steady-state tracking error \cite{Li.2016d}.
Further, the introduced FEL neural control scheme can be applied to a second-order system, i.e. $\ddot{x}=F(x)+B(x) u+ \Delta$ by substituting $e$ by a filtered tracking error as $s=\dot{e}+\lambda e$ \cite{Zeghlache.2018,Song.2019}.

On the other hand, as the designed controller should be programmed on a digital processor in real applications, the development of a control system in the discrete-time domain makes more sense. Using a discrete-time controller, the dependence of the closed-loop \emph{performance} on the sampling rate can also be eliminated. This would be more beneficial in the case of NN-based control systems in which the differential equations for updating the NN weights change to difference equations. Furthermore, in the case of discrete-time controllers, the NN weights' updating rate that guarantees the convergence of the training rule can be computed analytically \cite{DongHoShin.2006}. To illustrate the fundamental structure of a discrete-time FEL scheme, consider the equivalent discrete-time model of (\ref{Model_1}) as follows:
\begin{equation}
  x(k+1)=F_d(x(k))+B_d(x(k))u(k)+\Delta_d(k).
\end{equation}
Defining
\begin{gather}
u=B(k)^{-1}\left(-F_d(k)-{\hat{W}}^T \mu(k) +c e(k) +x_d(k+1) \right), \\
\Delta_d(k)={W^*}^T \mu(x(k))+\varepsilon,\\
e(k)=x(k)-x_d(k),
\end{gather}
where $0<c<1$ leads to the following equation.
\begin{equation}\label{Eq:15}
  e(k+1)=c e(k) - \tilde{W}^T \mu(k) + \varepsilon.
\end{equation}
By multiplying both side of (\ref{Eq:15}) by $e^T(k+1)$, we have
\begin{align}
\begin{split}
  e^T(k+1)\tilde{W}^T \mu(k)= c e^T(k+1)e(k)
  - e^T(k+1)e(k+1) + e^T(k+1)\varepsilon.
\end{split}
\end{align}
Using Cauchy–Schwarz and Young's inequalities, it is obtained that 
\begin{align}
\begin{split}
  e^T(k+1)\tilde{W}^T \mu(k) \leq  \|e(k+1)\|^2 \Big( -1 + \rho_1 + \rho_2 \Big)
  + \frac{c^2}{4\rho_1} \|e(k)\|^2  + \frac{1}{4\rho_2}\varepsilon_M^2,
\end{split}
\end{align}
with $\rho_1,\rho_2 > 0$.
Thus, if a Lyapunov function is defined as (\ref{Lyap_0}) (without the coefficient $1/2$), using the following updating rule,
\begin{equation}\label{Eq:15_1}
  \hat{W}(k+1)=\hat{W}(k)+\Gamma \mu(k) e^T(k+1),
\end{equation}
the first difference of $V(k)$ is obtained as follows:
\begin{align}
\begin{split}
\Delta V (k) & = V(k+1) - V(k)= \|e(k+1)\|^2- \|e(k)\|^2 \\
& \quad + tr \left(\tilde{W}^T (k+1) \Gamma^{-1} \tilde{W}^T (k+1) -\tilde{W}^T (k) \Gamma^{-1} \tilde{W}^T (k)\right)\\
& \leq \|e(k+1)\|^2 \Big( 1 + 2(-1 + \rho_1 + \rho_2) \Big) + \|e(k)\|^2 \Big( \frac{c^2}{4\rho_1} -1\Big)\\
& \quad + \frac{1}{4\rho_2}\varepsilon_M^2 + \|\mu\|_{\Gamma}^2\|e(k+1)\|^2.
\end{split}
\end{align}
Accordingly, we have
\begin{equation}
 \Delta V (k) \leq -k_1 \|e(k+1)\|^2 -k_2 \|e(k)\|^2 + c_1,
\end{equation}
where,
\begin{gather}
  k_1= 1 -2 \rho_1 -2 \rho_2 - \|\mu\|_{\Gamma}^2,\\
  k_2= 1- \frac{c^2}{4\rho_1}, \\
  c_1= \frac{1}{4\rho_2}\varepsilon_M^2.
\end{gather}
Thus, assuming the boundedness of $\mu(x)$, it is possible to determine $\rho_1$, $\rho_2$, $c$, and $\Gamma$ such that $k_1,k_2 > 0$ thereby guaranteeing  $\Delta V (k)<0$ for $ \|e(k+1)\|^2 > c_1/k_1$.
As seen, although the updating rule (\ref{Eq:15_1}) is similar to that of continuous-time systems, i.e. (\ref{Learn_1}), the stability analysis of the discrete-time FEL neural control is quite different and more complex compared to that of continuous-time systems.
Consequently, in the following, we mainly focus on the continuous-time formulation of control systems, while their discrete-time equivalent would be obtained in a similar manner as discussed above.

It should be noted that, in the introduced (continuous/discrete) adaptive control scheme, the convergence of the NN weights to their ideal values is not trivial, and it requires a Persistent Excitation (PE) \cite{Kacprzyk.2010b}. More precisely, in the absence of persistent exciting input signals, the NN weight estimates might drift to very large values, which will result in a variation of high-gain control \cite{Zhang.1999,Ge.2002}.
Different approaches have been proposed in the literature to prevent parameter drift in such conditions. Some of the more common methods are briefly introduced in the following.
\begin{enumerate}
  \item Dead-zone: In this straightforward method, the previously mentioned updating rule is only used when the tracking error exceeds a predefined threshold \cite{VijayaKumar.2009}. Otherwise, the NN weights remain constant. Although such a method can successfully prevent parameter drift, as discussed in \cite{TzirkelHancock.1992,Fabri.1996}, the determination of an appropriate threshold requires the bounds of the control gain function and the NN estimation error ($\varepsilon_M$), which may not be generally known.
  \item Projection: The second simple method is to limit the NN weights to a predefined interval. It means that the time derivative of the parameters is set to zero when they reach the given bounds \cite{Boskovic.2004}. The main drawback of this method is the requirement of the lower and upper bounds of the NN parameters.
  \item Sigma-modification: The third method, which has been introduced by Ioannou and Kokotovic \cite{Ioannou.1983,Ioannou.2012} is a more useful approach. In this method, a modification term is incorporated in the updating rule of the NN parameters as $\dot{\hat{W}}=\Gamma \left(\mu e^T -\sigma W\right)$, where $\sigma$ is a positive constant \cite{Shin.2004}. Such an approach has been employed in many NN-based flight control systems such as \cite{Hovakimyan.2001,Ge.2002,Hovakimyan.2002,S.S.Ge.2009}.
  \item e-modification: Another popular approach has been introduced in \cite{Narendra.1987}, where the constant parameter $\sigma$ in the previous technique is replaced by a term proportional to $|e|$ \cite{Zhang.1999,Rysdyk.2005}. The boundedness of the NN parameters using e-modification has been presented in \cite{RolfRysdyk.1998}. Further, as a major advantage of the e-modification technique over the $\sigma$-modification, such a modification term is effectively attenuated by approaching the tracking error to zero, and (in the lack of the estimation error $\varepsilon$,) this method does not affect the convergence of the NN weights to their ideal values in the presence of persistent exciting training signals.
  \item Alternate weights: This approach has been first proposed in \cite{ChrisJ.B.Macnab.2006}. The e-modification method may not achieve acceptable performance in the presence of large oscillatory disturbances \cite{C.Nicol.2008}. The basic idea of this method is that different sets of NN weights are capable of uniformly approximating the same nonlinear function. An alternate set of weights with a smaller magnitude than $\hat{W}$ can be used to improve the training. By keeping the NN weights close to the smaller alternate weights, it is possible to provide a more efficient compromise between the approximation performance and keeping the NN weights bounded, while there is a need for two distinct sets of NN weights and their corresponding updating rules. This method has been employed in \cite{Coza.2011} to design a flight control system for a quadrotor air vehicle under the wind buffeting.
\end{enumerate}

Although the aforementioned approaches result in bounded NN parameters, satisfactorily, they do not ensure the convergence of the NN weights to their ideal values. Recently, a variety of modified learning approaches have been proposed in the literature, which causes the improvement in training the NN parameters. Some of the more attractive methods are as follows:
\begin{enumerate}
  \item Composite learning: Different composite learning approaches have been introduced in the literature where their fundamental idea is to include the estimation performance into the updating law \cite{Xu.2014b,Xu.2017,Xu.2021b}. This can lead to faster learning speed as well as higher precision \cite{Xu.2019}. More specifically, the state estimation can be constructed as $\dot{\hat{x}}=F(x)+B(x) u+\hat{W}^T \mu (x) - \beta \tilde{x}$, where $\tilde{x}=\hat{x}-x$, and $\beta$ is a positive constant. Thus, the updating rule (\ref{Learn_1}) can be modified as $\dot{\hat{W}}=\Gamma \mu (e^T -\Gamma_1 \tilde{x}^T)$ where $\Gamma_1$ is a positive definite matrix \cite{Xu.2019}. An improved learning method has been presented in \cite{Xu.2018}, where the basic updating rule (\ref{Learn_1}) is augmented by a novel prediction error signal constructed using online recorded data within a time interval $[t-\tau,t]$, which is equal to $\tilde{W}^T \int_{t-\tau}^{t} \mu (x) + \int_{t-\tau}^{t} \varepsilon$. As shown in \cite{Xu.2018}, the proposed approach, which has been applied to the longitudinal model of a hypersonic aircraft, can lead to better tracking with less chattering.
  \item Concurrent learning: A beneficial approach, which has been introduced in \cite{Chowdhary.2011,Chowdhary.2014}, utilizes a set of recorded data points concurrently with instantaneous data to improve the convergence of both parameter and tracking errors. The main benefit of the concurrent learning method is that PE or high adaptation gains are not required. More precisely, in the case of nonlinear systems with \emph{parametric} uncertainty (which can be modeled as $\Delta(x)={W^*}^T\mu(x)$), it has been proved that, if the training input signal is exciting in a sufficiently large finite time interval, both tracking error ($e$) and NN weight's estimation error ($\tilde{W}$) converge exponentially to zero. However, this approach requires precise estimation of the time derivative of the system states, which may be impractical in some cases.
  \item Reinforced learning: Another approach to improve the training performance is to reinforce the learning signal \cite{Lin.2007,Lin.2009}. This can be done by modifying the training rule (\ref{Learn_1}) using the output of another NN (commonly known as the critic network) as $\dot{\hat{W}}=\Gamma \mu \left(e + \|e\| \hat{W}_c^T \mu_c \right)^T$ where $\hat{W}_c$ represents the output weights of the critic network, which is tuned in such a way that guarantees the closed-loop stability \cite{Bu.2019}. The provided learning signal is more informative than the basic training rule (\ref{Learn_1}) thereby strengthening the control performance \cite{Luo.2015b}.
\end{enumerate}

The above-mentioned formulation corresponds to the \emph{indirect} FEL-based control where a NN attempts to identify model uncertainties, and then the control command is constructed using the estimated uncertainty \cite{HaojianXu.2003}.
FEL can also be satisfactorily employed in the framework of direct adaptive control systems. In this regard, it is possible to directly estimate the entire control command $u$ or the uncertain term $B^{-1}\Delta$ in (\ref{Ctrl_0}) by a NN. As a result, the updating rule of $\hat{W}$ will include the control gain matrix $B$. However, there is no fundamental difference between the direct and indirect approaches regarding the formulation of the updating rules and the stability analysis of the closed-loop system.
Direct methods may also be preferred in cases where the control gain function is entirely unknown (see Section \ref{Sec:DirectNeuralBackstepping}).


\subsection{Pseudocontrol strategy}\label{Sec:PseudoControl}
In a somewhat similar manner to the introduced direct FEL-based neural control, in the case of \emph{nonaffine} models, traditionally, the output of a baseline controller (such as a PID controller) may be used to train a NN, which augments the output of the baseline controller to learn the inverse dynamics of the system, while there is considerable complexity in the closed-loop stability analysis \cite{Gomi.1993}. Accordingly, an auto-landing scheme has been proposed in \cite{Li.2004} for an aircraft under external disturbances using a FEL-based neural aided $H_{\infty}$ control. Similarly, a combination of a classical trajectory tracking control (using the loop shaping technique) with a FEL-based neural controller has been employed in \cite{Pashilkar.2006b} as a fault-tolerant auto-landing control method. Such a method has also been adopted in \cite{Li.2001} to control the attitude of a simplified model of a fighter aircraft using fully-tuned growing RBFNNs.


By incorporating a similar framework, type-2 Fuzzy Neural Networks (T2-FNN) have been employed in \cite{Khanesar.2015,Kayacan.2016} to augment a classical PD controller in the case of a set of SISO systems. FEL algorithm has been adopted, where the updating rule corresponding to the consequent part of the T2-FNN has been derived by minimizing $\int \left(\dot{e}+\lambda e \right)^2 dt$. Although in \cite{Khanesar.2015}, it has been assumed that the intended system has a second-order stabilizable dynamic model, the stability of the closed-loop system has been analyzed without making any assumption on the system characteristics (even the system's stabilizability!). Apparently, this is a consequence of estimating an \emph{explicit function of time} by a neural network as ${W^*}^T\mu(x)$, which is not generally feasible. This is a common issue in NN-based identification schemes (see Section \ref{Sec:NNDO}).
The proposed control system has been applied to the trajectory tracking control of a quadrotor UAV. A self-organizing neuro-fuzzy-based control has been introduced in \cite{Ferdaus.2019} in which the consequent part of the fuzzy rules has been trained using a similar FEL scheme, where the designed controller has been applied to a hexacopter and a flapping-wing Micro Aerial Vehicle (MAV) to control the altitude and the attitude of the system.
It has been claimed that the controller's performance does not depend on \emph{any} features for the system. This is clearly an exaggerated statement since the most obvious feature required in a controlled system is the system's controllability. Again, it seems that the generality of the stability analysis is due to the aforementioned concern regarding the NN-based identification schemes.
Different from \cite{Ferdaus.2019}, in \cite{Ferdaus.2020}, the stability analysis has been provided for a $n$th order, affine, SISO model. A FEL scheme has been used to train the consequent parameters of a neuro-fuzzy control system, which augments a PID controller, while the updating rule is subject to the parameter drift issue.

On the other hand, a simpler and popular approach to FEL-based direct adaptive control of nonaffine systems, known as the \emph{pseudocontrol} strategy, has been widely employed in IFCSs. Generally speaking, in this approach, the control command is determined using a model inversion block where a neural network is utilized to cancel out the inversion error \cite{Kim.1997}. To be more precise, consider a generic nonaffine nonlinear model of the system as follows:
\begin{equation}\label{Eq:16_0}
  \dot{x}=F(x,u).
\end{equation}
As seen, unlike the previous subsection, here, there is no need for an affine model of the controlled system. Despite the possible complexities in the control of nonaffine systems, the following design would be more effective in the case of nonconventional air vehicles with highly nonlinear dynamics, which could not be modeled as an affine model, satisfactorily. In particular, such a method could be an optimal choice in the case of an HFV, which possesses a completely nonaffine model \cite{Bu.2018}. Indeed, although, using some simplifications, HFVs are typically modeled as an approximate affine model (and the remaining nonlinear terms are treated as model uncertainty) to facilitate the control design, such an approach results in a conservative control system.
It should be noted that, in the case of a flight control problem, the dynamic model of the system is typically formulated as $\ddot{x}=F(\dot{x},x,u)$.
However, the following design can be applied to such a control problem, as well, using a simple change of variables and employing a composite error function consisting of both $e$ and $\dot{e}$.
Now, assuming the availability of an approximate inversion model, the control command can be computed as follows:
\begin{equation}\label{Eq:16_01}
  u=\hat{F}^{-1}(x,\nu),
\end{equation}
where $\nu$ denotes the pseudocontrol input, which should be designed. Notice that, although there is no need for an accurate inversion model, the chosen inversion model should capture the control assignment structures. It means that, for example, the inversion model should include the fact that the elevator deflection affects the pitch rate.
In addition, it is assumed that $\hat{F}^{-1}(x,\nu)$ is a one-to-one function. This assumption can be realized if $dim(u)=dim(x)$ \cite{Chowdhary.2013}, which is reasonable in a typical flight control problem.
Accordingly, the pseudocontrol input $\nu$ can be designed as follows \cite{Lee.2005}:
\begin{equation}\label{Eq:16_02}
  \nu = \dot{x}_d - k e - \nu_{ad},
\end{equation}
where $k$ is a positive constant and $\nu_{ad}$ denotes an additional command to alleviate the inversion error. More precisely, defining $\Delta(x,u)= F(x,u) - \hat{F}(x,u)$, we have:
\begin{equation}
  \dot{e}= - k e - \nu_{ad} + \Delta(x,u).
\end{equation}
Thus, if it is possible to have $\nu_{ad} = \Delta(x,u)$, the tracking error converges asymptotically to zero. However, $\Delta(x,u)$ is unknown. So, we estimate it using the feedback error learning scheme. In this regard, using a NN to identify $\Delta(x,u)$, we have $\Delta(x,u)={W^*}^T \mu (x,u) + \varepsilon$. Subsequently, $\nu_{ad}$ can be determined as $\nu_{ad} =\hat{W}^T \mu (x,u)$. Introducing a Lyapunov function $V$ as
\begin{equation}
  V=\frac{1}{2}e^T e + \frac{1}{2} tr \left(\tilde{W}^T \Gamma^{-1} \tilde{W}\right),
\end{equation}
and using the updating rule $\dot{\tilde{W}}\approx \dot{\hat{W}}=\Gamma \mu (x,u) e^T$, the time derivative of $V$ is obtained as the following equation.
\begin{align}
\begin{split}
\dot{V} & = - e^T \left(k e + \tilde{W}^T \mu (x,u) - \varepsilon\right)+ tr \left(\tilde{W}^T \Gamma^{-1} \dot{\tilde{W}}\right) \\
 & = - e^T \left(k e - \varepsilon \right).
\end{split}
\end{align}
Accordingly, the introduced control strategy can satisfactorily ensure the bounded tracking error. Again, one of the above-mentioned modification techniques can be adopted in the proposed updating rule to prevent parameter drift.
Notice that the proposed control framework results in a control law as $u=\hat{F}^{-1}\left(x,\dot{x}_d - k e - \hat{W}^T \mu (x,u)\right)$, thereby requiring the contraction assumption. A modification to the introduced strategy has been given in \cite{Bu.2018} by taking advantage of the Mean Value theorem to relax this assumption, while the sign of $\partial F/\partial u$ should be known, and there are some concerns with the provided stability analysis.
Besides, authors in \cite{Bu.2018} have employed the pseudocontrol approach in the case of a SISO system in the normal feedback form with $dim(x)>dim(u)$, while there is a requirement for the availability of all the system states in the proposed control scheme. To this end, if we have $z_1=e=x_1-x_{1d}$, $z_2=\dot{z}_1$, and $\ddot{x}_1=f(x,u)$, one can define a filtered tracking error $s=\dot{e}+\lambda e$, which results in $\dot{s}=f(x,u)-\ddot{x}_{1d}+\lambda \dot{e}$. Thus, by replacing the real tracking error $e$ with the filtered tracking error $s$ in the introduced method, it is possible to design a similar pseudocontrol framework.

The pseudocontrol strategy has been employed in different flight control systems \cite{Rahideh.2012} such as the attitude control of a tailless fighter aircraft \cite{Calise.2000,Brinker.2000,Calise.2001,Brinker.2001}, the trajectory tracking control of a helicopter \cite{Lee.2005}, the attitude control of a tilt-rotor aircraft \cite{Rysdyk.2005}, etc.
A similar direct adaptive control has been utilized in \cite{NhanNguyen.2006,Nguyen.2008} to control the trajectory of a conventional fixed-wing aircraft under structural damages. A hybrid direct-indirect adaptive control has also been developed in them in which parallel FEL algorithms attempted to provide both the control augmentation signal and estimated uncertain dynamics.
In \cite{Chowdhary.2013}, an inner loop attitude control block based on the pseudocontrol strategy has been employed within a fault-tolerant guidance and control system for a conventional fixed-wing air vehicle. An acceleration (outer loop) guidance loop has been designed, which attempts to provide feasible acceleration command in the presence of structural damages and actuator faults. The performance of the proposed approach was verified in the presence of severe structural and actuator damages.

As an alternative to the above-mentioned approach to control nonaffine systems, authors in \cite{KengPengTee.2008} attempted to directly estimate the desired control command rather than the inversion model error using a NN (in the case of a SISO system with stable zero dynamics). Under conservative assumptions on the value of $\partial F/\partial u$ (and its time derivative) and employing the implicit function theorem, one can assume that there is an ideal control command $u^*$ which ensures the closed-loop system stability, i.e. $F(x,u^*)=\dot{x}_d- k e$.
Subsequently, in lieu of utilizing an approximate inversion model, the Mean Value theorem has been adopted to provide an expression for $F(x,u)$ in terms of $F(x,u^*)$. Using such a formulation, a NN with a typical FEL scheme can be employed to estimate $u^*$. Although in this method, there is no need for an approximate inversion model of the system, different restrictive assumptions are required in the control design, which may not be satisfied in a practical flight control problem.
A somewhat similar approach has been utilized in \cite{DaoxiangGao.2014} in the framework of indirect adaptive control, where the singular perturbation theory has been adopted to move $u$ towards $u^*$ as $\epsilon \dot{u}= F(x,u^*)-F(x,u)$ with $\epsilon$ denotes a small positive constant. Such a method has been employed to control the longitudinal model of an HFV, where a set of NNs has been incorporated to estimate unknown dynamics. In this regard, there is no need for a strict feedback model and a backstepping design, while, again, restrictive assumptions should be made to ensure closed-loop stability.

\subsection{Neural backstepping control} \label{Sec:Backstepping}
In the basic FEL scheme, it was assumed that $dim(x)=dim(u)$. Also, in both the above-mentioned control structures, the entire dynamic model of the system is assumed invertible. However, in many cases, the dimension of the system inputs is less than that of the system states. The backstepping control method can be effectively employed in such circumstances when the dynamic model can be formulated in a strict feedback form. For simplicity, consider an uncertain nonlinear SISO system as follows:
\begin{gather}
  \dot{x}_i=f_i(\bar{x}_i)+g_i(\bar{x}_i)x_{i+1}+\bar{\Delta}_i+d_i, \quad 1\leq i \leq n-1, \label{Eq:16_1}\\
  \dot{x}_n=f_n(\bar{x}_n)+g_n(\bar{x}_n)u+\bar{\Delta}_n+d_n, \label{Eq:16_2}\\
  y = x_1 \label{Eq:16_3},
\end{gather}
where, $\bar{\Delta}_i$ and $d_i$ stand, respectively, for model uncertainties and external disturbances and $\bar{x}_i=[x_1,\ldots,x_i]^T$.
Without loss of generality, in the following, we assume that $n=2$. The introduced control method can be simply applied to higher-order systems. Defining $\Delta_i=\bar{\Delta}_i+d_i$ and the desired output as $y_d$, we have:
\begin{equation}
  \dot{e}_1=\dot{y}-\dot{y}_d = f_1(\bar{x}_1)+g_1(\bar{x}_1)x_{2}+\Delta_1 - \dot{y}_d.
\end{equation}
Thus, a virtual control can be defined for $x_2$ as
\begin{equation}\label{Eq:17}
  x_{2d}=g_1^{-1}(\bar{x}_1)\left(\dot{y}_d - k_1 e_1 - f_1(\bar{x}_1) - \hat{\Delta}_1 \right),
\end{equation}
where $\hat{\Delta}_1$ represent the estimation of $\Delta_1$ and $k_1$ is a positive constant. Therefore, defining $e_2=x_2-x_{2d}$, we have:
\begin{equation}
  \dot{e}_2 = \dot{x}_2-\dot{x}_{2d}= f_2(\bar{x}_2)+g_2(\bar{x}_2)u+ {\Delta}_2 - \dot{x}_{2d}.
\end{equation}
Finally, the control command can be defined as
\begin{equation} \label{Eq:18}
  u = g_2^{-1}(\bar{x}_2)\left(\dot{x}_{2d} - g_1(\bar{x}_1) e_1 - k_2 e_2 - f_2(\bar{x}_2) - \hat{\Delta}_2 \right),
\end{equation}
where $\hat{\Delta}_2$ denotes the estimation of $\Delta_2$ and $k_2$ is a positive constant.
Using feedforward NNs to estimate $\Delta_i$s, it is obtained that:
\begin{equation}\label{Eq:18_1}
  \Delta_i={W^*}_i^T \mu_i (\bar{x}_i) + \varepsilon_i,
\end{equation}
such that $\|\varepsilon_i\|\leq \varepsilon_{M_i}$. To derive the updating rules of the NNs' parameters, one can define a Lyapunov function as follows:
\begin{equation}
  V = \frac{1}{2}\left(e_1^2+e_2^2 + \tilde{W}_1^T \Gamma_1^{-1} \tilde{W}_1 + \tilde{W}_2^T \Gamma_2^{-1} \tilde{W}_2\right). \label{Eq:19_0}
\end{equation}
The time derivative of $V$ is obtained as
\begin{align}
\begin{split}
   \dot{V} = & e_1 \dot{e}_1 + e_2 \dot{e}_2 + \tilde{W}_1^T \Gamma_1^{-1} \dot{\tilde{W}}_1 + \tilde{W}_2^T \Gamma_2^{-1} \dot{\tilde{W}}_2 \\
     = & e_1 \left(g_1(\bar{x}_1)e_2 - \tilde{\Delta}_1 - k_1 e_1 \right) + e_2 \left(- g_1(\bar{x}_1) e_1 \right. \\
     & \left. - \tilde{\Delta}_2 - k_2 e_2 \right) + \tilde{W}_1^T \Gamma_1^{-1} \dot{\tilde{W}}_1 + \tilde{W}_2^T \Gamma_2^{-1} \dot{\tilde{W}}_2 \\
     = & - k_1 e_1^2 - k_2 e_2^2 + \tilde{W}_1^T \left( \Gamma_1^{-1} \dot{\tilde{W}}_1 - \mu_1 (x_1) e_1 \right) \\
     & + \tilde{W}_2^T \left( \Gamma_2^{-1} \dot{\tilde{W}}_2 - \mu_2 (\bar{x}_2) e_2 \right) + e_1 \varepsilon_1 + e_2 \varepsilon_2.
\end{split}
\end{align}
Thus, assuming $\dot{\tilde{W}}_i=\dot{\hat{W}}_i$, the updating rules of $W_i$s can be defined as follows:
\begin{equation}
  \dot{\hat{W}}_i = \Gamma_i \left( \mu_i (\bar{x}_i) e_i - \sigma_i \hat{W}_i \right), \label{Eq:19}
\end{equation}
where the second term on the right-hand side of the equation corresponds to the $\sigma$-modification. Using the updating rules (\ref{Eq:19}), it is easy to show that $\dot{V}\leq -k V + C$ with $k$ and $C$ denote positive constant. As will be discussed in Section \ref{Sec:Stability}, this can ensure that all signals in the closed-loop system are uniformly ultimately bounded.

Such a control method can similarly be employed in cases where $x_i$s are some vectors rather than scalar functions. A neural backstepping controller for an uncertain MIMO dynamic model of a helicopter has been introduced in \cite{Ge.2010} to control the attitude of the vehicle considering actuator dynamics, while each step of the design process deals with the control of a four-dimensional state vector.
A neural backstepping controller has been adopted in \cite{Zheng.2020b} to control a planar VTOL air vehicle, where a gradient descent training algorithm has been replaced the updating rule (\ref{Eq:19}).
Although it has been claimed that such a training method results in better control performance, there is a need for the exact value of the uncertain term that is estimated by the NN, while in this paper, it has been computed by approximating the time derivatives of the system states and using the dynamic equations of the air vehicle.

It should be noted that although the proposed adaptive backstepping control leads to bounded tracking error in the presence of model uncertainties and external disturbances, it suffers from the \textit{explosion of terms}. More precisely, the control command (\ref{Eq:18}) includes the time derivative of $\dot{x}_{2d}$, which requires the time derivative of $g_1(x_1)$, $f_1(x_1)$, and $\hat{\Delta}_1$. This issue becomes more problematic by increasing the relative degree of the system.

\subsubsection{Dynamic surface control}
To solve the above-mentioned issue, Dynamic Surface Control (DSC) has been introduced in \cite{Swaroop.2000} in which the virtual control is passed through a first-order filter. More precisely, if $x_{2c}$ is defined by (\ref{Eq:17}), then the desired value of $x_2$ is obtained as
\begin{equation}\label{Eq:19_1}
  \tau \dot{x}_{2d}+x_{2d}=x_{2c}, \quad x_{2d}(0) = x_{2c}(0),
\end{equation}
where $\tau$ represents the filter time constant. Subsequently, the filtering error is also incorporated in the Lyapunov function of the system to be compensated by the designed control commands.
Using such a technique, the problem of the explosion of terms in the traditional backstepping control can be effectively avoided, though at the cost of reducing the \emph{global} stability of the system obtained using the backstepping control to the \emph{semi-global} stability in the case of DSC \cite{Swaroop.2000}.

Several NN-based DSC methods have been introduced in the literature for different aerial vehicles \cite{Butt.2013,Zong.2014,Xu.2016d,Fu.2018}.
Such an approach has been proposed in \cite{Butt.2013b} to control the flight path angle and velocity of a flexible HFV, where the employment of the integral of the tracking error in the control law improves the tracking performance.
DSC has been employed in \cite{MouChen.2014} to control the attitude of a Near-Space Vehicle (NSV) in which recurrent wavelet NNs have been utilized at each step and trained using a composite learning method to compensate for external disturbances and model uncertainties.
Also, such a scheme has been adopted in \cite{Zhou.2015} to control the longitudinal dynamics of an air-breathing HFV considering model uncertainties and external disturbances compensated by fully tuned RBFNNs.
In addition, DSC has been applied to the longitudinal mode of an HFV in \cite{Xu.2015}. In comparison with conventional DSC, which results in a semi-globally uniformly ultimately bounded stability, global tracking has been achieved through aggregating the neural function approximation and a robust term (using a switching function), which brings the system states into the neural approximation domain from outside. The robust term has been designed to estimate the upper bound of uncertain terms in a similar way as discussed in Section \ref{Sec:CombinedNNDO}. However, the determination of the \emph{active region} of NNs (which is required in designing the switching functions \cite{Xu.2021b,Xu.2021}) is not trivial.

\subsubsection{Command filtered backstepping} \label{Sec:ComFilBackstepping}
To simplify the stability analysis of DSC, a command filtered backstepping has been proposed in \cite{Farrell.2009} for a nonlinear system without uncertainty. The introduced method attempts to eliminate the filter effects using a set of compensating signals. This idea has been extended to nonlinear systems with parametric uncertainties in \cite{Dong.2012}. To clarify the main idea, consider again the aforementioned control problem. Assuming that the virtual control signals $x_{2c}$ and $x_{2d}$ are defined, respectively, by (\ref{Eq:17}) and (\ref{Eq:19_1}), and by defining the auxiliary variable $\xi_1$ as
\begin{equation}
  \dot{\xi}_1=-k_1 \xi_1 + g_1(\bar{x}_1)\left(x_{2d}-x_{2c}\right), \quad \xi_1(0)=0,
\end{equation}
a compensated tracking error can be defined as $\epsilon_1=y-y_{d}-\xi_1$. Thus, we have:
\begin{align}
\begin{split}
  \dot{\epsilon}_1 & = \dot{y}-\dot{y}_{d}-\dot{\xi}_1 \\
                   & =  f_1(\bar{x}_1)+g_1(\bar{x}_1)x_{2}+\Delta_1 - \dot{y}_d + k_1 \xi_1 - g_1(\bar{x}_1)\left(x_{2d}-x_{2c}\right) \\
                   & = - k_1 \epsilon_1 + g_1(\bar{x}_1) e_2 - \tilde{\Delta}_1.
\end{split}
\end{align}
Accordingly, the control command can be defined as
\begin{equation}
    u = g_2^{-1}(\bar{x}_2)\left(\dot{x}_{2d} - g_1(\bar{x}_1) \epsilon_1 - k_2 e_2 - f_2(\bar{x}_2) - \hat{\Delta}_2 \right),
\end{equation}
which leads to
\begin{align}
\begin{split}
  \dot{e}_2= \dot{x}_2-\dot{x}_{2d}= - g_1(\bar{x}_1)\epsilon_1 - k_2 e_2 - \tilde{\Delta}_2.
\end{split}
\end{align}
Thus, using the following updating rules
\begin{gather}
  \dot{\hat{W}}_1 = \Gamma_1 \left( \mu_1 (\bar{x}_1) \epsilon_1 - \sigma_1 \hat{W}_1 \right), \\
  \dot{\hat{W}}_2 = \Gamma_2 \left( \mu_2 (\bar{x}_2) e_2 - \sigma_2 \hat{W}_2 \right),
\end{gather}
and defining a Lyapunov function as
\begin{equation}
  V = \frac{1}{2}\left(\epsilon_1^2+e_2^2 + \tilde{W}_1^T \Gamma_1^{-1} \tilde{W}_1 + \tilde{W}_2^T \Gamma_2^{-1} \tilde{W}_2\right),
\end{equation}
it can be shown that $\dot{V}\leq -k V + C$, where $k$ and $C$ are positive constants. This results in bounded $\epsilon_1$ and $e_2$.
As discussed in \cite{Dong.2012,Xu.2016c}, assuming that $g_1(x_1)$ is bounded, it can be simply proved that $\xi_1$ is also bounded, thereby resulting in a bounded tracking error.
A command filtered backstepping control has been designed in \cite{Xu.2016c} for the longitudinal dynamics of an HFV considering input constraints and additive actuator faults. The control gain functions ($g_i$) have also been considered unknown, where it has been assumed that model uncertainties, as well as the control gain functions, can be written into a parametric form with partially unknown parameters. Considering the neural network-based representation, the above assumption means that the residual terms $\varepsilon_i$ in (\ref{Eq:18_1}) are equal to zero, while in the case of complex air vehicles with nonparametric uncertainties, such an assumption becomes infeasible.
Similarly, a command filtered backstepping control has been adopted in \cite{Sonneveldt.2007,Sonneveldt.2009} to control the trajectory of an F-16 fighter aircraft model with parametric uncertainties, where second-order filters have been used to impose both the magnitude and rate limits on the system states (see Section \ref{Sec:InputConst1}).
Another analogous formulation has been presented in \cite{Xu.2014b} in which the time derivative of ${\xi}_i$ consists of ${\xi}_{i+1}$, where $i$ represents the step of the backstepping control design process. Using this formulation, the control command $u$ can be written in terms of $e_1$ rather than $\epsilon_1$.

\subsubsection{Backstepping augmented by the First-Order Sliding Mode Differentiators (FOSMD)}
Another improved approach to approximate the time derivative of the virtual control signal $x_{2d}$ is to employ a first-order sliding mode differentiator rather than employing a first-order filter. Using the FOSMD, the differentiation error tends to zero or a compact neighborhood of zero (depending on the signal's characteristics) after a finite-time transient process \cite{Levant.1998}.
Considering a known function $l(t)$, the FOSMD formulation is obtained as follows:
\begin{gather}
  \dot{\varsigma}_0=-\varrho_0 \left|\varsigma_0-l(t)\right|^{0.5} sign\left(\varsigma_0-l(t)\right)+\varsigma_1, \\
  \dot{\varsigma}_1=-\varrho_1 sign\left(\varsigma_1-\dot{\varsigma_0}\right),
\end{gather}
where $\varsigma_0$ and $\varsigma_1$ represent the states of the differentiator, and $\varrho_0$ and $\varrho_1$ denote design parameters. Therefore, $\dot{\varsigma}_0-\dot{l}(t)$ remains bounded if $\dot{\varsigma}_0(0)-\dot{l}(0)$ and $\varsigma_0(0)-l(0)$ are bounded.

This approach has been adopted in the backstepping control design in \cite{Xu.2018,Xu.2019} to control the longitudinal mode of an HFV.
As shown in \cite{Xu.2018}, using the FOSMD, the stability analysis is more concise compared to the traditional backstepping, DSC, and command filtered design. Besides, a neural backstepping control approach using FOSMD has been proposed in \cite{Wu.2017} for the longitudinal dynamic model of a sweep-back wings morphing aircraft subject to input–output constraints.
It is notable that higher-order sliding mode differentiators (HOSMD), which result in superior performance compared to FOSMDs \cite{Levant.2003}, can also be employed in the structure of the neural backstepping scheme \cite{Yu.2020}.

\subsubsection{Direct neural-backstepping control}\label{Sec:DirectNeuralBackstepping}
In addition to the above techniques, there are a variety of direct adaptive backstepping flight control systems in the literature, which can satisfactorily prevent the problem of the explosion of terms.
To be more precise, consider again the nonlinear model (\ref{Eq:16_1})-(\ref{Eq:16_3}) with $n=2$. Defining
\begin{gather}
  x_{2d}^{*}=g_1^{-1}(\bar{x}_1)\left(\dot{y}_d - k_1 e_1 - f_1(\bar{x}_1) - \Delta_1 \right), \\
  u^*= g_2^{-1}(\bar{x}_2)\left(\dot{x}_{2d} - g_1(\bar{x}_1) e_1 - k_2 e_2 - f_2(\bar{x}_2) - \Delta_2 \right), \label{Eq:20_0}
\end{gather}
and using two distinct neural networks to identify them as $x_{2d}^{*}={W^*}_1^T \mu_1(\bar{x}_1)+\varepsilon_1$ and $u^{*}={W^*}_2^T \mu_2(\bar{x}_2)+\varepsilon_2$, we have:
\begin{gather}
 x_{2d}={\hat{W}}_1^T \mu_1(\bar{x}_1)=x_{2d}^{*}-\varepsilon_1+{\tilde{W}}_1^T \mu_1(\bar{x}_1), \\
 u={\hat{W}}_2^T \mu_2(\bar{x}_2)=u^*-\varepsilon_2+{\tilde{W}}_2^T \mu_2(\bar{x}_2). \label{Eq:20_01}
\end{gather}

Thus, considering a Lyapunov function candidate as (\ref{Eq:19_0}), we have:
\begin{align}
\begin{split}
   \dot{V} = & \quad e_1 \dot{e}_1 + e_2 \dot{e}_2 + \tilde{W}_1^T \Gamma_1^{-1} \dot{\tilde{W}}_1 + \tilde{W}_2^T \Gamma_2^{-1} \dot{\tilde{W}}_2 \\
     = & \quad e_1 \left(g_1(\bar{x}_1)\left(e_2-\varepsilon_1+{\tilde{W}}_1^T \mu_1(\bar{x}_1)\right) - k_1 e_1 \right) + \\
     & e_2 \left(- g_1(\bar{x}_1) e_1 - k_2 e_2 + g_2(\bar{x}_2)\left({\tilde{W}}_2^T \mu_2(\bar{x}_2)-\varepsilon_2\right)\right) \\
     & + \tilde{W}_1^T \Gamma_1^{-1} \dot{\tilde{W}}_1 + \tilde{W}_2^T \Gamma_2^{-1} \dot{\tilde{W}}_2 \\
     = & - k_1 e_1 \left(e_1+\frac{g_1(\bar{x}_1)}{k_1}\varepsilon_1 \right) - k_2 e_2 \left(e_2+\frac{g_2(\bar{x}_2)}{k_2}\varepsilon_2 \right) \\ & + \tilde{W}_1^T \left( \Gamma_1^{-1} \dot{\tilde{W}}_1 + \mu_1 (x_1) g_1(\bar{x}_1) e_1 \right) \\
        & + \tilde{W}_2^T \left( \Gamma_2^{-1} \dot{\tilde{W}}_2 + \mu_2 (\bar{x}_2) g_2(\bar{x}_2) e_2 \right).
\end{split}
\end{align}

Accordingly, by introducing the following updating rules,
\begin{gather}
  \dot{\hat{W}}_i = - \Gamma_i \left(\mu_i (\bar{x}_i) g_i(\bar{x}_i) e_i + \sigma_i \hat{W}_i\right), \quad i=1,2, \label{Eq:20_1}
\end{gather}
and assuming that $g_i$s are nonzero and \emph{bounded}, again, it can be concluded that $\dot{V}<-kV+C$, which ensures that all signals in the closed-loop system remain bounded (see Section \ref{Sec:Stability}).
As seen, despite the simpler formulation of the direct method compared to previously proposed indirect backstepping schemes, the boundedness of control gain functions ($g_i$s) is necessary to guarantee the closed-loop stability.


It is notable that, using the aforementioned direct neural backstepping scheme, the control \emph{singularity} problem in the control of dynamic systems with unknown $g_i$s (induced by approaching $\hat{g}_i$s to zero) is also avoided \cite{Bu.2015}.
A direct neural backstepping control has been designed in \cite{Bu.2015} to control the longitudinal mode of an air-breathing HFV with unknown $g_i$s, where it is necessary to have $\overline{g}_i \geq g_i>0$ ($\overline{g}_i$ denotes a positive constant).
To this end, concerning the appropriate Lyapunov function candidate, $1/g_i$ is multiplied by $1/2 e_i^2$ to eliminate the requirement for $g_i$ in updating rules, and extra terms in $\dot{V}$ raised by this reformulation have been compensated by defining an appropriate ideal control command ($u^*$).
Besides, a filtered tracking error including the integral of the error has been considered as the error function to remove the steady-state error, while the tracking error corresponding to the second step of the backstepping design was not considered in the first step.

On the other hand, by introducing an output feedback form and utilizing High Gain Observers (HGOs) to estimate the time derivatives of the system output (Section \ref{Sec:OFB}), it is possible to derive the control command with no requirement for a backstepping scheme.  In this regard, a filtered tracking error is defined (as discussed in Section \ref{Sec:PseudoControl}) to provide a unified error dynamic model. Such a method has been adopted in \cite{Xu.2011,Xu.2015d} to control an HFV, where only one neural network is required in the altitude control block to determine the actual control command.
In a somewhat similar manner, a direct neuroadaptive control scheme has been integrated with the funnel control method in \cite{Bu.2018b} to control an air-breathing HFV considering non-affine dynamics. The altitude subsystem has been transformed into a simplified normal output feedback model, where only one NN is required to determine the control command.
Further, the non-affine dynamics of the vehicle have been handled by incorporating a low-pass filter in the last step of the design to define a new virtual control input in affine form.

In addition to the above-mentioned continuous-time backstepping control methods, several studies in the literature have addressed the design of a discrete-time neural backstepping controller. As discussed in Section \ref{Sec:FEL}, the general formulation of the NN updating rules in the discrete-time domain is similar to that of the continuous-time controller, while the stability analysis of the closed-loop system is quite different \cite{Xu.2011b}. A discrete-time direct neural backstepping control has been proposed in \cite{Xu.2012b} by incorporating HONNs to estimate uncertain terms in control commands. A similar control formulation has been given in \cite{Xu.2014} using Extreme Learning Machines (ELMs). To simplify the control structure, in \cite{Xu.2013,Xu.2013b}, dynamic equations corresponding to the altitude dynamics of an HFV have been aggregated into a prediction model as $x_1(k+n)= \bar{f}(x)+ \bar{g}(x) u$, where $\bar{f}$ and $\bar{g}$ represent uncertain nonlinear functions and $n$ denotes the system order. Subsequently, a single NN has been employed to tackle uncertain terms in the control command.
Such a method can be considered as the discrete-time equivalent of the above-mentioned approach to control output feedback models in the continuous-time domain.
Alternatively, an equivalent prediction model of an HFV has been defined in \cite{Xu.2015c} in which $x_i(k+n-i+1)$ is obtained as a function of $x_{i+1}(k+n-i)$. Using such a change in the formulation of system dynamics, all the information of the desired trajectory in future $n$ steps is involved in designing the controller, thereby improving the closed-loop performance. The designed controller in all these papers has been applied to the longitudinal mode of an HFV model.

\subsection{How to analyze the closed-loop stability?} \label{Sec:Stability}
As mentioned before, most of the current feedback error learning-neural control schemes in the literature can only guarantee the Uniformly Ultimately Bounded (UUB) stability of the closed-loop system.
To be more precise, it is not typically possible to prove the negative definiteness of the time derivative of the Lyapunov function, but it can be proved that
\begin{equation}\label{Lyap_3}
  \dot{V}\leq -k V + C,
\end{equation}
where $k$ and $C$ denote positive constants. As a result, it is obtained that \cite{Wang.2014b,Ge.2010}
\begin{equation}\label{Lyap_4}
  V (t) \leq \left(V(0)-\frac{C}{k}\right)e^{-k t}+\frac{C}{k}.
\end{equation}
In this regard, many studies in the literature attempted to propose a flight control system, which satisfies the local \cite{Lee.2001,Shin.2004}, semi-global \cite{Xu.2011,Xian.2015,Chen.2016,Ge.2010,Chen.2015c}, or global \cite{Xu.2015} UUB tracking.
There are fewer works that have addressed more stringent stability criteria including asymptotic, exponential, or finite-time stability.
\subsubsection{Asymptotic stability}\label{Sec:AsymptoticStability}
A variety of flight control systems are given in the literature, which can prove the convergence of the tracking error to zero as time tends to infinity.
As a straightforward approach, if we can assume that the estimation error $\varepsilon$ in (\ref{Lyap_2}) is zero (or negligible), it is obtained that $\dot{V}=-k_1 e^T e$, which guarantees the asymptotic convergence of the tracking error to zero. Such an assumption is reasonable in the case of dynamic systems with \emph{parametric} uncertainties. More precisely, in such a circumstance, it is possible to estimate uncertain terms as $\Delta(x)={W^*}^T \mu(x)$, where $\mu(x)$ and $W^*$ represent the vector of appropriate basis functions and the matrix of unknown weights, respectively.
According to this, a constrained adaptive backstepping control scheme has been proposed in \cite{Sonneveldt.2007} for a strict feedback system with parametric uncertainties. A set of modified tracking errors ($\bar{z}_i$) have been defined (Section \ref{Sec:InputConst1}), and it has been proved that the time derivative of the Lyapunov function is obtained as $\dot{V}=-\sum c_i \bar{z}_i^2$ with $c_i$ denotes positive constants. This leads to convergence of the modified tracking errors to zero as time tends to infinity \cite{Krstic.1992}, while the actual tracking error may increase if the control inputs are saturated. Finally, the proposed control scheme was employed in \cite{Sonneveldt.2007} to control the attitude of a simplified model of an F-16 aircraft with multi-axis thrust vectoring, considering actuator faults and symmetric structural damages (only in the simulation phase).
A similar approach has been presented in \cite{Sonneveldt.2009} to provide a trajectory tracking control for an F-16 fighter aircraft under parametric uncertainties and system constraints. An adaptive backstepping controller has been introduced in \cite{Shi.2014} for the attitude control of an NSV in the presence of model uncertainties and multiplicative actuator faults. For this purpose, first, an adaptive neural state observer has been proposed (Section \ref{Sec:NNDO}), and subsequently, the estimated states have been utilized in a backstepping control scheme. The asymptotic stability of the system has been proved assuming that the NN estimation error is negligible.
Obviously, such control approaches can not ensure the asymptotic stability of the system in the presence of \emph{nonparametric} uncertainties and external disturbances, which are usually present in practical flight control problems.

Several studies have been reported in the literature in which NNs are combined with \emph{discontinuous} feedback control methods such as variable structure or Sliding Mode Controllers (SMC) to guarantee the asymptotic stability of the closed-loop system. The fundamental idea of the combination of NN function approximation and robust terms (such as in SMC), which can result in the asymptotic stability of the closed-loop system, is given in Section \ref{Sec:CombinedNNDO}.
A robust output feedback control with neural network function approximation has been designed in \cite{Xian.2015} for the attitude and altitude control of a quadrotor UAV in the presence of model uncertainties and external disturbances, where the attitude dynamics are constructed in terms of the unit quaternion. Although the asymptotic stability of the closed-loop system has been proved, the proposed control command leads to high-amplitude and oscillatory thrust forces, and the chattering phenomenon due to the employment of the signum function in the control command.
Indeed, such discontinuous controllers suffer from well-known limitations including a requirement for an infinite control bandwidth and chattering. Unfortunately, ad hoc fixes for these effects result in a loss of asymptotic stability \cite{Patre.2008}.
An adaptive SMC has been proposed in \cite{Razmi.2019}, where the parameters of the sliding surface were trained by NNs through error back-propagation learning. The hyperbolic tangent function has replaced the signum function to eliminate the chattering phenomenon, while the asymptotic stability of the system has been achieved by neglecting model uncertainties in the control design process.

To overcome the above-mentioned issues, a trajectory tracking control system has been introduced in \cite{JonghoShin.2012} for a rotorcraft UAV using Robust Integral of the Signum of the Error (RISE) feedback \cite{Patre.2008}, where a NN has been adopted to compensate for uncertain dynamics. The RISE control scheme is a differentiable control method that can compensate for additive disturbances and parametric uncertainties. By combining it with an NN-based FEL method, there is no need for linearity in the parameters to ensure the asymptotic stability of the system.
The proposed approach was employed in \cite{JonghoShin.2012} in a multi-loop control structure, where the desired attitude is determined in the outer loop and the attitude tracking control has been addressed in the inner loop. The semi-global asymptotic stability of the inner loop in tracking the desired attitude has been proved assuming that the first four time derivatives of the reference trajectory are bounded.
Another alternative has been given in \cite{Wang.2014b,Li.2020}, where the signum function of $e$ has been substituted by $e/\left(\sqrt{e^2+\omega^2}\right)$ with $\omega(t)$ denotes a vanishing positive function satisfying $\int_{0}^{\infty}\omega^2(t) dt<\infty$. Accordingly, asymptotic tracking can be achieved, while the updating rules are subject to possible parameter drift.
A class of NN-based optimal control methods with guaranteed asymptotic stability has also been introduced in the literature, which will be addressed in Section \ref{Sec:Optimal}.

\subsubsection{Exponential stability}
As stated in \cite{Mo.2018}, in the case of a flight control problem, few studies have claimed to achieve exponential convergence \cite{Zhang.2018c}. This is even less so regarding uncertain dynamic systems. As mentioned earlier, an improved learning method has been introduced in \cite{Chowdhary.2014}, which can ensure the exponential parameter and tracking error convergence for a specific class of single-input nonlinear systems with parametric uncertainties assuming that a precise estimation of the time derivative of system states is available.

\subsubsection{Finite-time stability}
Moreover, a variety of indirect NN-based flight control systems has been developed in the literature employing SMC-based methods, which can guarantee the finite-time or \emph{practical finite-time} stability of the closed-loop system. The practical finite-time stability means that the tracking error converges to a small neighborhood of the origin in finite time \cite{Zhu.2011}.
To clarify the principal idea of the mentioned control methods, consider again the first control problem given in Section \ref{Sec:FEL}. Let's define a sliding manifold as
\begin{equation}\label{Eq:20_10}
  s=e+\eta \int_{0}^{t} sig^r(e) dt,
\end{equation}
where $sig^r(e)=[sign(e_1)|e_1|^r,\cdots,sign(e_n)|e_n|^r]^T$, $\eta>0$, and $0<r<1$. Now, if the control command is computed as
\begin{align}
\begin{split}
     u= B^{-1}  \big(\dot{x}_d-F(x)-\eta sig^r(e)-\hat{\Delta}(x)
                       -k_1 s -k_2 sig^r(s) \big), \label{Eq:20_11}
\end{split}
\end{align}
where the same NN function approximation as (\ref{NN_1}) with an updating rule as $\dot{\hat{W}}=\Gamma \left(\mu s^T-\sigma \hat{W}\right)$ is incorporated, using a Lyapunov function as
\begin{equation}\label{Eq:20_12}
  V=\frac{1}{2}s^T s + \frac{1}{2} tr \left(\tilde{W}^T \Gamma^{-1} \tilde{W}\right),
\end{equation}
it is easy to prove the satisfaction of (\ref{Lyap_3}), thereby guaranteeing the boundedness of all signals in the closed-loop system. As a consequence, we have
\begin{equation}\label{Eq:20_13}
  \dot{s}\leq -k_1 s - k_2 sig^r(s) + \rho_M I
\end{equation}
where $I$ and $\rho_M$ denote the identity matrix and a positive constant satisfying
\begin{equation}\label{Eq:20_14}
  \left\|\varepsilon-\tilde{W}^T\mu(x)\right\|_\infty \leq \rho_M.
\end{equation}
Thus, by appropriate choice of $k_1$ and $k_2$, one can simply show the convergence of $s$ to a compact neighborhood of the origin in finite time \cite{Xu.2018b}.
Such a method is extensively discussed within the framework of adaptive Terminal SMC (TSMC).
Further, by incorporating a robust term into the control command (in a similar manner as discussed in Section \ref{Sec:CombinedNNDO}) to compensate for $\rho_M$, and replacing the first term in the Lyapunov function (\ref{Eq:20_12}) by $\|s\|$, it is possible to guarantee the finite-time stability of the closed-loop system \cite{Yu.2018}.
An FTC has been introduced in \cite{Yu.2018} to control the longitudinal mode of a conventional aircraft using a similar SMC, which ensures the finite-time convergence of $s$ to zero under appropriate control gains. Self-constructing fuzzy neural networks have been utilized to estimate the bound of uncertain dynamics caused by actuator faults and model uncertainties, while the minimum estimation error $\varepsilon$ under optimal network weights ($W^*$) has been neglected.
A TSMC augmented by neural approximation and disturbance observers (as given in Section \ref{Sec:CombinedNNDO}) has been presented in \cite{Xu.2018b} for the trajectory tracking control of a quadrotor aerial robot under model uncertainties, input dead-zone, and external disturbances, where the practical finite-time stability of the system has been ensured. A sliding surface, which is equal to the time derivative of (\ref{Eq:20_10}) due to the second-order dynamics of the controlled system, has been utilized in the proposed design.
A formation flight control problem for a group of helicopter UAVs has been dealt with in \cite{Wang.2018b} using an analogous control scheme. TSMC combined with neural approximation and the above-mentioned robust term has been adopted in both the position and attitude control loops, which can ensure, respectively, the finite-time and practical finite-time stability of the position and attitude tracking error, while the control loops have been decoupled based on the multiple time-scale assumption.
The same problem has been addressed in \cite{Wang.2019c} using TSMC in the inner control loop and an adaptive NN-based control scheme in the outer loop, though, again, the control loops have been analyzed separately. The inter-vehicle collision avoidance has also been solved by incorporating an exponential potential function into the design process. The proposed control structure can guarantee the practical finite-time stability of the closed-loop system, while it requires only the \emph{relative} position of UAVs to their adjacent.

In addition to the aforementioned indirect adaptive control schemes, Direct T2-FNNs have been employed in \cite{Kayacan.2012,Kayacan.2016b,Camci.2018} as an augmentation for a PD controller in the case of SISO nonlinear systems, where the network parameters are updated by an SMC-based algorithm (with a sliding surface as $s=\dot{e}+\eta e$) using the output of the PD controller as the learning signal. The practical finite-time stability of the closed-loop system has been proved using a simple Lyapunov function as $V=1/2s^2$ assuming that a PD controller can stabilize the system \cite{Kayacan.2016b}. Six T2-FNNs have been used in \cite{Camci.2018} to control the trajectory of a 6-DOF quadrotor air vehicle, where each T2-FNN corresponds to a distinct system state. Due to the complexity of computing the gradient of the cost function with respect to the antecedent parameters of the FNN, particle swarm optimization has been adopted as a gradient-free approach to train them, while the consequent parameters have been updated using the mentioned SMC-based algorithm.
Notice that, despite the superior convergence properties of the above-mentioned approaches, as discussed earlier, they are subject to considerable  limitations of discontinuous control systems.

\section{Supplementary features in model-based IFCSs}\label{Sec:SumplementaryFeatures}

Several additional features may be required in different flight control systems due to different additional requirements.
In this section, we will address various supplementary features, which have been widely incorporated in IFCSs.
Again, the focus of this section is on model-based IFCSs, while some of the introduced elements such as self-organizing NNs can be effectively employed in model-free approaches, as well.

\subsection{Output Feedback (OFB) control}\label{Sec:OFB}
The basic feedback error learning method has been developed assuming all the system states are measurable. This assumption is not feasible in many applications. Accordingly, different modifications have been introduced in the literature to effectively control an uncertain nonlinear system using only the system inputs-outputs.
Such an intention is typically fulfilled by employing \textit{state observers} \cite{Seshagiri.2000}, where a composite Lyapunov function is subsequently incorporated to compensate for both the tracking error and the state estimation error.

A common assumption in the design of OFB control methods is that the system is input-output linearizable with a specified relative degree  \cite{KengPengTee.2008}. More precisely, consider a SISO dynamic system as $\dot{x}=f(x)+g(x)u$, $y=h(x)$. Thus, we have:
\begin{equation}\label{Eq:20}
  \dot{y}=\frac{\partial h}{\partial x}(f(x)+g(x)u)=L_f h(x)+L_g h(x) u,
\end{equation}
where $L_f h(x)=\frac{\partial h}{\partial x}f(x)$ is the Lie derivative of $h$ along $f$ \cite{Khalil.2002}. Assuming the system has a relative degree $\rho$, we have $L_g L_f^{\rho-1} h(x) \neq 0$. Therefore, defining
\begin{equation}\label{Eq:21}
  u=\frac{1}{L_g L_f^{\rho-1} h(x)}(-L_f^{\rho} h(x) + \nu),
\end{equation}
the dynamic model reduces to $y^{(\rho)}=\nu$ \cite{Shin.2011} (a similar definition can also be provided for a nonaffine system).
Using such an assumption and assuming that the system is globally exponentially minimum phase, an adaptive OFB control has been introduced in \cite{Hovakimyan.2002} for a nonaffine SISO system with an unknown (but bounded) dimension. A linear observer of dimension $2\rho-1$ has been developed to estimate the time derivatives of the tracking error signal. The estimated vector is then used as the training signal for a  Single-Hidden-Layer (SHL) NN that attempts to compensate for the model inversion error in the framework of the pseudocontrol strategy.
Under the same assumption, a backstepping control scheme has been designed in \cite{KengPengTee.2008}, where the time derivatives of $y$ have been estimated using High Gain Observers (HGOs). Also, an adaptive neural network has been developed to construct the control command in the presence of model uncertainties. Finally, the proposed control scheme has been applied to a helicopter model to control the altitude of the vehicle in vertical flight. Similarly, HGOs have been used in \cite{He.2017} to provide the estimation of the time derivatives of Euler angles, which are required in the attitude control of a flapping-wing micro aerial vehicle.
Analogously, a beneficial approach to control the longitudinal model of HFVs has been introduced in \cite{Xu.2011,Xu.2015d}. As discussed earlier, typically, a backstepping control scheme is designed for an HFV in which the longitudinal dynamics are transformed into the strict feedback form. Alternatively, a new formulation has been given in \cite{Xu.2011,Xu.2015d} to transform the altitude subsystem into a normal output feedback form. More precisely, considering the longitudinal dynamics of an HFV and defining $z_1=y=\gamma$, $z_2=\dot{z_1}$, and $z_3=\dot{z_2}$, we have
\begin{equation}\label{Eq:22}
  \dot{z_1}=z_2, \quad \dot{z_2}=z_3,\quad \dot{z_3}=a(X)+b(X)\delta_e, \quad y=z_1,
\end{equation}
where $X=[\gamma,\theta,q]^T$ and $a$ and $b$ are unknown.
Here, $\gamma$, $\theta$, and $q$ represent the flight path angle, the pitch angle, and the pitch rate, respectively.
Accordingly, $z_2$ and $z_3$ are the time derivatives of the system output and are unknown. Utilizing an HGO, the system states $Z=[z_1,z_2,z_3]^T$ can be estimated by $\hat{Z}=[z_1,\frac{\xi_2}{\varepsilon},\frac{\xi_3}{\varepsilon^2}]^T$, where
\begin{gather}\label{Eq:23}
  \dot{\xi_1}=\frac{\xi_2}{\varepsilon},\\
  \dot{\xi_2}=\frac{\xi_3}{\varepsilon},\\
  \dot{\xi_3}=\frac{-d_1\xi_3-d_2\xi_2-\xi_1+y(t)}{\varepsilon},
\end{gather}
and $\varepsilon$ is a small design constant and $d_1$ and $d_2$ are chosen such that $s^3+d_1s^2+d_2s+1$ is Hurwitz. Consequently, there exist positive constants $h_s$ and $t_s$ such that $\forall t > t_s$, we have $|\hat{Z}-Z|\leq \varepsilon h_s$ \cite{BEHTASH.2007}. Afterward, an NN-based control command has been developed in \cite{Xu.2011,Xu.2015d} to ensure the convergence of a filtered tracking error to a small neighborhood of zero.

An NN-based observer (see Section \ref{Sec:NNDO}) has been designed in \cite{Dierks.2010} to estimate the angular and translational velocities of a quadrotor air vehicle, which are subsequently utilized in the control loop. On the other hand, non-model-based filters have been employed in \cite{Xian.2015} to provide an estimation of the unknown angular velocity, which was required in the proposed OFB control.

\subsection{Minimal-learning parameter}\label{Sec:MLP}
One of the major drawbacks of neural networks in the structure of the feedback error learning scheme is the excessive computational burden of the training process due to the high number of parameters that should be identified. An efficient identification technique with significantly fewer training parameters, called the \emph{Minimal-Learning Parameter} (MLP), has been widely employed by researchers in recent years.
MLP has been first introduced by Yang and his colleagues and employed in the traditional backstepping control combined with T-S fuzzy systems \cite{Yang.2003,Yang.2004} or RBFNNs \cite{YanshengYang.2006}. Subsequently, it was effectively integrated with DSC \cite{Li.2010b} (to solve the problem of the explosion of complexity in classical backstepping control) and with direct adaptive fuzzy control \cite{Chen.2009b} to directly approximate the desired control input signals rather than unknown system's nonlinearities.

Generally speaking, this technique attempts to estimate the \textit{norm} of the unknown weight vector (or matrix) rather than estimating its elements \cite{Lai.2016}. To be more precise, consider again the control problem given in Section \ref{Sec:FEL} with the dynamic model as (\ref{Model_1}). Suppose that $\|W^*\|^2 \leq \varpi$, where $\varpi$ denotes an unknown constant. Defining $\hat{\varpi}$ as the estimation of $\varpi$, a Lyapunov function can be defined as:
\begin{equation}\label{Eq:24}
  V=\frac{1}{2}e^Te+\frac{1}{2\lambda}\tilde{\varpi}^2,
\end{equation}
where $\tilde{\varpi}=\hat{\varpi}-\varpi$ and $\lambda$ is a positive constant. Thus, we have:
\begin{equation}\label{Eq:25}
  \dot{V}=e^T\left(F(x)+B(x)u-\dot{x}_d+{W^*}^T\mu+\varepsilon\right)+\frac{1}{\lambda}\tilde{\varpi}\dot{\hat{\varpi}}.
\end{equation}
Using Cauchy–Schwarz and Young's inequalities, it is obtained that
\begin{gather}
  e^T{W^*}^T\mu \leq \frac{a^2 \varpi e^Te \mu^T \mu}{2} + \frac{1}{2a^2}, \\
  e^T\varepsilon \leq \frac{a^2e^Te}{2}+\frac{\varepsilon_M}{2a^2},
\end{gather}
where $a$ represents a positive design constant.
Therefore,
\begin{align}
\begin{split}\label{Eq:26}
  \dot{V} & \leq  e^T\left(F(x)+B(x)u-\dot{x}_d \right)+\frac{\varepsilon_M}{2a^2}+\frac{1}{2a^2}
   +\frac{a^2 \varpi e^T e \mu^T \mu}{2}+\frac{a^2e^Te}{2}+ \frac{1}{\lambda} \tilde{\varpi} \dot{\hat{\varpi}} \\
  & =  e^T\left(F(x)+B(x)u-\dot{x}_d \right)+ \frac{\varepsilon_M}{2a^2} +\frac{1}{2a^2}
   +\frac{a^2 \hat{\varpi} e^T e \mu^T \mu}{2}+ \frac{a^2e^Te}{2} \\
  & \quad + \tilde{\varpi}\left( \frac{1}{\lambda} \dot{\tilde{\varpi}}- \frac{a^2 e^T e \mu^T \mu}{2}\right).
\end{split}
\end{align}
Thus, it is possible to define the control command and the updating rule of $\hat{\varpi}$ as follows:
\begin{gather}
  u=B^{-1}\left(-F(x)+\dot{x}_d-\left(\frac{a^2}{2}+k_1 \right)e- \frac{a^2 e \hat{\varpi} \mu^T \mu}{2} \right), \label{Eq:26_1}\\
  \dot{\hat{\varpi}}=\frac{a^2 \lambda e^T e \mu^T \mu}{2}-\sigma \lambda \hat{\varpi} \label{Eq:27},
\end{gather}
where $k_1$ and $\sigma$ denote positive design parameters, and the second term on the right-hand side of (\ref{Eq:27}) represents the $\sigma$-modification term. Substituting (\ref{Eq:26_1}) and (\ref{Eq:27}) in (\ref{Eq:26}) yields
\begin{equation}\label{Eq:28}
  \dot{V} \leq - k_1 e^T e - \sigma \tilde{\varpi}\hat{\varpi}+ \frac{\varepsilon_M}{2a^2} +\frac{1}{2a^2}.
\end{equation}
So, setting $\sigma=\frac{2k_1}{\lambda}$ and knowing that
\begin{equation}
  \tilde{\varpi}\hat{\varpi} \geq \frac{\tilde{\varpi}^2}{2}-\frac{\varpi^2}{2},
\end{equation}
finally, (\ref{Eq:28}) can be written as follows:
\begin{equation}\label{Eq:29}
  \dot{V} \leq - k_1 \left(e^T e + \frac{\tilde{\varpi}^2}{\lambda}\right) + \frac{k_1 \varpi^2}{\lambda}+ \frac{\varepsilon_M}{2a^2} +\frac{1}{2a^2}=-kV+C,
\end{equation}
where $k=2k_1$ and $C= \frac{k_1 \varpi^2}{\lambda}+ \frac{\varepsilon_M}{2a^2} +\frac{1}{2a^2}$. As discussed in Section \ref{Sec:Stability}, (\ref{Eq:29}) leads to the convergence of both the tracking error and the norm estimation error to a small neighborhood of zero, where the appropriate value of $a$ is determined considering the tradeoff between larger steady-state error and more control effort. Accordingly, we should train only a scalar parameter ($\hat{\varpi}$) rather than a matrix ($\hat{W}$), thereby considerably reducing the computational burden corresponding to the online tuning of the NN parameters.
However, by comparing (\ref{Eq:27}) with (\ref{Learn_1}), it can be understood that such an achievement is obtained at the cost of less efficient use of the error vector $e$ in the MLP technique. More precisely, here, we use only the norm of the tracking error (in a scalar updating rule) instead of using all its elements, separately, which results in a conservative design.
A similar formulation can also be given for an MLP technique in the case of \emph{direct} adaptive control designs.

This approach can be employed, in a similar way, in the structure of the backstepping control method. The design has been enhanced in \cite{Zong.2014} by updating only one parameter in the attitude control block that corresponds to the maximum of the norm of all the three RBFNNs employed in the control system. However, this results in a more conservative design, thereby requiring more control effort.
On the other hand, as mentioned earlier, authors in \cite{Xu.2015d} have transformed the longitudinal dynamics of an HFV into the output feedback form in which the new system states were approximated using HGOs. Thus, there is a need for only one neural network in the proposed scheme, where the MLP technique has been employed in the training phase.
A similar formulation has been utilized in \cite{Bu.2018} for a nonaffine model of an HFV. Both the velocity and attitude control blocks have been designed using the pseudocontrol strategy. Also, fuzzy wavelet neural networks have been employed to compensate for model uncertainties, where, owing to the employment of the MLP technique, only the norm of the weight matrix was tuned.
Besides, in \cite{Xu.2016d}, DSC has been integrated with the MLP technique in the case of an HFV, which is subject to actuator bias fault.
Considering unknown control gain functions ($g_i$s), the MLP technique has been utilized to estimate $\lambda_i={g_{i_{min}}}^{-1} \|W^*\|^2$ to avoid the control singularity problem, while due to the unusual formulation of the employed NNs, the small-gain theorem \cite{Jiang.1994,Li.2010b} has been involved to ensure the UUB stability of the system.

Also, regarding the application of the MLP technique in the backstepping control of \emph{VTOL UAVs}, such an approach has been utilized in a backstepping trajectory tracking control scheme for a quadrotor air vehicle in \cite{Wang.2014b} in which the MLP technique has been applied to each of the six RBFNNs employed to estimate model uncertainties. Further, authors in \cite{Fu.2018} have used the DSC along with the MLP technique in the trajectory tracking control of a multi-rotor UAV considering output constraints. A robust term has also been incorporated to estimate the neural approximation error (Section \ref{Sec:NNDO}).
Besides, a neuroadaptive control approach has been proposed in \cite{Song.2019} for a quadrotor UAV under model uncertainties and actuator faults. Compared to previously-mentioned studies, a NN has been employed in this paper to estimate \emph{an upper bound for the norm} of model uncertainties (instead of estimating the uncertainty itself), where the MLP technique has been adopted to estimate an upper bound for the norm of the weight vector of that NN.

The MLP technique has also been employed in discrete-time neural backstepping controllers \cite{Liu.2010b}, where the updating rule of the NN weights' norm and the stability analysis can be obtained similarly to the basic discrete-time FEL in Section \ref{Sec:FEL}. Such an MLP scheme has been used in \cite{Xu.2012b} in a discrete-time neural backstepping controller applied to the longitudinal mode of an HFV. However, the introduced updating rule for $\varpi$ in \cite{Liu.2010b,Xu.2012b} may result in negative values in some time intervals. Thus, the design has been improved in \cite{Xu.2013b,Xu.2014} by assuming that $\|W^*\| \leq \varpi sign(\varpi)$, which allows $\varpi$ to be either positive or negative.

\subsection{Systems with unknown control direction} \label{Sec:UnknownDirection}
The design of a controller for a dynamic system with unknown control direction is a challenging problem. This is due to the fact that a control command with incorrect direction can simply make the system unstable. In such cases, an interesting idea would be to alternately change the control direction. Accordingly, if the control command is applied in the wrong direction, the systems states get away from the desired trajectory until the control direction changes. Subsequently, the amplitude of the control command should increase by increasing the tracking error to get the system back to the desired trajectory.
Such an idea has been first introduced in \cite{Nussbaum.1983}, and a function with the above-mentioned characteristics is known as a \emph{Nussbaum} function. Nussbaum function has been employed in different studies to provide acceptable closed-loop performance in the case of complex systems with unknown control direction \cite{Chen.2019}.
To clarify the control design procedure using the Nussbaum function, consider a SISO dynamic model as $\dot{x}=f(x)+g(t)u$, where $g(t)$ is a time-varying control gain with unknown direction. To ensure the stabilizability of the system, assume that $g(t) \in I=[\underline{g},\overline{g}]$, where $\underline{g}$ and $\overline{g}$ denote unknown constants and $0 \notin I$. Defining $e=x-x_d$, we have
\begin{equation}
  \dot{e}=\dot{x}-\dot{x}_d=f(x)+g(t) u - \dot{x}_d.
\end{equation}
If $g(t)$ was available, the control command could be computed as $u=g^{-1}(-f(x)+\dot{x}_d-k e)$, where $k$ is a positive constant. However, due to the unknown control direction, such a command is not feasible. Thus, we define a control command as follows:
\begin{gather}
  u=N(\zeta)\eta, \\
  \dot{\zeta}=e \eta, \\
  \eta= f(x)-\dot{x}_d+k e, \label{Eq:29_1}
\end{gather}
where $N(\zeta)$ represents a Nussbaum function like $N(\zeta)=\exp(\zeta^2)\cos (\pi \zeta /2)$. Defining a Lyapunov function as $V=1/2 e^2$, we have
\begin{equation}
  \dot{V}=e \dot{e}=e \left(f(x)+g N(\zeta) \eta - \dot{x}_d \right).
\end{equation}
By adding and subtracting $\dot{\zeta}$ to the right side of the equation, $\dot{V}$ is obtained as
\begin{align}
\begin{split}
 \dot{V} & =  e f(x) + g N(\zeta) e \eta + \dot{\zeta} -\dot{\zeta} - e \dot{x}_d \\
             & =  g N(\zeta) \dot{\zeta} + \dot{\zeta} - k e^2.
\end{split}
\end{align}
Now, by multiplying both sides of the equation by $\exp(c t)$, where $c=2k$, the following equation is obtained.
\begin{equation}
  \frac{d}{dt}\left(V e^{c t} \right) = \left(g N(\zeta) \dot{\zeta} + \dot{\zeta}\right) e^{c t}.
\end{equation}
Thus, we have
\begin{equation} \label{Eq:30}
  V  = e^{-c t} \int_{0}^{t} \left(g N(\zeta)  + 1 \right) \dot{\zeta} e^{c \tau} d\tau.
\end{equation}
Consequently, according to Lemma 2 in \cite{Ge.2003}, it is proved that $V(t)$ and $\zeta(t)$ are bounded, thereby guaranteeing the bounded tracking error. In cases where $f(x)$ is also an unknown function, as discussed earlier, we can simply substitute $f(x)$ in (\ref{Eq:29_1}) by its estimation as $\hat{f}(x)=\hat{W}^T \mu(x)$ and include an additional term $1/2 \tilde{W}^T \Gamma_1^{-1} \tilde{W}$ in the Lyapunov function to ensure the closed-loop stability.
A similar approach has been employed in \cite{Xu.2015b} in the framework of DSC to control the longitudinal mode of an HFV considering dead-zone input nonlinearity, where a set of NNs have been used to estimate uncertain terms in the control command.

\subsection{Neural networks and Disturbance Observers (DOs)} \label{Sec:NNDO}
\subsubsection{Neural disturbance observer} \label{Sec:NDO}
As discussed earlier, NNs can be effectively employed in the closed-loop control to estimate and compensate for model uncertainties, external disturbances, and also complex parts of the control command. In addition to the above-mentioned control systems, NNs can also be utilized as a powerful DO in an open-loop identification problem. To this end, consider again the nonlinear model (\ref{Model_1}) where $\Delta(x,u)$ corresponds to the effect of model uncertainties, actuator faults, and external disturbances on the system dynamics \cite{MouChen.2014}. Notice that an external disturbance is generally an explicit function of time (not the system states and inputs). Thus, the identification of $\Delta$ as a function of system states (and inputs) requires an implicit assumption that external disturbances can be formulated as a function of the system states (and inputs). Although such an assumption makes sense in the case of some types of external disturbances, in a general case, it is not reasonable.
In such circumstances, it may be possible to estimate $\Delta$ by a NN with time-dependent weights (or even time-dependent structure). This brings new challenges to the convergence analysis of the NN, which would be an interesting research direction.
Another idea would be assuming that external disturbances are smaller than an unknown bounded function of system states, i.e. $|d(t)|\leq {W^*}^T \mu(x)$ \cite{YanshengYang.2006}, while it may lead to a conservative design, thereby significantly increasing the control effort in the case of control problems.

Here, assuming that the uncertain terms in the dynamic model can be formulated as a function systems states (and inputs), we can introduce a new state-space model as
\begin{equation}
  \dot{\hat{x}}=F(x)+B(x)u+\hat{\Delta}+\kappa (x-\hat{x}), \label{Eq:30_01}
\end{equation}
where $\kappa$ represents a positive constant (which is tuned according to the compromise between the convergence rate of the introduced observer and its sensitivity to measurement noises \cite{Castaldi.2014b}), and $\hat{\Delta}=\hat{W}^T\mu(x,u)$ denotes the estimation of $\Delta$. Notice that different types of feedforward and recurrent NNs can be formulated in such a compact form \cite{MouChen.2014}. Now, by defining $e_D=\hat{x}-x$ and $\tilde{W}=\hat{W}-W^*$, a Lyapunov function can be proposed as
\begin{equation}
    V=\frac{1}{2}e_D^T e_D + \frac{1}{2} tr \left(\tilde{W}^T \Gamma^{-1} \tilde{W}\right).
\end{equation}
Thus, we have
\begin{equation}
  \dot{V}=e_D^T\left(-\kappa e_D +\tilde{W}^T\mu(x,u)-\varepsilon\right)+ tr \left(\tilde{W}^T \Gamma^{-1} \dot{\hat{W}}\right),
\end{equation}
where $\varepsilon$ denotes the bounded estimation error of the NN. As a consequence, if the NN's parameters are updated as $\dot{\hat{W}}=\Gamma \mu e_D^T$, it is obtained that $\dot{V}=-e_D^T (\kappa e_D + \varepsilon)$, which results in $\dot{V} < 0$ for $\|\kappa e_D\|>\|\varepsilon\|$, thereby guaranteeing the bounded estimation error. The employment of one of the modification techniques introduced in Section \ref{Sec:FEL} in the updating rule is also recommended to avoid parameter drift.
It is notable that, in the proposed DO, there is no need for an affine model, and we can simply substitute $F(x)+B(x)u$ in (\ref{Eq:30_01}) with $F(x,u)$.
A similar method has been used in \cite{Lungu.2020} to estimate uncertain terms in the dynamics model of a UAV including external disturbances induced by different types of atmospheric disturbances, i.e. the wind shear, wind gust, and atmospheric turbulence.
Owing to the presence of both $v$ and $\dot{v}$ in the disturbance term (where $v$ represents the wind velocity vector), a dynamic equation is then derived (using the estimated uncertainty) as $\dot{v}=\chi(x,u,v)$ to estimate the total wind velocity. Subsequently, an auto-landing control system has been proposed in \cite{Lungu.2020} for the Sekwa UAV in the presence of external disturbances using a combination of the backstepping control and the dynamic inversion, while the designed scheme attempted to control six independent outputs by only four system inputs, which is not generally feasible. More precisely, the pseudo-inverse operator employed to compute the control command may result in inappropriate commands in the case of inconsistent control objectives.

Moreover, it should be noted that an analogous formulation can be developed to provide a neural \emph{state} observer. A neural observer has been proposed in \cite{Dierks.2010} by incorporating both the kinematic and dynamic equations of the system to estimate the translational and angular velocities of a quadrotor knowing the position and attitude of the vehicle.
Such a DO can also be utilized in the closed-loop control by substituting $e_D$ in the updating rule of the NN by $e_D+e$ with $e$ represents the tracking error. Indeed, this would be a variant of the composite learning method introduced in Section \ref{Sec:FEL}.

\subsubsection{Combination of NN function approximation and DOs} \label{Sec:CombinedNNDO}
There are a variety of robust control approaches in the literature in which a combination of DOs and NNs has been adopted to simultaneously compensate for external disturbances and model uncertainties, respectively. Using such an identification scheme, it is possible to distinguish between external disturbances, which are explicit functions of time, and internal disturbances, which can be modeled as a function of system states (and inputs). In addition, using a combination of DOs and NN-based estimators, DOs will be capable of compensating for the estimation error of the NN. More specifically, consider a nonlinear dynamic model as
\begin{equation}
  \dot{x}=F(x)+B(x)u+\Delta(x,u)+d(t),
\end{equation}
where $\Delta(x,u)$ and $d(t)$ represent model uncertainties and external disturbances, respectively.
Now, considering the following definitions:
\begin{gather}
  \dot{\hat{x}}=F(x)+B(x)u+\hat{\Delta}(x,u) + \hat{D}+ \kappa (x-\hat{x}), \label{Eq:31_00}\\
  \Delta=W^{*^T}\mu(x,u)+\varepsilon, \quad \hat{\Delta}=\hat{W}^T \mu(x,u), \\
  D(t)=d(t)+\varepsilon, \\
  e_D=\hat{x}-x, \quad e=x-x_d,
\end{gather}
and assuming $dim(x)=dim(u)=n$, an appropriate control command can be formulated as follows (the introduced approach can be used in the case of other types of dynamic systems and control methods in a similar manner):
\begin{equation}
  u=B(x)^{-1}(x)\left(\dot{x}-F(x)-\hat{\Delta}-\hat{D}-k_1 e\right).
\end{equation}
Subsequently, a Lyapunov function can be formulated as
\begin{equation}
 V=\frac{1}{2}\left(e^T k_2 e + e_D^T k_3 e_D + \tilde{D}^T \tilde{D} + tr(\tilde{W}^T \Gamma^{-1} \tilde{W}) \right),
\end{equation}
where $k_i$ represents positive constants, and $\tilde{D}=\hat{D}-D$. Accordingly, we have
\begin{align}
\begin{split}
 \dot{V} & = k_2 e^T \left(-k_1 e - \tilde{W}^T \mu - \tilde{D} \right) + tr(\tilde{W}^T \Gamma^{-1} \dot{\tilde{W}}) + \\
             & \tilde{D}^T \left(\dot{\hat{D}}-\dot{D}\right) + k_3 e_D^T \left(-\kappa e_D + \tilde{W}^T \mu + \tilde{D} \right).
\end{split}
\end{align}
Thus, the following updating rules can be defined:
\begin{gather}
  \dot{\tilde{W}}=\dot{\hat{W}}=\Gamma  \left(\mu\left(k_2 e^T -k_3 e_D^T \right) - \sigma_W \hat{W}\right), \label{Eq:31_0} \\
  \dot{\hat{D}}= \left(k_2 e - k_3 e_D \right) - k_4 \left[\dot{\hat{x}}-\dot{x}+\kappa e_D \right], \label{Eq:31}
\end{gather}
where $k_4$ denotes a positive constant.
Concerning the second term on the right-hand side of (\ref{Eq:31}), using (\ref{Eq:31_00}), we have
\begin{equation}\label{Eq:31_01}
  \dot{\hat{x}}-\dot{x}+\kappa e_D=\tilde{W}^T \mu +\tilde{D}.
\end{equation}
Consequently, assuming that $\mu(x,u)$ and $\dot{D}$ are bounded, one can simply prove the satisfaction of (\ref{Lyap_3}), thereby ensuring the boundedness of all signals in the closed-loop system.
Notice that, although the updating rule (\ref{Eq:31}) consists of $\dot{x}$, there is no need for it to compute $\hat{D}$ because the estimated disturbance ($\hat{D}$) is obtained as the integral of (\ref{Eq:31}) (the integral of other terms in the updating rule can be calculated using an auxiliary state variable \cite{Li.2017c}).
Such a combination has been employed in \cite{Li.2017c} in a backstepping design. The same approach has also been utilized in \cite{Yu.2020} to provide a decentralized attitude synchronization tracking of multi-UAVs in the presence of actuator faults and wind effects. Similarly, the trajectory tracking control of multiple trailing UAVs has been addressed in \cite{Yu.2020b} using DSC, where an NN+DO has been adopted to compensate for unknown aerodynamic parameters, actuator faults, and wake vortices.
A partially analogous scheme has been utilized in \cite{He.2017} to provide a trajectory tracking control for a flapping-wing micro aerial vehicle considering model uncertainties and external disturbances. In a similar manner, a combined NN and DO has been incorporated in \cite{Chen.2016} in the framework of an FTC to control the attitude of a 3-DOF helicopter.
Further, an analogous scheme has been employed in \cite{Chen.2015c} in a backstepping controller designed to control the attitude of an NSV.
The same identification approach has also been adopted in \cite{Xu.2018b}, where a DO is utilized to compensate for external disturbances, the estimation error of NNs, and the effect of unknown input dead-zone.

Another effective combination of NNs and DOs with less complexity and no requirement to use the boundedness of $\mu(x,u)$ and $\dot{D}$ in the stability analysis, relies on the estimation of the upper bound of $D$ rather than that of the exact value of it.
Such a method, which results in a \emph{conservative} design, can be classified as a \emph{robust adaptive control}.
For this purpose, consider again the above-mentioned dynamic model and the following definitions (for simplicity, suppose that $x,u \in \mathbb{R}$, while the introduced approach can be extended to MIMO systems with a similar formulation):
\begin{gather}
  \dot{\hat{x}}=F(x)+B(x)u+\hat{\Delta}(x,u) + \upsilon + \kappa (x-\hat{x}), \\
  \Delta=W^{*^T}\mu(x,u)+\varepsilon, \quad \hat{\Delta}=\hat{W}^T \mu(x,u), \\
  D(t)=d(t)+\varepsilon, \quad \|D\| \leq D_M,\\
  e_D=\hat{x}-x, \quad e=x-x_d, \\
  u=B(x)^{-1}(x)\left(\dot{x}_d-F(x)-\hat{\Delta}-\upsilon -k_1 e\right),
\end{gather}
where $\upsilon$ should be designed. Now, redefining the Lyapunov function as
\begin{align}
\begin{split}
  V =  \frac{1}{2}\left(k_2 e^2 + k_3 e_D^2 + k_4 \tilde{D}_M^2+ \tilde{W}^T \Gamma^{-1} \tilde{W}\right) ,
\end{split}
\end{align}
we have
\begin{align}
\begin{split}
 \dot{V} & = k_2 e \left(-k_1 e - \tilde{W}^T \mu +D - \upsilon \right) + \tilde{W}^T \Gamma^{-1} \dot{\tilde{W}} + \\
             & \quad k_4 \tilde{D}_M \dot{\hat{D}}_M + k_3 e_D \left(-\kappa e_D + \tilde{W}^T \mu + \upsilon - D \right).
\end{split}
\end{align}
Thus, using the updating rule (\ref{Eq:31_0}) and
\begin{gather}
  e_a = k_2 e - k_3 e_D , \\
  \dot{\hat{D}}_M=1/k_4 \left(e_a \tanh (e_a/\epsilon)-\sigma_M \hat{D}_M \right), \label{Eq:31_1}\\
  \upsilon = \hat{D}_M \tanh (e_a / \epsilon),
\end{gather}
where $\epsilon$ denotes a positive constant, it is obtained that
\begin{align}
\begin{split}
 \dot{V} & = - k_2 k_1 e^2  - k_3 \kappa e_D^2 + \tilde{D}_M \left(e_a \tanh (e_a/\epsilon)-\sigma_M \hat{D}_M \right) \\
             & \quad - \sigma_W\tilde{W}^T\hat{W} + e_a D - e_a  \hat{D}_M \tanh (e_a / \epsilon).
\end{split}
\end{align}
Having the following inequality \cite{Polycarpou.1996} for any $\epsilon > 0$ and $z \in \mathbb{R}$,
\begin{equation}
 0 \leq |z| - z \tanh (z/\epsilon) \leq 0.2785 \epsilon,
\end{equation}
it is easy to show that (\ref{Lyap_3}) is satisfied.
The utilization of the hyperbolic tangent function instead of the signum function in the presented formulation is an effective way to avoid the chattering phenomenon, while the possible asymptotic stability of the closed-loop system reduces to UUB stability.
To be more specific, if we simply employ the signum function in the introduced control scheme and eliminate the $\sigma$-modification terms from (\ref{Eq:31_0}) and (\ref{Eq:31_1}), one can simply prove the asymptotic convergence of the tracking error to zero, though at the cost of possible parameter drift and previously discussed limitations of discontinuous control systems (Section \ref{Sec:AsymptoticStability}).
Such an approach has been utilized in \cite{Lai.2016} to control the position and attitude of a helicopter with unknown inertia matrix considering aerodynamic frictions. Accordingly, the unknown aerodynamic forces and moments have been estimated using RBFNNs, where the upper bound of the estimation error corresponding to NNs, as well as external disturbances, has been compensated by the introduced DO.

The introduced identification scheme can be similarly employed in the backstepping control design.
It has been employed in \cite{Zou.2015b} in a backstepping trajectory tracking control applied to a model-scaled helicopter in order to deal with the NN's estimation error, where a switching function has been adopted to integrate the NN and the introduced DO.
Further, a similar approach has been used in the framework of DSC to control the longitudinal mode of an HFV in the presence of model uncertainties, dead-zone input nonlinearity \cite{Xu.2015b}, and actuator faults \cite{Xu.2016d}.
Analogously, in \cite{MouChen.2014}, recurrent wavelet neural networks have been integrated with such a DO in a DSC to compensate for external disturbances, model uncertainties, and the effect of input constraints in the case of the attitude control of an NSV.
An adaptive neural backstepping control has been proposed in \cite{Xu.2019} for an HFV, where a similar approach has been utilized in each step of the backstepping control to deal with model uncertainties and estimation error of NNs.
In \cite{Fu.2018}, DSC has been applied to a multi-rotor UAV to provide an attitude control system. A similar identification method has been employed to compensate for model uncertainties and external disturbances.

On the other hand, a reverse combination of NN+DO has been introduced in \cite{Liu.2006b}, where first, a DO attempts to estimate the entire model uncertainties and external disturbances as a lumped disturbance, and subsequently, a NN has been employed to compensate for estimation error of the DO.
The proposed  scheme has been utilized to control the roll angle of an air vehicle considering the wing rock phenomenon.
However, using such an approach, the estimation error of the NN is not identified and thus remains uncompensated.

\subsection{Fault-tolerant control}\label{Sec:FTC}
As mentioned earlier, the introduced (direct and indirect) NN-based adaptive controllers have been applied to faulty systems, as well. They include but not limited to the basic FEL-based control \cite{Li.2004,Suresh.2005,Pashilkar.2006b}, the pseudocontrol strategy \cite{Calise.2001,Brinker.2001,Johnson.2000b,Chowdhary.2013}, the neural backstepping design \cite{Shin.2004,Pashilkar.2006,Sonneveldt.2007,Sonneveldt.2009,Fu.2016,Chen.2016,Li.2020}, and hybrid direct-indirect adaptive controllers \cite{NhanNguyen.2006,Nguyen.2008}.
In this regard, a typical approach to deal with operational faults is to incorporate the nonlinear terms induced by the actuator faults (or structural damages) into the model uncertainty and estimate (and compensate for) them as a lumped uncertainty by FEL-based NNs \cite{Xu.2016d,Yu.2018,Song.2019,Cheng.2020,Zeghlache.2018}.
Although the actuator faults (or structural damages) suffer from the same issue as external disturbances (i.e. the explicit dependence on time), by considering a sequence of abrupt faults, the coefficients corresponding to system faults can be deemed as time-independent functions between two sequential faults. Thus, the stability analysis can be performed for a specific time interval between two sequential faults (see the following subsection).
Accordingly, the aforementioned combination of NNs and DOs can also be utilized in fault-tolerant flight control systems \cite{Chen.2016,Yu.2020}.


To provide more efficient FTC systems with less conservativeness, in addition to the aforementioned generic adaptive neural control methods, there are various NN-based controllers in the literature that have been customized to specifically deal with actuator/sensor faults and structural damages. Some of the more commonly used schemes in this field are given in the following.

\subsubsection{FEL-based fault identification} \label{Sec:FELFault}
It is possible to employ the FEL method to directly identify the coefficients corresponding to actuator faults, while simultaneously estimating model uncertainties in the closed-loop control.
To clarify the basic idea, consider again the dynamic model (\ref{Model_1}), and suppose that the actual plant input is determined as
\begin{equation}\label{Eq:32}
  u(t)=\xi(t) u_c(t) + \delta(t),
\end{equation}
where $u_c \in \mathbb{R}^n$ represents the computed control command, and $\xi(t)$ and $ \delta(t)$ denote an unknown diagonal matrix and an unknown vector corresponding to multiplicative and additive actuator faults, respectively. Such a formulation can represent different types of actuator faults, such as the stuck type fault and the loss of effectiveness \cite{Liu.2019c}. Considering a sequence of sudden actuator faults, $\xi$ and $\delta$ can be considered as piecewise constant functions. Accordingly, defining $t_i$ as the time of the occurrence of the $i$th actuator fault, one can assume that $\xi$ and $\delta$ remain constant for $t \in (t_i,t_{i+1})$.
In the following, we will focus on this time interval, while, due to the finite number of such time intervals, the design can be extended to the entire flight time assuming that the occurrence of the fault at $t_i$ does not violate the controllability of the system.
Now, if the ideal system input ($u^*$) is defined as (\ref{Ctrl_1}) with $\hat{\Delta}=\hat{W}^T\mu$, then a control command can be determined as follows:
\begin{equation}
  u_c= k_2 u^* +k_3,
\end{equation}
where $k_2$ and $k_3$ represent unknown constants satisfying
\begin{equation}
  \xi k_2 = I, \quad \xi k_3 + \delta =0,
\end{equation}
which require $\xi$ to be invertible. Knowing that $\xi$ is a diagonal matrix, the invertibility implies that no actuator should be completely stuck.
Owing to the unknown value of $k_2$ and $k_3$, their estimations are employed in the control command. Thus, we have:
\begin{equation}
  u= \xi(\hat{k}_2 u^* +\hat{k}_3)+\delta = u^* + \xi (\tilde{k}_2 u^* +\tilde{k}_3).
\end{equation}
Thus, using the following updating rules
\begin{gather}
  \dot{\hat{k}}_2= \Gamma_2 \left(u^* e^T B(x)\right)^T,  \label{Eq:32_1}\\
  \dot{\hat{k}}_3= \Gamma_3 \left(e^T B(x)\right)^T, \label{Eq:32_2}
\end{gather}
and employing (\ref{Learn_1}), one can define the following Lyapunov function
\begin{equation}
  V= \frac{1}{2} \left(e^T e + tr \left(\tilde{W}^T \Gamma \tilde{W} + \tilde{k}_2^T \xi \Gamma_2^{-1} \tilde{k}_2+ \tilde{k}_3^T \xi \Gamma_3^{-1} \tilde{k}_3 \right) \right)
\end{equation}
Here, we have assumed that $\xi$ is a positive definite matrix, which is a reasonable assumption due to the fact that $\xi$ corresponds to the effectiveness ratio of actuators ($0< \xi_{ii} \leq 1$).
Utilizing the above-mentioned updating rules, the time derivative of $V$ is obtained as (\ref{Lyap_2}), thereby ensuring the bounded tracking error. As discussed previously, it is recommended to incorporate a modification term in the updating rules (\ref{Eq:32_1}) and (\ref{Eq:32_2}) to avoid the parameter drift in the absence of the PE condition.

A similar approach has been used in \cite{Tang.2003} for a SISO system in the framework of traditional backstepping control. Similarly, in \cite{Xu.2016c}, such an approach has been utilized in a command filtered backstepping considering parametric uncertainty in both internal dynamics and the control gain function, while in \cite{Xu.2019}, the prediction error has also be involved in the NN updating rules. All these controllers have been applied to the longitudinal mode of an HFV. An analogous method has been employed in \cite{Peng.2020}, where the designed controller has been applied to an HFV considering flexible dynamics and state constraints. Alternatively, by considering a similar plant in the framework of DSC, authors in \cite{Yuan.2020} attempt to estimate $\frac{1}{\inf|\xi_{ii}|}$, which allows to deal with time-varying actuator faults at the expense of employing a conservative design.
DSC has been utilized in \cite{Liu.2019b} to control the skid-to-turn missile in the presence of partial state constraints and actuator faults. Compared to the above-mentioned studies, the additive fault $B(x)\delta$ has been aggregated with the model uncertainty $\Delta$ into a single term, where the upper bound of it has been estimated by a NN.
Further, instead of estimating $k_2$, the matrix $\xi$ has been estimated directly, while there is no basic difference between these two design methods.

\subsubsection{Using a separate Neural fault detection and identification (FDI) block} \label{Sec:SeparateFDI}
Traditionally, NNs were utilized as a separate Fault Detection and Isolation (FDI) scheme in the framework of FTCs. The main idea in it comes from the comparison of the output of the system and pre-trained NNs, where the residuals are interpreted as a fault if they exceed predefined thresholds \cite{Napolitano.2000}. However, such an approach cannot ensure closed-loop stability, and it may also lead to false alarms in the presence of severe external disturbances or unexpected damages.

There are other types of indirect fault identification approaches in the literature (which have been designed separately from the control system), as well.
The main concern about such a decentralized design is the challenges in analyzing the closed-loop stability considering the estimation error of the fault identification block (which is commonly neglected in the stability analysis).
An NN-based fault identification block has been proposed in \cite{Shi.2014} to estimate multiplicative actuator faults and model uncertainties, distinctly.
The introduced method is similar to a FEL-based fault identification scheme, while the tracking error $e$ is substituted by the estimation error of a neural observer (see Section \ref{Sec:NDO}).
The estimated model uncertainties and actuator faults have been subsequently employed in the structure of a backstepping attitude controller applied to an NSV.
Further, authors in \cite{Talebi.2009}, have attempted to identify the combination of fault dynamics and model uncertainties as a lumped uncertainty using a neural state observer. Such an approach has been employed in the paper to tackle sensor and actuator faults in the case of a satellite.
The updating rules are obtained using a FEL method considering the estimation error of the neural observer as the learning signal.

In a somewhat similar fashion, a neural observer has been employed in \cite{Baldi.2013,Baldi.2016} within the framework of \emph{nonlinear geometric approach} for fault detection and identification \cite{Persis.2001}. The fundamental assumption (which could be a restrictive assumption in flight control problems) in such an approach is the existence of a \emph{coordinate change} in the state space and the output space that provides an observable subsystem, which is affected by a specific fault but not affected by external disturbances and other faults. By exploiting such subsystems and using the same neural observer as introduced in Section \ref{Sec:NDO}, authors in \cite{Baldi.2013,Baldi.2016} have attempted to detect and identify different (but not simultaneous) sensor and actuator faults in a satellite, while considering external disturbances. Finally, the proposed scheme in \cite{Baldi.2013} has been employed in an attitude control system based on a typical LQG controller designed for a linear model of the satellite.

Alternatively, an RLS optimization has been adopted in several studies to identify different operational faults of an air vehicle.
In \cite{Emami.2019c}, multiplicative actuator faults have been identified using a generalized Online Sequential Extreme Learning Machine (OS-ELM) algorithm (for MIMO systems), which is based on the RLS optimization (see Section \ref{Sec:Model-free-Indirect}).
The model uncertainties and external disturbances have been neglected at this stage, while they have been compensated by a robust model predictive control, which is applied to a quadrotor UAV.
A neural state observer has been introduced in \cite{Abbaspour.2017,Abbaspour.2018} in which the NN weights have been updated using an Extended Kalman Filter (EKF), which is formulated by a similar formulation to the RLS optimization.
The proposed approach has been evaluated in the presence of different faults such as abrupt and intermittent faults.
Besides, such a method has been adopted in a dynamic inversion control in \cite{Abbaspour.2018} to control the attitude of a fixed-wing aircraft considering actuator faults.

According to the obtained results in \cite{Abbaspour.2017,Emami.2019c,Nguyen.2008}, the use of an RLS optimization-based updating rule results in faster convergence of the NN weights and higher accuracy in comparison with FEL-based approaches (which are developed based on Lyapunov's direct method), particularly in the case of an abrupt actuator fault.
In addition, unlike the RLS optimization-based approaches, a fault identification block, which is developed using Lyapunov's direct method (such as in \cite{Talebi.2009}), may result in severe changes in its estimation at the moment of an abrupt actuator fault  \cite{Abbaspour.2017}. This phenomenon, which has not been addressed in typical stability analyses, can be a challenging issue in FEL-based FTCs.

\subsubsection{Multimodel approaches}
A number of studies attempted to identify the dynamic model of the system using an online identification problem employing recurrent NNs (such as NARX NNs), and then design a controller for the identified model \cite{Savran.2006}. The challenging issue with such a control system is analogous to that of previously mentioned indirect FDI schemes. More precisely, the identification error is typically neglected in the stability analysis of the closed-loop system.

As a more reliable solution, a multimodel approach has been developed in \cite{Emami.2019} to identify a 6-DOF model of a fixed-wing aircraft in the presence of different actuator faults. To be more precise, a set of local NARX NNs have been first trained considering different fault conditions, i.e the elevator, aileron, and rudder faults, where each local model corresponds to a specific fault condition. Subsequently, the output of the entire model is computed by a weighted average of the outputs of local NNs, where the relative weight of each local model is determined using an OS-ELM-based optimization.
It means that each local model can be considered as a hidden node of an extended ELM, where the output layer of this extended ELM is trained using the OS-ELM algorithm. Accordingly, the entire model can be considered as a deep neural network with two hidden layers, where the first layer (corresponding to local models) is trained offline, and the second layer is trained by the OS-ELM approach.
Such an identification algorithm has been adopted in \cite{Emami.2019b} to provide a reliable prediction model for the system. The obtained model has been used in a Model Predictive Control (MPC) to provide a trajectory tracking control for a fixed-wing aircraft. As illustrated in \cite{Emami.2019b}, the proposed approach not only can deal with actuator faults that have been considered in the offline training of local NNs, but it can compensate for all the actuator faults and structural damages that can be modeled as a combination of the local models. In this regard, the local NNs can be considered as the basis vectors of a multidimensional space, which are capable of representing all vectors in that space.
Also, the prediction error of the model has been tackled by a DO in the proposed model predictive controller.
The stability of the closed-loop system has been analyzed using a terminal constraint in the MPC framework, while the feasibility of such a constrained optimization problem is not trivial \cite{Mayne.2013}.

\subsection{Consideration of input constraints}
Similar to FTC systems, a typical approach to overcome the input constraints is to consider nonlinear terms induced by input constraints (such as dead-zone or saturation function) as an uncertain term, which is estimated and compensated by NNs \cite{Xu.2013,Shao.2019}.
Nevertheless, the same issue with neural DOs, i.e. the estimation of an explicit function of time using a NN that is a function of system states, exists here as well.
In addition to such a control approach, other types of NN-based control designs have been proposed, which can deal with input constraints. The most commonly used approaches to this goal are given in the following.

\subsubsection{Pseudocontrol Hedging (PCH)}
A traditional approach to deal with input constraints in the framework of the pseudocontrol strategy called the Pseudo-Control Hedging (PCH), is to prevent the adaptive elements in the control system from \emph{seeing} the effects of input constraints by manipulating the reference trajectory \cite{Johnson.2000b,NakwanKim.2003}. For this purpose, consider again the dynamic model (\ref{Eq:16_0}). Considering the desired trajectory $x_d$, a reference trajectory is defined for the system as $\dot{x}_r=\dot{x}_d+\nu_h$, where $\nu_h$, which represents a residual term induced by input constraints, should be designed. Defining $\bar{e}=x-x_r$, we have:
\begin{align}
\begin{split}\label{Eq:33_0}
   \dot{\bar{e}} & =F(x,u)-\dot{x}_r=\hat{F}(x,u)-\dot{x}_r + \Delta(x,u) \\
               & =\hat{F}(x,u_c)-\dot{x}_r + \Delta(x,u)+\left(\hat{F}(x,u)-\hat{F}(x,u_c)\right),
\end{split}
\end{align}
where $u_c$ denotes the desired control command and $ \Delta(x,u)=F(x,u)-\hat{F}(x,u)$. Thus, if we define
\begin{equation}
  \nu_h=\hat{F}(x,u)-\hat{F}(x,u_c),
\end{equation}
using the control command $u_c$ defined by (\ref{Eq:16_01})-(\ref{Eq:16_02}), and employing the same procedure as given in Section \ref{Sec:PseudoControl} (by substituting $e$ by $\bar{e}$), it can be concluded that both the tracking error $\bar{e}$ and the estimation error of the weight matrix $W$ are bounded. In this regard, due to the substitution of $e$ by $\bar{e}$ in the updating rule of the NN weights, they can be satisfactorily updated even at the time of input saturation owing to the elimination of the effect of input constraints from $\bar{e}$ using the introduced term $\nu_h$. However, concerning the boundedness of the real tracking error $e=x-x_d$, there is a need for a restrictive assumption, i.e.
\begin{equation}
  \left\|\int_{0}^{t} \nu_h(\tau) d\tau \right\| \leq \nu_M,
\end{equation}
with $\nu_M$ is a positive constant.

This approach has been utilized in \cite{Johnson.2000,Johnson.2003} to control the attitude of a Reusable Launch Vehicle (RLV) considering actuator faults. As discussed in \cite{Johnson.2003}, even in the lack of system controllability, the adaptation mechanism is satisfactory, which results in a rapid recovery once the system controllability is retrieved.
Similarly, PCH has been adopted in \cite{Hovakimyan.2001,Johnson.2002,Johnson.2005,Abaspour.2015} in a trajectory tracking control problem applied to a helicopter, where the PCH technique has been employed in both the inner and outer control loops.
As a result, the interaction between adaptive elements in the outer loop and the characteristics of the inner loop can also be avoided.
The same approach has been utilized in \cite{Johnson.2006} to control the trajectory of a ducted-fan VTOL UAV.
Further, PCH has been employed in \cite{Lungu.2016} to overcome actuators' nonlinearities in the landing control of a fixed-wing aircraft.

\subsubsection{Employment of a modified tracking error} \label{Sec:InputConst1}
Another effective approach to handle different types of input constraints with less restrictive assumptions is to introduce an auxiliary state variable corresponding to a \emph{filtered} version of the effect of input constraints. More precisely, consider the dynamic model (\ref{Model_1}). Suppose that the real system input ($u$) is obtained as $h(u_c)$, where $u_c$ and $h$ represent the computed (desired) control command and a known nonlinear function, respectively.
Notice that $h(.)$ can represent different types of input nonlinearities such as the saturation function, dead-zone nonlinearity, etc, or user-defined filters \cite{Farrell.2003} to generate feasible control commands according to the physical constraints of the system.
Now, we define an auxiliary variable $\gamma$ as follows:
\begin{equation}
  \dot{\gamma}=-k \gamma +B(x) \delta u, \label{Eq:33_01}
\end{equation}
where $\delta u= u - u_c$. Accordingly, a \emph{modified tracking error} can be defined as
\begin{equation}
  z=x-x_d-\gamma=e-\gamma.
\end{equation}
Notice that, in the absence of input constraints, $\gamma$ tends to zero, and so the introduced modified tracking error reduces to the real tracking error.
Besides, the introduced modified tracking error has a similar formulation to the compensated tracking error used in the command filtered backstepping control (Section \ref{Sec:ComFilBackstepping}).
Considering the following definitions,
\begin{gather}
  u_c= B(x)^{-1}\left(-F(x)-\hat{\Delta}+\dot{x}_d-k e \right), \\
  \Delta={W^*}^T \mu(x)+\varepsilon, \quad \hat{\Delta}=\hat{W}^T \mu(x),\\
  \dot{\hat{W}}=\Gamma \mu z^T, \label{Eq:33}
\end{gather}
and by defining a Lyapunov function as
\begin{equation}
  V=\frac{1}{2}z^T z + \frac{1}{2} tr \left(\tilde{W}^T \Gamma^{-1} \tilde{W}\right),
\end{equation}
one can prove the boundedness of $z$. Thus, assuming that $B(x)$ and $\delta u$ are also bounded (the boundedness of $\delta u$ is a consequence of the system controllability), it is easy to see that $\gamma$ is also bounded, thereby resulting in a bounded real tracking error ($e$).
Besides, even if the system controllability is lost at some time intervals, the updating rule of the NN is still stable thanks to the utilization of the bounded term $z$ rather than the real tracking error in (\ref{Eq:33}).

By comparing the above-mentioned approach with the PCH technique, it can be found that both methods attempt to eliminate the effects of input constraints from the tracking error that is involved in the updating rule of the NN parameters, while the employment of the low-pass filter (\ref{Eq:33_01}) in the current scheme relaxes the necessary assumption on the residual term induced by input constraints.

Such an approach has been employed in \cite{Sonneveldt.2007,Sonneveldt.2009}  in the framework of the command filtered backstepping control.
The same technique has been adopted in \cite{Butt.2013b} to control the longitudinal mode of an HFV using a DSC design.
Second-order filters have been utilized in these papers ($h(.)$ is defined as a linear second-order transfer function) to deal with the magnitude, rate, and bandwidth limits of the control commands.
Notably, as discussed in \cite{Sonneveldt.2007}, the constraints on the system states can also be similarly taken into account by filtering the virtual control commands in the backstepping control.
Modified tracking errors have also been utilized in \cite{Chen.2015c} and \cite{Cheng.2020}, respectively, in a backstepping control and an SMC to deal with input saturation, where the designed controller in \cite{Cheng.2020} has been applied to the longitudinal model of an air-breathing flexible HFV.
Similarly, a modified tracking error has been adopted in \cite{Wang.2015b} to tackle input saturation in a backstepping control scheme applied to the longitudinal dynamic model of an HFV considering additive faults, which has been estimated and compensated as a disturbance term using a NN.
In addition, an analogous approach has been employed in \cite{Wu.2017} in a DSC design to control the longitudinal dynamic model of a morphing aircraft in the presence of input saturation.
Besides, a somewhat similar scheme, borrowed from \cite{Chen.2011c}, has been employed in \cite{Liu.2019b} to deal with partial state constraints in an integrated guidance and control design for skid-to-turn missile using DSC.
More precisely, an auxiliary state variable $\gamma$ has been defined in \cite{Liu.2019b} based on $\delta u$, where $\gamma$ has been involved in the desired virtual control command instead of the tracking error, while the given stability analysis in the paper requires some revision.

\subsubsection{Neuro-predictive control}
Model predictive control (MPC) is an advanced control method that can satisfactorily deal with input, state, and output constraints. More precisely, an optimization problem is constructed to minimize the tracking error within a prediction horizon, as well as the control effort within a control horizon, while considering the system constraints. The optimization problem is solved at each time step. The first element in the computed control sequence is applied to the system, and the entire process is repeated in future steps. Despite the numerous advantages of MPC in dealing with nonlinear, MIMO, and constrained system dynamics, there are significant concerns regarding the stability analysis of the system and the high computational burden of MPC. The stability of the closed-loop systems can be ensured using terminal costs and terminal constraints \cite{Mayne.2013}. However, such stabilizing terminal conditions can make the optimization problem infeasible. In this regard, the \emph{recursive feasibility} problem has been extensively addressed by researchers to provide a feasible control design with guaranteed stability \cite{Besselmann.2012,He.2014}. On the other hand, different practical MPC schemes have been introduced in the literature to provide a computationally efficient control system for real applications \cite{awrynczuk.2014}.

Concerning the application of MPC in IFCSs, it should be noted that NNs can be adopted in the framework of MPC in different ways.
A straightforward way is the employment of a (typically recurrent) NN to learn the system dynamics as a prediction model and utilizing it in the structure of MPC.
A NARX NN, with an RLS optimization-based online adaptation, has been used in \cite{Akpan.2011b} as the prediction model of a 6-DOF F-16 fighter aircraft, and afterward, it has been incorporated in an MPC to control the vehicle's attitude considering input constraints.
In \cite{Garcia.2015}, an adaptive feedforward NN has been employed to estimate the translational acceleration of a fixed-wing aircraft in a moving time window, where the identified model has been adopted in an MPC-based trajectory tracking scheme in the presence of input constraints and model uncertainties. However, the closed-loop stability has not been analyzed in these studies owing to the complicated structure of the proposed nonlinear MPC.
Multiple model-based MPC using a set of local NARX NNs as the prediction model of the system has also been introduced in \cite{Emami.2019b}, where both the system constraints and actuator faults have been considered in the control design process.

On the other hand, regarding model-based approaches, a linear MPC has been proposed in \cite{Yan.2012b}, where the linearization error and unmolded dynamics have been estimated by a feedforward NN in an offline identification problem. The deigned control system has been employed in the altitude control of a helicopter, while the estimation error of the NN has not been considered in the design process, and the closed-loop stability has not been analyzed.
In addition, a fault-tolerant MPC has been introduced in \cite{Emami.2019c}, where an OSELM-based algorithm has been adopted to identify actuator faults. Also, the input constraints have been considered in the given design, and the estimation error of the identification block has been compensated by a DO. Further, the closed-loop stability has been proved using terminal constraints, while there are concerns regarding the feasibility problem.

\subsubsection{Using Nussbaum function}
In \cite{Wen.2011}, Nussbaum functions have been employed in a backstepping control scheme to overcome the input saturation. To this end, the saturation function is approximated by a smooth function $g(v)$, and the approximation error is included in an unknown disturbance term. Subsequently, a Nussbaum function has been utilized to handle $\partial g/ \partial v$, which is created in the last step of the backstepping control design. However, there are concerns about the stability analysis of the proposed approach given in the paper.
To be more precise, although the Input-to-State Stability (ISS) assumption has been employed in the paper, the boundedness of the introduced Lyapunov function has been proved considering the input saturation without using the ISS condition, while this is an irrational result.
Based on such a method, A DSC has been proposed in \cite{MouChen.2014} to control the attitude of an NSV considering the input saturation and external disturbances. Surprisingly, there is no assumption on the stabilizability of the system to ensure closed-loop stability in the presence of input saturation. Apparently, this is due to the employment of the aforementioned theorem in \cite{Wen.2011}.
Similarly, considering a more stringent problem, a backstepping control has been developed in \cite{Li.2017c} for a SISO model of a helicopter to control the pitch angle of the vehicle in the presence of input and output constraints. Neural networks have been employed to identify model uncertainties, while disturbance observers attempt to compensate for unknown external disturbances. Again, Nussbaum functions have been used to deal with the input saturation, where there is no assumption on the stabilizability of the air vehicle. In this regard, further investigations should be conducted by researchers to evaluate the application of the Nussbaum function in the control of constrained systems.
But, similar to the discussion presented in Section \ref{Sec:UnknownDirection}, Nussbaum functions have been successfully adopted in \cite{Xu.2015b} to deal with dead-zone input nonlinearity as an unknown control gain function.

\subsection{Consideration of state/output constraints}
As discussed previously, the use of modified tracking errors and the MPC framework can be beneficial in dealing with state constraints, as well.
Meanwhile, there are other approaches in the literature to cope with state/output constraints in the structure of IFCSs.
The most commonly used method for this purpose is the employment of Barrier Lyapunov Functions (BLFs).
A barrier function is defined as a function, $f(z)$, which tends to infinity as its variable, $z$, tends to a predefined bound \cite{Ngo.2005}.
Accordingly, considering a desired bound $k_b$ for the tracking error $e=x-x_d$, a BLF can be defined as follows \cite{Tee.2009}:
\begin{equation} \label{Eq:34}
  V_0=\frac{1}{2}\ln \left(\frac{k_b^2}{k_b^2-e^T e}\right),
\end{equation}
which is a positive definite function for $\|e\|<k_b$ (it is assumed that $\|e(0)\|<k_b$). Thus, considering the dynamic model (\ref{Model_1}) and using the same NN function approximation as given in Section \ref{Sec:FEL}, a Lyapunov function can be defined for the system as
\begin{equation}
  V=V_0+\frac{1}{2} tr \left(\tilde{W}^T \Gamma^{-1} \tilde{W}\right).
\end{equation}
This results in the following equation.
\begin{equation}
 \dot{V}=\frac{e^T}{k_b^2-e^T e}\left(F(x)+B(x)u+\Delta-\dot{x}_d\right)+tr \left(\tilde{W}^T \Gamma^{-1} \dot{\tilde{W}}\right).
\end{equation}
Consequently, using the control command (\ref{Ctrl_1}) and by defining the following updating rule,
\begin{equation}
  \dot{\hat{W}}=\Gamma \mu \frac{e^T}{k_b^2-e^T e},
\end{equation}
we have:
\begin{equation}
   \dot{V}=\frac{e^T}{k_b^2-e^T e}\left(-k_1 e + \varepsilon \right),
\end{equation}
which ensures the negative definiteness of $\dot{V}$ for $\|k_1 e\|>\|\varepsilon\|$, thereby guaranteeing the satisfaction of $\|e\|<k_b$ (assuming that $k_b>\|\varepsilon/k_1\|$).

Such a control formulation can also be employed in the backstepping control design to impose both the state and output constraints on the controlled system.
In \cite{Wang.2014b}, a BLF has been adopted in a backstepping control scheme in the position control loop to keep the trajectory tracking error of a quadrotor UAV in a desired bound. Similarly, a BLF has been utilized in \cite{Li.2017c} within a backstepping controller to control the pitch angle of a 3-DOF helicopter considering output constraints. The constraint on the angle of attack (AOA) has also been dealt with by a BLF in \cite{Xu.2019} in a backstepping controller applied to the longitudinal mode of an HFV. In addition, in \cite{Wu.2017}, both the velocity and altitude constraints have been considered in a backstepping design using BLFs, where the designed control system is applied to the longitudinal dynamic model of a morphing aircraft.

As discussed in \cite{Fu.2018}, the satisfaction of output constraint using the BLF is achieved at the expense of excessive control effort in the case of approaching the tracking error to the boundaries of the permissible region.
Accordingly, there is a trade-off between choosing a narrow range for the outputs' tracking error and reducing the control effort.
More specifically, as a typical BLF imposes a constant upper bound on the system output, it may lead to large control inputs at initial times.
An asymmetric BLF with time-varying bounds has been employed in \cite{Fu.2018} to deal with time-varying output constraints in which the constant parameter $k_b$ in (\ref{Eq:34}) is substituted by a function of time. The introduced scheme has been utilized in a DSC design in case of the attitude control of a multi-rotor UAV.

Another effective approach to tackle output constraints using a time-varying funnel-like bound is known as the \emph{funnel control}.
The key point of the funnel control is to construct a time-varying gain to control a dynamic system in such a way that the (norm of the) tracking error falls within a funnel boundary $\frac{1}{\varrho(t)}$, where $\varrho(t)$ is a continuous bounded function \cite{Ilchmann.2007}. To be more specific, the funnel control is somewhat similar to the BLF-based approach, where the Lyapunov function (\ref{Eq:34}), in the case of a single-output system, is changed to
\begin{equation}
  V_0= \frac{1}{2}\left(\frac{e}{\Phi(t)-|e|}\right)^2,
\end{equation}
where $\Phi(t)=\frac{1}{\varrho(t)}$. Such an approach has been employed in \cite{Bu.2018b} in a backstepping design to deal with both the velocity and altitude constraints in the case of the longitudinal mode of an air-breathing HFV with a nonaffine model.
Similarly, a Lyapunov function has been introduced in \cite{Li.2020} as
\begin{equation}
  V_0= \frac{1}{2}\tan^2\left(\frac{\pi e}{2 \xi(t)}\right),
\end{equation}
where $\xi(t)$ represents a funnel-like function. This method has been utilized in \cite{Li.2020} to control an HFV using a typical backstepping control in the presence of external disturbances and actuator faults.
Although such approaches suffer from the same issue as the BLF scheme, i.e. the excessive control effort in the vicinity of the permissible output boundaries, the initial large control inputs can be avoided due to the employment of a funnel boundary.

\subsection{Self-organizing neural networks}

Although due to the universal approximation property, NNs (or FNNs) can approximate almost all nonlinear functions with an acceptable estimation accuracy, the determination of the appropriate number of hidden nodes (or fuzzy rules) in the network is not an easy task.
In addition, the development of a variable structure NN rather than a fixed-structure NN (with only variable parameters) provides greater power to deal with time-varying characteristics of dynamics systems.
Several self-organizing NNs have been introduced in the literature to deal with such issues. Further, a self-organizing FNN can eliminate the requirement for prior knowledge about the system \cite{Ferdaus.2019b}.
Typically, a set of growing and pruning algorithms are defined in a self-organizing network to add or remove hidden nodes to (/from) the NN when necessary. As a result, a set of modifications may be required in the network's parameters (at the time of the change in the network's architecture) to ensure the continuity of the network output.

As a traditional approach in this field, Minimal Resource Allocation Network (MRAN) was introduced in \cite{Lu.1998}, which has been developed based on RBFNNs. The growing phase in MRAN is activated if 1) the incoming data is far away from the center of existing hidden nodes, 2) the estimation error in the current step is larger than a predefined threshold, and 3) the root mean estimation error over a moving window is larger than a predefined threshold.
Such an approach can be considered as a \emph{clustering} problem. In this regard, the center of the newly added node is set to the last incoming data, while the output weight of that neuron is equal to the current estimation error of the network. On the other hand, a hidden node is pruned from the network architecture if the normalized output of that neuron becomes less than a predefined threshold in a specific number of consecutive steps. In addition, the network parameters are trained using either a Least Mean Squares (LMS) optimization or an EKF algorithm. An extension to MRAN, called Extended MRAN (EMRAN), has also been introduced in \cite{Saratchandran.2000} in which only the parameters of the nearest neuron to the current input data are updated at each step. This leads to a significant reduction in the online computational burden of the algorithm.
Such a learning strategy has been adopted in several studies.
In \cite{Li.2004}, an MRAN-aided $H_\infty$ control is incorporated in an auto-landing control problem of a conventional aircraft considering external disturbances and actuator faults. The NN, which was trained using the FEL method, augments the control command of the baseline controller.
Similarly, an EMRAN-aided controller has been proposed in \cite{Pashilkar.2006b} to control a fighter aircraft in the landing phase considering actuator faults and severe winds, where the NN attempted to learn the inverse dynamics model of the system using a FEL scheme.
However, the closed-loop stability has not been analyzed in these two papers.
In a similar manner, EMRAN has been adopted in \cite{Ismail.2014} to augment a baseline controller, combined with an SMC to ensure the closed-loop stability, where the designed controller has been applied to an auto-landing problem considering actuator faults and severe winds.

The ambiguity in how to determine the parameters employed in an MRAN is a challenging issue, while there is no explicit relationship between the design parameters and the estimation error of the NN. As an alternative, the concept of the \emph{significance of a neuron} has been employed in \cite{Huang.2004} to provide more efficient growing and pruning rules for RBFNNs. The significance of a neuron is defined as the average of its output over all the input samples it has seen. Accordingly, a new neuron is added to the network only if its significance is greater than a chosen learning accuracy, while a neuron is pruned if its significance becomes less than the learning accuracy.
The main concern about such an approach is the complexity of computing the significance of a neuron, which has been determined in \cite{Huang.2004} assuming a uniform distribution for the input data.
Such a concept has been extended in \cite{Ferdaus.2019} to develop the growing and pruning rules within the framework of a Generic Evolving Neuro-Fuzzy Inference System (GENEFIS), which was first introduced in \cite{Pratama.2014}.
An $\epsilon$-completeness criterion has also been utilized in the paper to add a new rule when a new incoming sample cannot be covered by any existing rules. According to this criterion, which has been introduced in \cite{Wu.2001}, the firing strength of at least one fuzzy rule corresponding to each data in the operating range, should not be less than $\epsilon$.
Subsequently, to update the antecedent parameters of the fuzzy rules, the Generalized Adaptive Resonance Theory+ (GART+), which uses the Bayes decision theory, has been first employed to determine the winning rule corresponding to each newly added data, and then, a \emph{vigilance test} has been conducted to investigate the capability of the winning rule to accommodate the newest data.
Further, an SMC-based approach has been adopted to train the consequent parameters of the fuzzy rules.
Alternatively, hyperplane-based clusters have been employed in \cite{Ferdaus.2020}, which removes the antecedent parameters in the proposed neuro-fuzzy system. More precisely, the membership function of each fuzzy rule has been defined according to the distance between the current data point and the corresponding hyperplane, where the hyperplane parameters are considered as the consequent parameters of the network.
The main idea in the introduced self-organizing network has been borrowed from \cite{Ferdaus.2019b} in which a Parsimonious Learning Machine (PALM) has been developed for data regression. However, different from the basic PALM, which requires various predefined thresholds, the growing and pruning mechanisms in \cite{Ferdaus.2020} have been developed using the concept of bias-variance.
Accordingly, by defining the expected squared tracking error of the system as the Network Significance (NS), the NS has been derived as a sum of the bias and variance of the plant's expected output. Then, a high variance of the system outputs has been interpreted as the high complexity of the network, which in turn activates the rule pruning mechanism. On the other hand, the rule growing algorithm is activated in the presence of high output bias, which is induced by an oversimplified network.
Finally, similarly to \cite{Ferdaus.2019}, an SMC-based training method has been adopted to update the consequent parameters of the network.
The proposed schemes in both the above-mentioned research \cite{Ferdaus.2019,Ferdaus.2020} have been utilized in the altitude and attitude control blocks of a hexacopter and a flapping-wing micro aerial vehicle as an aid to a baseline controller.
Similarly, a self-constructing FNN has been introduced in \cite{Yu.2018}, where the distance between the incoming data and existing clusters has been considered as a measure to add a new rule, while the distance between the existing clusters has been analyzed to prune insignificant rules. The obtained self-constructing FNN has been utilized to approximate the upper bounds of model uncertainties, while it has been employed in an FTC applied to a longitudinal model of a fixed-wing aircraft.
Notice that, despite the development of various effective self-organizing NNs in the literature, significant concerns still remain about the optimality of the network's architecture. As a consequence, the development of a truly \emph{generic} approach to construct an \emph{optimal} network structure depending on different characteristics of the obtained data from a plant is an important research direction, which must be addressed in future studies as a critical step in developing a fully autonomous control scheme.

Finally, concerning multiple-model-based structures, a self-organizing multi-model ensemble has been given in \cite{Emami.2020}, in which a new local NN is added to the proposed multi-model structure if the estimation error of the entire model exceeds a predefined threshold. In addition, a local NN is considered as an insignificant model and pruned from the entire model if the normalized weight of the model in the entire scheme becomes less than a predefined threshold. The proposed approach has been employed in the paper to identify the time-varying dynamic model of a fixed-wing aircraft at different flight conditions.


\subsection{Concerns with air vehicle's characteristics} \label{Sec:Modelcharacteristics}

The position and attitude of an air vehicle can be determined using the kinematic equations based on the translational and angular velocities, respectively. As a result, the position and attitude can be controlled indirectly in a backstepping scheme, where in the first step, the position (or attitude) controller is designed, and the second step deals with the control of the translational (or angular) velocity. Besides, the measurement noises or the simplifications in the kinematic equations, which appear in the first step of the backstepping controller, can be estimated and compensated (as an uncertain term) by NNs \cite{Fu.2018}.

On the other hand, the dynamic equations of an aerial vehicle can be generally derived using either the Newton-Euler or Euler-Lagrange methods.
Regarding conventional multi-rotor VTOL UAVs (with no tilt-rotor), the system dynamics are divided into the rotational and translational equations, where, due to the under-actuated dynamics of the vehicle, the translational motion (typically in the horizontal plane) should be indirectly controlled by the vehicle's attitude.
Thus, a straightforward control method to deal with such a dynamic model would be a multi-loop control design wherein the desired attitude, which is controlled by the inner loop, is determined using the translational dynamics in the outer control loop \cite{Efe.2011b,Li.2016d,Song.2019}.
Such a framework may also be expressed within a backstepping control scheme.
More precisely, the first step of the backstepping controller would deal with the translational dynamics, while the attitude dynamics have been addressed in the second step \cite{Wang.2014b,Dierks.2010}.
In this regard, the desired attitude is indirectly determined (usually by employing the inverse kinematics method) to provide the desired forces required in the outer loop (or in the first step of the backstepping controller) \cite{Das.2009,Kayacan.2016,Xu.2018b}.
Accordingly, owing to the complicated relationship between the vehicles' attitude and the translational dynamics, it may be a requirement for a control law (in the inner loop) that can guarantee the asymptotic stability of the inner loop (rather than a bounded tracking error). Otherwise, the possible tracking error caused by the inner control loop should be considered in the outer loop, while it can complicate the stability analysis. Further, as discussed earlier, uncertain forces and moments induced by uncertain dynamics or external disturbances in the translational and rotational dynamic models can be estimated by distinct NNs.

Besides, a similar framework can also be employed in the case of a helicopter.
In addition to the inverse kinematics method to determine the desired attitude (or the desired angular velocity) \cite{Johnson.2005,Kang.2019b}, it is also possible to define a virtual control input for the attitude dynamics in a backstepping control scheme \cite{Zou.2015b}.
Such a virtual control would be computed according to the translational dynamics of the vehicle, which have been addressed in the previous steps of the backstepping design, by taking into account the kinematic equations (corresponding to the rotation matrix or the quaternion) in such a way that the closed-loop stability can be analyzed based on the Lyapunov stability theorem \cite{Kuo.2021}.
Again, NNs can be adopted to compensate for model uncertainties, external disturbances, or the inversion model error (in the pseudo-control strategy) \cite{Lee.2005} in each loop.
By employing a backstepping scheme in \cite{Ge.2010} for the attitude control of a helicopter, the uncertain control gain matrix ($g_i$) in the dynamic model has been estimated by a distinct NN, while the extra terms due to the minimum estimation error of that NN ($\varepsilon$) has been considered in the last step of the backstepping design corresponding to the actuator dynamics. However, the proposed design suffers from the chattering phenomenon.
Further, the issue of unknown inertia matrix has been dealt with in \cite{Lai.2016} wherein an additional adaptive rule has been defined to estimate it (while taking advantage of the Cholesky decomposition).

A somewhat similar control framework can also be designed in the case of conventional fixed-wing aircraft. A backstepping design has been employed in \cite{Sonneveldt.2009}, where the desired trajectory is first transformed to the desired velocity, flight path, and heading angles using the kinematic equations and subsequently, they are converted to the desired thrust force and angular velocity using dynamic equations.
Finally, in the last step, the control-surface deflections have been determined according to the desired angular velocity, where unknown forces and moments have been estimated using the FEL method. In such a framework, the outer loop is typically considered as the guidance loop, while the inner loops are known as the main control system.
A similar approach has been utilized in \cite{Emami.2019b}, where the desired trajectory is first transformed to the desired Euler angles (and subsequently to the desired angular velocity) in the guidance loop, while in the inner loop, the actuator deflections have been determined based on the desired angular velocity.
Another approach is to decouple the control problem of the longitudinal and lateral-directional modes of a fixed-wing air vehicle (using some simplifications) and address them separately \cite{Kim.1997}. In this regard, several studies have addressed only one of these two subtasks and skipped the other part \cite{Yu.2018}.

In the case of HFVs, almost all of the above-mentioned papers have investigated only the longitudinal model of the vehicle, where the velocity and altitude subsystems are typically decoupled, as well. To this end, the effect of thrust force on the altitude subsystem should be neglected, and the change rate of the velocity is considered much smaller than that of the altitude (known as \emph{time-scale decomposition}) \cite{Xu.2013}. As a result, the velocity subsystem is obtained as a simple SISO model with a single state, while the altitude subsystem includes four state variables including the altitude ($h$), the Flight Path Angle (FPA, $\gamma$), the pitch angle ($\theta$), and the pitch rate ($q$).
Further, the consideration of flexible modes results in introducing additional states, which are not directly involved in the control design process \cite{Butt.2013b,Zong.2014,Xu.2015b}, and they may be considered as a disturbance compensated by a DO.
In this regard, the wind effect, which results in an excessive angle of attack, can also be considered as an unknown disturbance \cite{Xu.2017}.
In such a framework, the main system inputs are the throttle setting and the elevator deflection, which directly influence the velocity and altitude subsystems, respectively. Other systems inputs such as the diffuser area ratio and the canard deflection can also be considered in the design, while there are typically assumed to be constant or a linear function of other system inputs \cite{Zong.2014}.
Besides, a coordinate change has been employed in \cite{Xu.2021b} to deal with the non-minimum phase behaviour of the attitude subsystem (due to the coupling between the lift force and the elevator deflection) that is typically eliminated by the canard deflection as an additional control input in most studies.
Knowing that $h=V \sin \gamma$, a typical method to design a controller for the altitude subsystem is incorporating an intermediate PID controller between $h$ and $\gamma$ to derive the desired FPA \cite{Xu.2016d}, and subsequently transforming the remaining system (including $x=[\gamma,\theta,q]^T$) into a strict feedback form, which can be controlled using a backstepping design \cite{Xu.2012b}.
In this regard, both the direct and indirect adaptive backstepping designs can be used to control such a strict feedback system \cite{Xu.2015}
On the other hand, the backstepping design can be avoided by deriving a normal output-feedback model (in the case of continuous-time systems) \cite{Xu.2011,Xu.2015d} or a prediction model (in the discrete-time domain) \cite{Xu.2013}. In such a circumstance, it may be possible to use only one NN to compensate for uncertain terms in the control command.
Such a normal feedback form has been employed in a pseudocontrol strategy in \cite{Bu.2018} to deal with nonaffine dynamics of the vehicle.
There also few works in the literature, which have addressed the coupled dynamics of the velocity and attitude subsystems. A combination of singular perturbation theory and implicit function theorem has been incorporated in \cite{DaoxiangGao.2014} to deal with the longitudinal model of an HFV in a unified manner, while conservative assumptions on the dynamic model are required in the proposed control scheme. In a more effective way, a neural backstepping controller has been proposed in \cite{Peng.2020} for a MIMO model of an HFV considering the coupling between the velocity and attitude subsystems, where a combined adaptive design and a DO (as discussed in Section \ref{Sec:CombinedNNDO}) has been adopted.

\section{Towards truly model-free control systems} \label{Sec:Model-free-approaches}

\subsection{Neural network-based system identification}\label{Sec:Model-free-Indirect}
As discussed earlier, in the basic indirect FEL-based control, a \emph{nominal} model of the system is derived and subsequently, a set of NNs are employed to identify model uncertainties and external disturbances as a \emph{single term} \cite{Lungu.2016}.
This type of dynamic modeling leads to a conservative control design with reduced efficiency.
To be more precise, most difficulties in modeling a complex dynamic system may originate from the existence of hidden states in the system, not from the model uncertainties caused by a lumped disturbance \cite{Abbeel.2010}.
In addition, in many of the above-mentioned papers, an initial controller was designed based on a nominal model of the system, and then a control augmentation was proposed considering instantaneous model uncertainties. However, in the case of severe structural damages or significant dynamic changes, such an approach may lead to excessive control effort or even closed-loop instability \cite{Suresh.2005}.

In this regard, the development of a valid dynamic model of the system is a crucial task in the control theory. Typically, this is performed by two different approaches: the first-principles modeling and the system identification. Obtaining an acceptable dynamic model of an air vehicle using the first-principles modeling requires detailed information about the aerodynamic and propulsive forces and moments acting on the vehicle, which may not be practical for complicated systems.
Alternatively, system identification attempts to fit a mathematical model on the obtained system inputs-outputs. It can be effectively employed to identify the system dynamics, particularly in the case of complex systems.
However, the employment of such a black-box model identification suffers from the lack of interpretability of the obtained model \cite{Gu.2019}.
There are various studies in the literature that have demonstrated the superior performance of the integration of two methods compared to that of only one method \cite{Li.2014b,Tang.2014,Hamel.2014}, while such an approach remains an open research topic in the framework of IFCSs \cite{Gu.2020}.

In contrast to the FEL method, there are a variety of intelligent controllers in the literature that include separate identification and control design processes.
In this regard, the previously mentioned unique capabilities of NNs make them an ideal candidate to be used in the identification process of such control systems. Different feedforward and Recurrent Neural Networks (RNNs) have been employed for this purpose, where the training process of the network would be performed in the framework of the supervised learning using either offline or online approaches (or a combination of them).
More precisely, the system identification and the control design processes can be fulfilled sequentially or simultaneously in an iterative manner, which is also known as \emph{iterative learning control}.
As a result, the iterative learning control can effectively deal with time-varying dynamic systems, and consequently, it can be classified as an intelligent control system, while the employment of a pre-trained NN in the flight control system may not be considered as an IFCS.

Further, the identification problem using a NN can be considered as an \emph{optimization} problem in which the NN's parameters are determined by minimizing the prediction error of the NN with respect to adjustable parameters of the NN.
Therefore, different optimization algorithms (from traditional approaches such as the gradient-descent method, the Gauss-Newton method, the Levenberg-Marquardt (LM) method, etc., to heuristic methods such as the genetic algorithm \cite{Leung.2003}) can be effectively employed to train the NN parameters. Such optimization algorithms have been thoroughly discussed in the literature \cite{Hagan.2016}, and thus, in the following, only the Online Sequential Extreme Learning Machine (OS-ELM) method will be briefly introduced as a conventional online training algorithm in the structure of indirect adaptive flight control systems.
Besides, concerning the network structure, some of the most commonly used NNs for identifying the system dynamics in the case of an IFCS will be discussed in the following.

\subsubsection{Single-hidden-layer neural networks}\label{Sec:SingleHiddenLayerNN}
As a special case of state-space modeling of a dynamic system, input-output representation is a simpler popular approach to model nonlinear systems \cite{Savran.2006}. Using such a formulation, the system output (in the discrete-time domain) can be represented as follows:
\begin{align}
\begin{split}\label{Eq:50}
      y(k) = h\left(y(k-1),\cdots,y(k-P), u(k-1),\cdots,u(k-M),d(k)\right),
\end{split}
\end{align}
where $u$ and $d$ denote the system input and the vector of disturbances, respectively. Also, $P$ and $M$ represent the number of past outputs and inputs employed in the modeling. Here, $h$ is an unknown function, which should be identified.
Also, considering (\ref{Eq:50}), the system states are $\left[y(k),\cdots,y(k-P)\right]^T$.

In this regard, the assumption on the influence of external disturbances and noises on the system dynamics results in introducing two conventional model structures.
More specifically, in the presence of the \emph{state noise}  (which is also known as the \emph{equation error}), the dynamic model (\ref{Eq:50}) can be simplified as
\begin{align}
\begin{split}\label{Eq:51}
      y(k) = f\left(y(k-1),\cdots,y(k-P), u(k-1),\cdots,u(k-M)\right)+d(k),
\end{split}
\end{align}
where $f$ and $d$ represent, respectively, an unknown nonlinear function and an additive disturbance term. Such a model is known as a Nonlinear Autoregressive with exogenous inputs (NARX) model, which can be identified by a NARX NN. According to (\ref{Eq:51}), a NARX NN employs the past \emph{measured system outputs} and system inputs as the network inputs.
Consequently, a NARX NN can be considered as a feedforward NN with taped delay lines. The use of the NARX structure in training the network is also known as \emph{series-parallel} identification \cite{daCostaLopes.2015}.

On the other hand, in the presence of the \emph{output noise}, the system dynamics model can be formulated as
\begin{align}
\begin{split}\label{Eq:52}
      x(k) & = f\left(x(k-1),\cdots,x(k-P), u(k-1),\cdots,u(k-M)\right), \\
      y(k)& = x(k)+d(k).
\end{split}
\end{align}
Accordingly, the system output at each step is a function of the disturbance that occurs at the same time step only. Nonlinear Output Error (NOE) NNs can be utilized to identify such dynamic models, where the NN employs the past \emph{network's outputs} and system inputs as the network input. Thus, an NOE NN would be a recurrent network, and employing such a scheme in the training process results in a \emph{parallel} identification method \cite{Dreyfus.2005,Zhang.2004}.
Further, a combination of (\ref{Eq:51}) and (\ref{Eq:52}) can be taken into account to consider both the state and output noises in the dynamic model, simultaneously. Such a method, which results in a Nonlinear Autoregressive with Moving average and exogenous inputs (NARMAX) model, requires both the past measured system outputs and model outputs in the dynamic model \cite{Dreyfus.2005}.
Besides, a class of identification techniques has been introduced in the literature to combine the advantages of both the parallel and series-parallel approaches, which are typically developed based on the average of the measured outputs and the predicted outputs.

Concerning the difference between these two types of NNs, the use of series-parallel identification prevents several complexities of training a recurrent network, thereby guaranteeing the training convergence. In addition, the series-parallel structure, which is also known as the teacher forcing method, leads to a faster training speed \cite{Talebi.2010b,Akpan.2011b}.
On the other hand, parallel identification suffers from stability problems and complicated training methods \cite{Patan.2012,Patan.2015}. However, it should be noted that a NARX NN can be used only in the case of the \emph{single-step ahead prediction}, while in multi-step ahead predictions, there is a need for an NOE network.
Although one can convert the NARX neural network after the training process to an NOE network, the use of the series-parallel approach in the training phase may lead to inaccurate predictions for long prediction horizons \cite{Zhang.2004}.

NARX NNs have been used in \cite{Savran.2006} to identify the nonlinear model of an F-16 aircraft using a hybrid offline-online training algorithm considering model uncertainties. The NN's parameters have been trained using the LM method. Subsequently, the obtained model has been employed in a fault-tolerant NN-based adaptive PID attitude control system.
A similar identification technique has been used in \cite{Akpan.2011b} for a similar aircraft model, where the identified model has been utilized in a predictive attitude controller. To train the network parameters, an adaptive updating rule with exponential forgetting has been derived based on a recursive formulation of the Gauss–Newton method, which is similar to the OS-ELM algorithm introduced in the following.

On the other hand, concerning the training algorithm of a NN in an identification problem, in contrast to the FEL method, which has been developed based on the Lyapunov stability theorem, a variety of online training algorithms have been introduced in the framework of an open-loop identification problem, which is typically developed based on the minimization of the mean squared prediction error \cite{Hagan.2016}.
As a simple and popular method, OS-ELM, which has been developed based on the Recursive Least Squares (RLS) optimization \cite{Liang.2006}, can be effectively employed in online identification problems. The use of such an approach to identify the system dynamics (based on the RLS optimization) can result in a significantly better performance (compared to the FEL-based method) in the structure of the trajectory tracking control of a damaged aircraft \cite{Nguyen.2008}.
In the following, a brief description of the OS-ELM algorithm is given.
Extreme learning machine (ELM), which can be considered as a single-hidden-layer feedforward neural network with random constant weights and biases in the hidden layer, has been employed in several studies as a part of the control system. This is due to the simple linear learning method of this type of NNs in which only the output weights of the NN are trained during the identification process \cite{Emami.2018b}.
Now, consider an ELM as $f(u)=W^T \mu (u)$ to identify the unknown mapping between system inputs ($u$) and outputs ($y$) in the case of a SISO system (a similar formulation can be presented for MIMO systems \cite{Emami.2019c}).
Considering a set of system inputs-outputs $\mathcal{D}=\{\left(u(k),y(k)\right)|k=1,...,K\}$ with $K$ distinct samples, the introduced ELM can be trained through the data set $\mathcal{D}$. As a result, ideally, we should have $\Phi W=Y$ where,
\begin{gather}
  \Phi=\left[\mu(u(1)) \quad \cdots \quad \mu(u(K))\right]^T, \\
  Y=\left[y(1) \quad \cdots \quad y(K)\right]^T.
\end{gather}

Assuming $K > N$, ${\Phi}_{K\times N}$ becomes a non-square matrix. In such cases, the optimal weights of ELM can be determined using the least-squares optimization as $\hat{W}={\Phi}^\dag{Y}$, where $\Phi^\dag=({\Phi}^T{\Phi})^{-1}{\Phi}^T$ denotes the pseudo-inverse of ${\Phi}$ \cite{Yan.2014}. However, in the case of online training problems, the incoming data are obtained one by one. Thus, $\hat{W}$ can be updated at each iteration using a recursive formulation as follows:
\begin{gather}
  \hat{W}(k+1)=\hat{W}(k)+\ell(k) e(k), \\
  P(k+1)=\left(I-\ell(k) \mu(u(k))^T\right)P(k),
\end{gather}
where,
\begin{gather}
  e(k)=y(k)- \hat{W}(k)^T\mu(u(k)), \\
 \ell(k)=\frac{P(k) \mu(u(k))}{1+\mu(u(k))^T P(k) \mu(u(k))}. \\
\end{gather}

As discussed in \cite{KarlJohanAstrom.2008}, there is a need for a persistent exciting regressor $\mu(u(k))$ to ensure the convergence of $\hat{W}$ to its optimal value. Various types of OS-ELM have been proposed in the literature for different identification and control purposes \cite{Jia.2016,Wang.2016}. OS-ELM with constant or variable forgetting mechanism has been widely utilized by researchers to identify time-varying system dynamics \cite{Zhao.2012,Soares.2016}.
Further, the OS-ELM algorithm can be adopted, in a similar manner, to identify the vector of unknown parameters corresponding to linear-in-parameters model uncertainty in the dynamic model or to relative weights of local models in a multi-model ensemble.
Such an approach has been utilized in \cite{Emami.2019c} to identify the unknown coefficients corresponding to actuator faults in the case of a quadrotor UAV. Subsequently, a trajectory tracking method has been proposed using an acceleration-based model predictive control, which ensures the bounded tracking error in the presence of system constraints.
Besides, a hybrid offline-online identification scheme has been presented in \cite{Emami.2019} for a generic transport model in the presence of actuator faults. A set of local NARX NNs has been first trained under specific flight conditions and actuator faults, and subsequently, they have been aggregated as a single model using a set of adaptive weights updated using an OS-ELM-like approach. A similar method has been adopted in \cite{Emami.2019b} to develop a fault-tolerant trajectory tracking control based on a modified model predictive control. The proposed approach leads to acceptable trajectory tracking even in the presence of unexpected actuator faults and flight conditions.

Although, due to the universal approximation property, NNs can estimate almost all continuous dynamic systems using a sufficient number of hidden nodes, increasing the hidden nodes may lead to the overfitting problem \cite{Srivastava.2014}.
To be more precise, the generalization capability of NNs in modeling the system dynamics is a crucial issue in utilizing them in a wide range of operating conditions, which are not necessarily covered in the training stage \cite{Bansal.2016}. This, in turn, may lead to different considerations about the training of a NN, such as employing PE input signals, selecting appropriate frequency range for input signals according to dynamic modes of the system, determining optimal network structure, etc. These concerns have been thoroughly addressed in the field of system identification \cite{Tischler.2006}, which are beyond the scope of this paper.

\subsubsection{Deep neural networks}
As an alternative, deep NNs utilize more hidden layers rather than increasing the hidden node in a single hidden layer.
In this regard, Convolutional Neural Networks (CNN) can be considered as one of the most important deep NNs.
CNN has a cascade connection structure. Each CNN cell has two layers: the convolution layer and the sub-sampling layer. Also, the last layer is fully connected. The output of a CNN can be formulated as $y(k) = V \Phi(x(k))$, where $\Phi$ represents the operation of hidden layers and $V$ is the weight vector of the final layer.
As discussed in \cite{Yu.2019}, CNN is an extremely powerful tool for the identification of nonlinear systems. This is due to the following facts: the convolution operation in CNN is the same as the input-output relation of the linear time-invariant systems; a CNN employs sparse connectivity and shared weights, thereby reducing the NN parameters and the risk of the over-fitting issue; the multi-level pooling results in a robust identification scheme against the measurement noises. However, despite the above-mentioned characteristics, few researchers have addressed the development of flight control systems using a CNN-based identified model.
CNN has been utilized in \cite{Kang.2019b} to identify uncertain terms induced by hidden states, varying inertia, and aerodynamic disturbances in the dynamic model of a helicopter UAV. More precisely, the dynamic model consists of a simple nominal dynamic model and a set of CNNs.
A two-step optimization process has been  adopted. First, the parameters of a nominal first-principles-based model have been optimized using the least-squares method, where model uncertainties have been neglected at this stage. Subsequently, the parameters of deep CNNs have been determined in an open-loop optimization using the Stochastic Gradient-Descent (SGD) method. The dynamic model has been trained and validated under different aerobatic maneuvers. Afterward, an adaptive backstepping controller has been designed for the air vehicle which ensures semi-global UUB stability. The use of CNNs in the dynamic model to identify different types of model uncertainties results in a less conservative control system compared to conventional FEL-based controllers, which attempt to compensate for only bounded uncertain terms.

Moreover, concerning the direct employment of CNNs in the control system, CNN is an appropriate choice for high-level control schemes (such as localization and path planning), due to its excellent capability in extracting useful information, particularly from images \cite{Giusti.2016,Kim.2015,Carrio.2017}.

\subsection{Neuroadaptive optimal control} \label{Sec:Optimal}
\subsubsection{Optimal control formulation (HJB vs. HJI equations)}
The feedback control law may be obtained using an $H_2$ or $H_\infty$ optimal control problem at each time step.
To be more precise, considering a nonlinear affine model as $\dot{x}=F(x)+B(x)u$, we can define a cost-to-go function as follows:
\begin{equation} \label{Eq:100}
  V(x)=\int_{t}^{\infty}L(x,u)d\tau,
\end{equation}
where $L(x,u)$ represents a running cost function. The introduced cost function is also known as a \emph{value} function if the running cost $L(x,u)$ is considered as a \emph{reward} function (this is the common notation in the framework of reinforcement learning). Notice that, here, it has been assumed that $x_d=0$. Thus, in the case of trajectory tracking problems, we should consider the system dynamics as $\dot{e}=F(x)+B(x)u-\dot{x}_d$ and substitute $x(t)$ in (\ref{Eq:100}) by $e(t)=x(t)-x_d(t)$.
Now, defining the Hamiltonian as
\begin{equation}\label{Eq:101}
  H(x,\lambda,u)=L(x,u)+\lambda^T(t) \left(F(x)+B(x)u\right),
\end{equation}
with $\lambda$ denotes the Lagrange multiplier, the optimal control law can be obtained within the framework of \emph{dynamic programming} by the following equation \cite{Bryson.1975},
\begin{equation}\label{Eq:102}
 0= \min_u H(x,\frac{\partial V^*}{\partial x},u),
\end{equation}
where the superscript $^*$ stands for the optimal solution. In the literature, (\ref{Eq:102}) is known as the Hamilton-Jacobi-Bellman (HJB) equation, while, in general, there is no analytic solution for it.
It is notable that, in the case of unconstrained linear time-invariant (LTI) systems and using a quadratic running cost $L$, the HJB equation reduces to the well-known algebraic Riccati equation \cite{Zhu.2015,Kalise.2020}.

Now, using a quadratic running cost as $L(x,u)=x^T Q x+u^T R u$ with $Q$ and $R$ denote positive definite matrices, one can obtain the optimal control law as
\begin{equation}\label{Eq:103}
  u^*(x)=-\frac{1}{2}R^{-1}B^T(x)\frac{\partial V^*}{\partial x}.
\end{equation}
By substituting (\ref{Eq:103}) in the HJB equation (\ref{Eq:102}), it is obtained that:
\begin{equation}\label{Eq:104}
  x^T Q x -\frac{1}{4} {\nabla V^*}^T B R^{-1} B^T \nabla V^* + {\nabla V^*}^T F(x) =0,
\end{equation}
where $\nabla V^*=\frac{\partial V^*}{\partial x}$. Accordingly, the optimal cost function is determined by solving the differential equation (\ref{Eq:104}) considering the boundary conditions, and subsequently, the optimal control law is computed using (\ref{Eq:103}) at each time.

A similar discussion can be provided in the framework of an $H_\infty$ control problem in which the control objective is to achieve closed-loop stability while attenuating external disturbances. More precisely, consider the nonlinear dynamic model $\dot{x}=F(x)+B(x)u+D(x)w$, where $w$ denotes external disturbances. Accordingly, considering a running cost function $L(x,u,w)$, the optimal control problem can be formulated as \cite{AbuKhalaf.2006}
\begin{equation}\label{Eq:105}
  0= \min_u \max_w H(x,\nabla V^*,u,w),
\end{equation}
where
\begin{equation}\label{Eq:105_1}
  H(x,\lambda,u)=L(x,u)+\lambda^T(t) \left(F(x)+B(x)u+D(x)w\right).
\end{equation}
Eq. (\ref{Eq:105}), which is known as the Hamilton-Jacobi-Isaacs (HJI) equation, represents a minimax optimization problem. It can be referred to as a two-player differential game, where the player $u$ attempts to minimize the cost function while the player $w$ tries to maximize it \cite{Dierks.2010b}. Again, by defining a quadratic running cost as
\begin{equation}\label{Eq:106}
  L(x,u,w)=x^T Q x+u^T R u-\beta^2 w^T P w,
\end{equation}
with $\beta$ and $P$ represent, respectively, a positive constant and a positive definite matrix, the optimal control law and the worst-case disturbance can be obtained, respectively, as (\ref{Eq:103}) and
\begin{gather}
  w^*(x)=\frac{1}{2\beta^2}P^{-1}D^T(x)\nabla V^* \label{Eq:108}.
\end{gather}
By substituting (\ref{Eq:103}) and (\ref{Eq:108}) in (\ref{Eq:105}), the HJI equation becomes as \cite{vanderSchaft.1992,Dierks.2010b}
\begin{gather}
  x^T Q x + \frac{1}{4} {\nabla V^*}^T E \nabla V^* + {\nabla V^*}^T F(x) =0, \label{Eq:109} \\
  E = \frac{1}{\beta^2}D P^{-1}D^T - B R^{-1} B^T.
\end{gather}
In this regard, $V^*$ should be determined by solving the HJI Partial Differential Equation (PDE) (\ref{Eq:109}) considering the boundary conditions, and the optimal control law is then computed using (\ref{Eq:103}) at each time. Unfortunately, finding the solution of HJ PDEs (\ref{Eq:104}) or (\ref{Eq:109}) is not generally an easy task at all. Another challenging issue in such optimal control problems is that they require the complete system dynamics model, which may not be available in real applications.

\subsubsection{Approximate dynamic programming (continuous-time systems)}
Different approaches have been introduced in the literature to provide a numerical approximation for these control problems \cite{Kalise.2020}.  These approaches are typically addressed within the framework of approximate (or sometimes adaptive) dynamic programming (ADP) \cite{Bea.1998,AlTamimi.2008,Si.2004}.
The principal difference between such adaptive controllers and the previously proposed control methods in Section \ref{Sec:Model_based_IFCS} is that here, we attempt to determine the approximate \emph{optimal} control law, adaptively, while the previous control methods do not necessarily satisfy the optimality condition.
\emph{Policy iteration} and \emph{Value iteration} are the well-known methods in the literature to determine the approximate solution of HJ equations \cite{Liu.2014}.
The policy iteration method consists of a policy evaluation and a policy improvement step at each iteration.
In the $i$th iteration, first, the value function $V^{(i)}(x)$ corresponding to the current control law $u^{(i)}(x)$ is computed by solving $H\left(x,\nabla V^{(i)}(x),u^{(i)}(x)\right)=0$, while in the second step, the control law is updated using (\ref{Eq:103}) (a similar approach can also be taken into account in the case of the HJI equation).
Such an iterative method will continue until the convergence of the policy function $u(x)$.
As discussed in \cite{Beard.1997}, the policy iteration algorithm will converge to the optimal solution by having an initial stabilizing control law (policy).
On the other hand, the value iteration method includes an iterative approach for finding the optimal value function, and once the optimal value function is determined, the optimal policy can be explicitly computed using (\ref{Eq:103}) \cite{Sutton.1998}. Unlike the policy iteration, the value iteration does not require an initial stabilizing control law.
In a more general view, however, both methods can be expressed within the framework of the \emph{generalized policy iteration} \cite{Sutton.1998,Vamvoudakis.2010}. The concept of the generalized policy iteration can be defined as a set of interacting approximate policy evaluation and policy improvement steps, in which in the first step, we do not completely evaluate the cost of a given control law, but only update the current cost estimate \emph{towards} that value. Similarly, in the policy improvement step, the control policy is not fully updated to the minimizing policy for the new cost estimate, but we only update the policy \emph{towards} that policy. Nevertheless, the convergence analysis of such an ADP scheme, in a general case, is not trivial.

Owing to the unique capabilities of NNs in learning different nonlinear functions, traditionally, two NNs were employed as the actor and critic network to approximate the optimal policy and value function, respectively \cite{Liu.2014}. Such an approach can be categorized as a Heuristic Dynamic Programming (HDP) scheme \cite{Werbos.1992,AlTamimi.2008}. In the following, we focus on solving the HJB equation, while a similar discussion can be provided in the case of the HJI equation.
Accordingly, we have
\begin{gather}
  V(x)={W_v^*}^T \mu_v(x) + \varepsilon_v,  \hat{V}(x)=\hat{W}_v^T \mu_v(x),\\
  u(x)={W_u^*}^T \mu_u(x) + \varepsilon_u,  \hat{u}(x)=\hat{W}_u^T \mu_u(x),
\end{gather}
where $V(x)$ represents the corresponding value function of $u(x)$, which satisfies $H\left(x,\nabla V(x),u(x)\right) =0$. Thus, it is obtained that
\begin{gather}
  x^T Q x+u^T R u + \left(\nabla V(x)\right)^T \left( F(x)+B(x)u \right)=0, \label{Eq:110} \\
  u(x)=-\frac{1}{2}R^{-1}B^T(x)\nabla V \label{Eq:110_1}.
\end{gather}
Consequently, knowing that $H\left(x,\nabla V^*,u^*\right) =0$, we can define
\begin{multline}\label{Eq:111}
  e_c= H(x,\nabla \hat{V},u) - H\left(x,\nabla V^*,u^*\right) =  x^T Q x  \\
   + u^T R u + \hat{W}_v^T \nabla \mu_v(x) \left( F(x)+B(x)u \right),\\
\end{multline}
where the last equality is obtained using $\nabla \hat{V}(x)=\left(\nabla \mu_v(x)\right)^T \hat{W}_v$. Therefore, an appropriate training rule may be obtained by minimizing $E_c=1/2 e_c^2$. Using a normalized gradient descent algorithm, we have
\begin{align}
\begin{split}\label{Eq:112}
  \dot{\hat{W}}_v & = \dot{\tilde{W}}_v = -\alpha_c \frac{\partial E_c / \partial \hat{W}_v}{\left(1 + \phi^T \phi \right)^2} = -\alpha_c \frac{\phi}{\left(1 + \phi^T \phi \right)^2}e_c \\
   & = -\alpha_c \frac{\phi \phi^T}{\left(1 + \phi^T \phi \right)^2} \tilde{W_v} + \alpha_c \frac{\phi}{\left(1 + \phi^T \phi \right)^2} \left(\nabla \varepsilon_v (x)\right)^T \left(F(x)+B(x)u\right),
\end{split}
\end{align}
where $\phi= \nabla \mu_v(x) \left( F(x)+B(x)u \right)$ and $\tilde{W}_v=\hat{W}_v - W_v^*$.
However, due to the unknown value of $u$, it should be substituted by $\hat{u}$.
As can be observed in (\ref{Eq:112}), such a training algorithm requires the PE condition to ensure $\lambda_{min}(\phi \phi^T) >0$ \cite{Wang.2017c}.
Further, the updating rule of $\hat{W}_u$ would be obtained according to (\ref{Eq:110_1}).
However, there is a need for a nonstandard modification term in this updating, which consists of the cross-product of the actor and the critic networks' weights to ensure the closed-loop stability. The obtained training rule as well as the proof of the UUB stability of the system (under conservative assumptions) can be found in \cite{Vamvoudakis.2010}.

Alternatively, event-triggered optimal control schemes have been introduced in the literature, where the control law is updated only at the time instants that a triggering condition is satisfied, while it remains constant in other times. Such a control scheme can significantly reduce the online computational cost of a controller. An event-triggered optimal control has been introduced in \cite{Vamvoudakis.2014}, where the updating rule of the critic network has been derived similarly to (\ref{Eq:112}). In addition, the actor network's updating rule has been obtained using (\ref{Eq:110_1}) by defining
\begin{equation}\label{Eq:112_01}
  e_u=\hat{W}_u^T \mu_u(x)+\frac{1}{2}R^{-1}B^T(x) \left(\nabla \mu_v(x)\right)^T \hat{W}_v.
\end{equation}
Accordingly, by defining $E_u=1/2e_u^2$, the updating rule of $\hat{W}_u$ at the $j$th triggering instant $t_j$ is obtained as
\begin{align}
\begin{split}\label{Eq:112_02}
  \hat{W}_u(t_j^+) & =\hat{W}_u(t_j)-\alpha_u \frac{\partial e_u}{\partial \hat{W}_u} e_u^T \\
                              & =\hat{W}_u(t_j)-\alpha_u \mu_u(x_j) e_u^T(t_j),
\end{split}
\end{align}
where $\alpha_u$ denotes a positive constant. Similar to the above-mentioned design, there is a need for a robustifying term in the control law to ensure the closed-loop stability while requiring several conservative assumptions. Such an approach has been extended in \cite{Vamvoudakis.2017} to a trajectory tracking control problem by defining an augmented state, which consists of both the tracking error and the desired trajectory. The designed controller has been subsequently applied to a linear model of the elevation of a Quanser helicopter.

Another alternative training rule for the critic network can be derived based on the method of weighted residuals \cite{AbuKhalaf.2005,Vrabie.2009}. More precisely, at each step, $\hat{W}_v$ can be obtained by projecting $e_c$ onto $\partial e_c / \partial \hat{W}_v$ and setting the result to zero, i.e.
\begin{equation}\label{Eq:112_1}
  \left\langle \frac{\partial e_c}{\partial \hat{W}_v},e_c \right\rangle=0,
\end{equation}
where $\langle f,g\rangle =\int f^T g$. Thus, we have
\begin{equation}\label{Eq:112_2}
   \left\langle \phi , \phi  \right\rangle \hat{W}_v +  \left\langle L(x,u) , \phi  \right\rangle=0,
\end{equation}
which leads to
\begin{equation}\label{Eq:112_3}
  \hat{W}_v= - \left\langle \phi , \phi  \right\rangle^{-1} \left\langle L(x,u) , \phi  \right\rangle.
\end{equation}
Indeed, such an approach leads to the solution of the least-squares optimization. Subsequently, an improved control law can be obtained using  $\hat{u}(x)=-\frac{1}{2}R^{-1}B^T(x) \left(\nabla \mu_v(x)\right)^T \hat{W}_v$. This process will continue until convergence.
However, as discussed in \cite{Luo.2015}, such an iterative optimization process still requires rich input signals to ensure the existence of $\left\langle \phi , \phi  \right\rangle^{-1}$. In addition, computing the necessary integrals in (\ref{Eq:112_3}) may be a complicated task in practice. Thus, they are typically approximated by discretization. This method has been employed in \cite{Luo.2015} to successively solve the HJI equation, where the designed controller has been applied to a linear model of a fighter aircraft.

It should be noted that the introduced actor-critic scheme can also be implemented using a single NN \cite{Dierks.2010b}. This can be performed by employing a critic NN to approximate the value function $V(x)$ and subsequently, computing the approximate optimal control law as
\begin{equation}\label{Eq:113}
  \hat{u}(x)=-\frac{1}{2}R^{-1}B^T(x) \left(\nabla \mu_v(x)\right)^T \hat{W}_v.
\end{equation}
Accordingly, there is a need for a modification term in the updating rule (\ref{Eq:112}) to ensure closed-loop stability.
The modification term is obtained by assuming that the optimal control law $u^*(x)$ can stabilize the system such that the following equation holds \cite{Xue.2021}.
\begin{equation}\label{Eq:113_1}
  \dot{J}_s(x)=\left(\nabla J_s\right)^T  \left(F(x)+B(x)u^*\right) = -\left(\nabla J_s\right)^T \Lambda \left(\nabla J_s\right),
\end{equation}
where $J_s(x)$ and $\Lambda(x)$ represent a Lyapunov function of the system (as a polynomial) and a positive definite matrix, respectively
\cite{Liu.2014}. Consequently, the modification term is obtained by preventing the function $J_s(x)$ from increasing as follows:
\begin{align}
\begin{split}\label{Eq:114}
 \Delta \dot{\hat{W}}_v & = - \alpha_s \frac{\partial \dot{J_s}(x)}{\partial \hat{W}_v} \\
                                     & =- \alpha_s \frac{\partial \left[\left(\nabla J_s\right)^T \left( F(x)+B(x) \hat{u}(x) \right)\right]}{\partial \hat{W}_v},
\end{split}
\end{align}
where $\alpha_s$ represents a positive constant.
A similar approach has been employed in \cite{Wang.2017c} in an event-triggered $H_\infty$ control problem to solve an HJI equation, where the proposed method has been applied to a linear model of an F-16 aircraft. Besides, such a scheme has been adopted in \cite{Nodland.2013} in combination with an NN-based state observer to provide a trajectory tracking controller for a helicopter UAV, where the NNs have been trained online by an \emph{on-policy} learning method. In the on-policy learning, the control law that is applied to the system (called the \emph{behavior policy}) is the same as the control law, which is evaluated and improved (called the estimation or \emph{target policy}). On the other hand, in the \emph{off-policy} learning scheme, the behavior policy and the target policy can be unrelated.
The employment of off-policy learning in the control design process provides considerable advantages in comparison with on-policy learning schemes \cite{Luo.2015}. More specifically, in the on-policy $H_\infty$ control, the external disturbance should be obtained by (\ref{Eq:108}), while specifying the disturbance term is typically impractical in real systems. In addition, the issue of the \emph{exploration} (which is partly related to the PE condition) is of significant importance to guarantee the convergence of the control law to the optimal policy. However, since in the on-policy learning, we should apply the target policy to the system, the exploration would be limited by the UAV trajectory.

Further, a remaining issue with all the above-mentioned designs is that they still depend on the system dynamics (i.e. $F(x)$ and $B(x)$). To tackle such a problem, model-free off-policy learning schemes have been introduced in the literature to provide an acceptable approximate solution for the optimal control problem. To this end, consider again the dynamic model of the system. By adding and subtracting the target policy $u^{(i)}(x)$ (at $i$th optimization iteration) to the model, it is obtained that
\begin{equation}\label{Eq:115}
 \dot{x}=F(x)+B(x)u^{(i)} +B(x) \left(u - u^{(i)}\right).
\end{equation}
Thus, considering the value function $V$ corresponding to $u^{(i)}(x)$, we can write
\begin{align}
\begin{split}\label{Eq:116}
   \dot{V}  =  \left(\nabla V\right)^T \dot{x} = \left(\nabla V\right)^T \left(F(x)+B(x)u^{(i)} \right) + \left(\nabla V\right)^T B(x) \left(u - u^{(i)}\right).
\end{split}
\end{align}
As a result, the policy evaluation equation (\ref{Eq:110}) can be reformulated as follows \cite{Luo.2015}:
\begin{equation}\label{Eq:117}
   \dot{V}  = \left(\nabla V\right)^T B(x) \left(u - u^{(i)}\right) - x^T Q x - {u^{(i)}}^T R u^{(i)},
\end{equation}
By integrating from both sides of (\ref{Eq:117}) in a specific time interval, a policy evaluation equation is obtained which is independent of the internal dynamics $F(x)$. Thus, we can redefine the design process by employing the policy evaluation equation (\ref{Eq:117}) rather than  (\ref{Eq:110}). Such an approach, which is similar to the Integral Reinforcement Learning (IRL) scheme \cite{Vrabie.2009b,Zhu.2015}, has been utilized in \cite{Luo.2015} to approximately solve an HJI equation, where the designed controller has been applied to a linear model of the longitudinal dynamics of an F-16 aircraft.
Meanwhile, the model-free approach to optimal control may be better expressed within the framework of reinforcement learning, which will be discussed in the following subsection.

Another concern with above mentioned ADP schemes is that the obtained information from the system inputs-outputs data is used to update only a \emph{scalar} function, i.e. the value function. This results in inefficient usage of data which may slow down the convergence. To deal with such an issue, another actor-critic ADP method has been introduced in the literature, which attempts to approximate the optimal value function derivative $\nabla V^*(x)$ (using the critic NN) rather than the value function itself. This method falls into the framework of Dual Heuristic Programming (DHP) \cite{Lewis.2009}.
Two sets of NNs have been employed in \cite{Han.2002} as the actor and critic networks in a constrained minimum-time optimal control problem, i.e. the control of the flight path angle of a missile given a final Mach number. Indeed, instead of utilizing a single NN, a set of NNs have been trained offline as the actor (and critic) NN, which have been employed sequentially to determine $u^*(x)$ and $\nabla V^*(x)$ during the time. Also, to deal with the free-final time, the dynamic equations of the system have been reformulated considering the flight path angle as the independent variable, thereby providing a fixed-final condition problem.
A similar actor-critic method has been utilized in \cite{Ferrari.2004}, where an offline training stage has been performed using the linearized model of the air vehicle, and an online training phase has been employed to improve the closed-loop performance. The designed controller has been applied to a fixed-wing aircraft considering model uncertainties, unmodeled dynamics, and actuator faults. However, the closed-loop stability was not analyzed in these papers.

\subsection{Direct adaptive control using Reinforcement learning (RL)}\label{Sec:RL}
The concept of adaptive optimal control, particularly for systems with unknown system dynamics, can be presented within the framework of Reinforcement Learning (RL) as well.
Although the notion of RL and the optimal control theory share a somewhat similar idea, i.e. moving towards the optimal solution over time, they possess different mathematical notations due to their different origins \cite{Khan.2012}.
In a conventional RL problem, which is typically formulated in the \emph{discrete-time} domain, the objective is to search for an optimal control law (policy) for a dynamic system (agent) while interacting with an uncertain environment that maximizes the total reward obtained during an episode.
Traditionally, the problem is formulated as a Markov Decision Process (MDP) described by a four-tuple $(x, u, F, R)$. Here, $x$ and $u$ represent, respectively, the current system state and inputs. The system inputs are obtained according to a policy $\pi$, which in turn, results in receiving a reward $R(x,u)$.
Further, $F$ denotes the system dynamics model or (in a probabilistic formulation) a stationary transition distribution $F \sim P\left(x(k+1)|x(k),u(k)\right)$, which satisfies the Markov property \cite{Lewis.2009}
\begin{align}
\begin{split}\label{Eq:119}
    P\left(x(k+1)|x(1),u(1),\cdots x(k),u(k)\right) = P\left(x(k+1)|x(k),u(k)\right).
\end{split}
\end{align}
Thus, the choice of appropriate states to satisfy the Markov property is of significant importance. However, although most of the theoretical achievements within the framework of RL have been obtained under such a property, many approaches can still work well for different practical problems which do not satisfy the Markov property \cite{Kober.2013}.
Now, similar to the common notation in the RL framework, consider a discrete-time optimal control problem as follows:
\begin{gather}
  \max_{\pi} \mathbb{E}_{\pi} \left[\sum_{k=t}^{\infty} \gamma^{k-t} R(x,u) \right], \label{Eq:120}\\
  \text{subject to }x(k+1)=F(x(k),u(k),d(k)). \nonumber
\end{gather}
Accordingly, the objective is to maximize the obtained accumulative reward, $R(x,u)$, and the expected value is computed considering the random external disturbance $d(k)$. Further, $\pi_k$ and $\gamma \in (0,1)$ represent the current control command (policy) and the discount factor, respectively.
Thus, the control command $u(k)$ is computed at each step using either the stochastic or deterministic policy $\pi$ (in the case of stochastic policies, $\pi$ represents the conditional probability distribution of the control command, i.e. $\pi(u|x)$, while concerning deterministic policies, we have $u(k)=\pi(x)$).
It is notable that a similar discussion can also be made on the basis of an average reward rather than a discounted reward, which eliminates the requirement for a discount factor (for more details, see \cite{Kober.2013}).

The focus of this section is on a model-free optimal control approach. Like other adaptive control methods, we can employ either an indirect or a direct control design procedure. To be more precise, it is possible to first derive an estimation of the system dynamics model $F$ and then attempt to (approximately) solve the optimization problem (\ref{Eq:120}), or try directly to develop an optimal control policy.
Within the framework of the RL, the former approach is known as the model-based RL, while the latter corresponds to the model-free RL.
On the other hand, owing to the unstable behavior of typical aerial vehicles and the inherent trial and error scheme employed in RL, commonly, the learning phase should be performed in a simulation environment on an existing \textit{model} of the system.
Thus, even in the model-free RL, there is a requirement for a (simple) dynamic model of the system to be used in the learning phase (in the simulation environment). We will give a short insight into the method of eliminating the requirement for a dynamic model in the RL-based flight control systems at the end of this section.
Concerning the model-based RL, however, the model would be obtained by the system identification method as discussed in Section \ref{Sec:Model-free-Indirect}.
Consequently, apart from the model identification phase (required in the model-based RL), the learning process in the flight control design in both the model-based and model-free RL schemes can be discussed in the same fashion.

Now, in a general view, we can solve an adaptive optimal control problem within the RL framework through two different approaches: ADP and direct policy updating \cite{Recht.2019}, which will be addressed in detail in the following.

\subsubsection{Approximate dynamic programming (discrete-time systems)}
Within the ADP framework, we first attempt to estimate the action-value function $Q(x,u)$, which is defined as follows \cite{Sutton.1998}:
\begin{equation}\label{Eq:121}
  Q(x,u)=\mathbb{E}_{\pi} \left[ \left. \sum_{k=t}^{\infty} \gamma^{k-t} R(x,u) \right| x(t)=x,u(t)=u  \right].
\end{equation}
Notice that, it is also possible to derive the RL-based control formulation using the value function $V(x)$ rather than the action-value function. Indeed, such an approach would result in the discrete equivalent of the previously discussed ADP scheme for continuous-time systems. However, as will be observed in the following, the employment of the introduced action-value function instead of the value function can help to develop an entirely model-free control system \cite{Lewis.2009}.
Now, using the concept of DP, one can obtain the Bellman optimality equation as follows:
\begin{equation}\label{Eq:122}
  Q^*(x,u)=R(x,u)+\gamma \mathbb{E}_{\pi} \left[\max_{u'}  Q^*\left(x(k+1),u'\right) \right],
\end{equation}
where the superscript $^*$ denotes the action-value function corresponding to the optimal policy. Such an approach can be utilized in both the on-policy and off-policy iterative learning schemes to estimate the action-value function.

A traditional on-policy learning approach to iteratively estimate the action-value function is known as \emph{Sarsa}. In this regard, by incorporating the \emph{Temporal difference} (TD) error, which is equivalent to the Hamiltonian introduced in the previous subsection for continuous-time systems, an on-policy learning rule can be derived as
\begin{align}
\begin{split}\label{Eq:124}
  Q_{k+1}(x,u) = Q_{k}(x,u) + \eta \left(R(x,u)+\gamma Q_{k} \left(x(k+1),u(k+1)\right)-Q_{k}(x,u)\right),
\end{split}
\end{align}
where $\eta \in \mathbb{R}^+$ denotes the learning rate. The second term on the right-hand side of the equation corresponds to the TD error.
Subsequently, an improved action is chosen at each step using
\begin{equation}\label{Eq:125}
  \pi_{k+1}(x)=\arg \max_u Q_{k+1}(x,u) .
\end{equation}
Such an approach (using the multi-step TD, which is discussed in the following) has been employed in \cite{Luo.2018} to control a 2-DOF Quanser helicopter. The optimization problem has been presented for a linear model of the system, which leads to the well-known algebraic Riccati equation.
Sarsa has also been adopted in \cite{Reddy.2016} for a quite complex control problem, i.e. the control of glider soaring in a turbulent environment (by taking advantage of turbulent fluctuations), while such a problem has been dealt with in \cite{Reddy.2018} by employing an off-policy value-iteration method.

Accordingly, if we estimate the action-value function corresponding to the current control policy using a NN as $\hat{Q}(x,u)=\hat{W}_q^T \mu_q(x,u)$, the network weights $\hat{W}$ can be updated at each step using the gradient descent method as follows \cite{Sutton.1998}:
 \begin{align}
\begin{split}\label{Eq:126}
  \hat{W}_q(k+1)= & \hat{W}_q(k)+\eta \mu_q(x,u) \big( R\left(x,u\right)+  \\
                              & \gamma \hat{Q} \left(x(k+1),u(k+1)\right) - \hat{Q} \left(x(k),u(k)\right)\big),
\end{split}
\end{align}
where  $\eta$ represents a positive learning rate, and $x$ and $u$ correspond to the current value of the system state and input. As seen, the proposed updating rule is similar to the training rule (\ref{Learn_1}) employed in the FEL scheme, where the tracking error has been substituted by the TD error. A notable point, however, is that (\ref{Eq:126}) is obtained using a \emph{semi-gradient} method rather than a true gradient descent scheme. This is due to the employment of $R\left(x,u\right)+\gamma \hat{Q}\left(x(k+1),u(k+1)\right)$ as the target value of the action-value function, which in turn is a function of $\hat{W}_q$, while the effect of it is not included in the gradient function.

Thereafter, in the policy improvement step, (\ref{Eq:125}) can be solved as $$\partial \hat{Q}(x,u) / \partial u =0,$$ which shows well a principal advantage of employing the action-value function instead of the value function, hence we can simply determine the improved policy at each step by maximizing the Q-function with respect to $u$ with no requirement for the system dynamics model.

On the other hand, concerning off-policy learning methods, an off-policy TD-based learning rule called the \emph{Q-learning} can be derived as
\begin{align}
\begin{split}\label{Eq:123}
 Q_{k+1}(x,u) = Q_{k}(x,u) + \eta \left(R(x,u)+\gamma \max_{u'}  Q_{k} \left(x(k+1),u'\right)-Q_{k}(x,u)\right),
\end{split}
\end{align}
while, again, the improved policy would be determined using (\ref{Eq:125}). Such a learning method has been adopted in \cite{Shi.2018} to learn the optimal servoing gain in an Image-Based Visual Servoing (IBVS) design for the trajectory tracking control of a quadrotor UAV, while the learning rate $\eta$ has been updated using a fuzzy controller.

In a similar manner to Sarsa, the NN-based estimation can also be adopted in the Q-learning algorithm. The corresponding updating rule is obtained by substituting  $\hat{Q}\left(x(k+1),u(k+1)\right)$ in (\ref{Eq:126}) by $\max_{u'} \hat{Q} \left(x(k+1),u'\right)$ or $$\sum_u \pi\left(u|x(k+1)\right) \hat{Q} \left(x(k+1),u\right)$$ in the case of stochastic target policy $\pi$.
Such a scheme has been employed in \cite{Nie.2019} to control an airship in a 3D environment, where the scale of the state space was reduced by a coordinate transformation. Different variants of the Q-learning method have been presented in the literature, which are beyond the scope of this paper (see for example \cite{Khan.2012,Jang.2019}).

In the Q-learning, a common choice for the behavior policy is to choose either the current improved target policy (with a probability of $1-\epsilon$) or a random action (with a probability of $\epsilon$, where $\epsilon$ denotes a small positive constant). Such a behavior policy results in a good exploration, which is critical in the convergence of off-policy algorithms.
As an alternative, an evolutionary exploration algorithm has been introduced in \cite{Won.2017} in which a set of random trajectories are generated at each step while the mean and the variance of them are updated considering the obtained reward corresponding to each trajectory in such a way that the resultant behavior policy moves toward better trajectories. Such an approach has been adopted in \cite{Won.2017} to train a flapping-wing aerial vehicle using the Q-learning.
Nevertheless, NN-based off-policy learning algorithms suffer from convergence issues in various problems. The updating rules, which are derived based on the true gradient descent method (such as the gradient-TD method) can address this issue at the expense of excessive computational complexity, while their performance in real applications is still not clear. In this regard, a comprehensive comparison between the performance of semi-gradient methods and TD methods those based on true gradient descent, in the case of intelligent flight control systems, is a necessity in future research.

It is also possible to derive an (on-policy) updating rule by attempting to eliminate the TD error at each time step using a least-squares optimization (or an RLS optimization similar to the OS-ELM approach introduced in Section \ref{Sec:Model-free-Indirect}) to solve the following equation.
\begin{equation}\label{Eq:127}
  \hat{W}_q(k+1)^T \left(\mu_q(x,u)-\gamma \mu_q \left(x(k+1),u(k+1)\right) \right) = R\left(x,u\right).
\end{equation}
To this end, there is a requirement for the regression vector $$\left(\mu_q(x,u)-\gamma \mu_q \left(x(k+1),u(k+1)\right) \right)$$ to be persistent exciting \cite{Lewis.2009}. Such a method has been utilized in \cite{Palunko.2013} to generate the desired trajectory for a quadrotor aimed to transport a suspended load.

As discussed, both the Q-learning and Sarsa have been developed based on the TD method, where its iterative updating rule bases in part on the \emph{current estimation} of $Q$. Thus, they are known as \emph{bootstrapping} methods. Further, notice that the proposed schemes are developed based on a simple \emph{one-step} TD error. More complex and effective learning rules can be derived by employing \emph{multi-step} TD. Multi-step TD learning is indeed a bridge between the simple one-step TD learning and the Monte Carlo method wherein the updating rule is derived using the entire sequence of rewards obtained from the current state until the end of the episode.
A detailed description of multi-step TD can be found in \cite{Sutton.1998}.
Compared to the TD method, the Monte Carlo approach could not be used in an online training scheme, because we should wait until the end of the episode to determine the obtained rewards corresponding to the current policy. On the other hand, there are some concerns with the convergence of the TD learning, which is a bootstrapping method, particularly under the usage of neural approximation.
The Monte Carlo method has been adopted in \cite{Ng.2006} to maximize a value function in order to develop a controller for a helicopter in low-speed aerobatic maneuvers, e.g. the inverted flight of the aerial vehicle, where the optimization process has been performed in the simulation environment using an identified stochastic, nonlinear model of the system.
Using the Monte Carlo method, a collision-avoidance control system has been proposed in \cite{Sadeghi.2017}. In this regard, after the training of the action-value function, which was modeled by a CNN, the control command, i.e. the velocity direction of the UAV, could be obtained by maximizing the Q-function at each step.
An intelligent trajectory generation approach has been proposed in \cite{Zhang.2018} for a UAV aimed to collect information from the environment considering the constraint on the total energy consumption of the vehicle. A CNN has been utilized to estimate the Q-function using an off-policy modified Deep RL (DRL) method.
In contrast to the TD and Monte Carlo methods, in the DRL method, a replay buffer has been utilized, which stores a finite number of tuples of $(x_k, u_k,r_k,x_{k+1})$ obtained under an exploration (behavior) policy. Subsequently, at each step, a mini-batch of samples is chosen uniformly from the entire buffer allowing for a set of \emph{uncorrelated} samples to be used in the training process. In addition, a copy of the main NN called the target network has been generated, where the target network, which provides the target value for training the main critic network, is trained with a significantly less learning rate, thereby avoiding the learning divergence.
The two above-mentioned high-level control systems can be considered as preliminary intelligent path planning designs, which could be integrated with conventional IFCSs to provide a completely intelligent flight control system.
The development of such a combination would be a critical step to develop a truly intelligent UAV, while, due to the complicated and high-dimensional nature of the problem, it has not been thoroughly addressed by researchers yet.

Despite the simplicity of introduced approaches to approximate the optimal action-value function (and subsequently, the optimal policy), they still face fundamental challenges to ensure the convergence to the optimal solution (particularly in the case of off-policy algorithms). More specifically, different impractical assumptions (such as the requirement for visiting all possible state-action pairs for an infinite number of times) have been adopted in the literature to achieve the convergence property \cite{Sutton.1998}.

\subsubsection{Direct policy updating}
Another approach to solve the optimization problem (\ref{Eq:120}) is to directly update the approximate optimal policy rather than employing an estimated action-value function to find the optimal policy.
More precisely, here, we attempt to directly find an appropriate updating rule for the approximate optimal policy, which is estimated by a NN as ${\pi}(x)= \hat{W}_{\pi}^T \mu_{\pi}(x)$, or ${\pi}(u|x)= \hat{W}_{\pi}^T \mu_{\pi}(x,u)$ in the case of a stochastic policy
(for the ease of notation, we do not use the $\hat{.}$ symbol for the estimated optimal policy in the rest of this section).
Such a direct policy parametrization brings a principal advantage into the control design process that we can incorporate the prior knowledge of the optimal policy in the parametrization of the estimated optimal policy.

In this context, the most commonly used approach, called the \emph{policy gradient} method, attempts to update the network weights $\hat{W}_{\pi}$ by moving in the direction of the gradient of a performance function in order to improve ${\pi}(x)$.
Typically, the value function $V_{\pi} (x)$ (the subscript ${\pi}$ indicates that the value function has been computed along the trajectory obtained by $\pi$) is chosen as the performance function.
In the following, we first give a brief introduction to the policy gradient theorem for stochastic policies and then address the corresponding theorem of deterministic policies as a special case.
Now, defining the advantage function as
\begin{equation}\label{Eq:128}
  A_{\pi} (x,u)=Q_{\pi}(x,u)-V_{\pi} (x),
\end{equation}
one can derive an equation for the difference between the value functions corresponding to two different policies as follows \cite{Schulman.2015,Kakade.2002}:
\begin{equation}\label{Eq:129}
  V_{\pi}(x_0)-V_{\varpi}(x_0)=- \sum_{x} \rho_{\varpi} (x) \sum_{u} \varpi(u|x) A_{\pi} (x,u),
\end{equation}
where,
\begin{equation}\label{Eq:130}
  \rho_{\varpi} (x)=\sum_k \gamma^{k} P\left(x(k)=x | x(0)=x_0\right),
\end{equation}
denotes the unnormalized discounted visitation frequency, where actions are determined according to $\varpi$. Let $\varpi$ be a fixed policy (which may be considered as the behavior policy in off-policy methods) and $\pi$ corresponds to the estimated optimal policy. Thus, using the fact that $\sum_u \pi(u|x) A_{\pi} (x,u) =0$ and $$\sum_u \left(\varpi(u|x)-\pi(u|x)\right) V_{\pi} (x)=0,$$ we have \cite{Pi.2020}:
\begin{align}
\begin{split}\label{Eq:131}
V_{\pi}(x_0) - V_{\varpi}(x_0) & = \sum_{x} \rho_{\varpi} (x) \sum_{u} \left(\pi(u|x)-\varpi(u|x)\right) Q_{\pi} (x,u) \\
                                               & = \sum_{x} \rho_{\varpi} (x) \sum_{u} \left(\pi(u|x)-\varpi(u|x)\right) A_{\pi} (x,u).
\end{split}
\end{align}
By estimating the target policy as ${\pi}(u|x)= \hat{W}_{\pi}^T \mu_{\pi}(x,u)$ and differentiating both sides of the (\ref{Eq:131}) with respect to $\hat{W}_{\pi}$, one can obtain the gradient of the performance function as follows:
\begin{align}
\begin{split}\label{Eq:132}
  \nabla V_{\pi}(x_0) =  \sum_{x} \rho_{\varpi} (x) \sum_{u} \nabla \pi(u|x) Q_{\pi} (x,u)+ \left(\pi(u|x)-\varpi(u|x) \right)\nabla Q_{\pi} (x,u),
\end{split}
\end{align}
where $\nabla=\frac{\partial}{\partial \hat{W}_{\pi}}$. The obtained result is analogous to the off-policy actor-critic algorithm proposed in \cite{Degris.2012}, while the second term on the right-hand side of (\ref{Eq:132}) is neglected in \cite{Degris.2012}.
A similar equation can also be derived by substituting $Q_{\pi} (x,u)$ in (\ref{Eq:132}) by $A_{\pi} (x,u)$.
Now, considering the special case $\pi=\varpi$ in (\ref{Eq:132}), it is obtained that
 \begin{gather}
 \nabla V_{\pi}(x_0) =  \mathbb{E}_{\rho_{\pi},\pi} \frac{\nabla \pi(u|x)}{\pi(u|x)} Q_{\pi} (x,u), \label{Eq:133}\\
 \nabla V_{\pi}(x_0) =  \mathbb{E}_{\rho_{\pi},\pi} \frac{\nabla \pi(u|x)}{\pi(u|x)} A_{\pi} (x,u). \label{Eq:134}
\end{gather}
The first equation is known as the fundamental equation of the policy gradient theorem, while the second one is called the policy gradient with baseline, which in turn reduces the variance of the algorithm, thereby improving the performance.
Now, the NN weights can be updated through either an off-policy or on-policy method using each data sample as follows:
\begin{equation}\label{Eq:135}
  \hat{W}_{\pi}(k+1)=\hat{W}_{\pi}(k)+\eta \rho(x,u)  \frac{\nabla {\pi}(u|x)}{{\pi}(u|x)} \hat{A}_{\pi} (x,u),
\end{equation}
where $\rho(x,u)=\frac{{\pi}(u|x)}{\varpi(u|x)}$, called the importance sampling ratio, is employed to compensate for the fact that the data samples have been collected under the behavior policy $\varpi(u|x)$ rather than the estimated target policy ${\pi}(u|x)$ (in the on-policy learning, we have $\rho=1$). Further,
\begin{equation}\label{Eq:136}
  \hat{A}_{\pi} (x,u)=R(x,u)+\gamma \hat{V}_{\pi} (x(k+1)) - \hat{V}_{\pi} (x),
\end{equation}
denotes an estimation of the advantage function. As seen, it requires the estimation of the value function, which can be obtained by a critic network using the introduced learning schemes in the previous section for the action-value function, while in the case of the off-policy learning of the value function, unlike the learning algorithm of the action-value function, we should again employ the importance sampling ratio in the updating rule \cite{Sutton.1998}.

A variety of \emph{conservative} approximated policy gradient approaches have been introduced in the literature to restrict the policy update at each step, thereby improving the performance of the method.
This is due to the great effect of the \emph{magnitude} of $\Delta\hat{W}_{\pi}$ (which can also be controlled by the learning rate) in each iteration of the policy gradient on the performance and the convergence of the algorithm. Trust Region Policy Optimization (TRPO) \cite{Schulman.2015} and Proximal Policy Optimization (PPO) \cite{Schulman.2017} are two common methods in this field. TRPO employs a constrained optimization problem in which an approximated value function is optimized (through updating the target policy) subject to a constraint on the KL divergence of the old policy and the new policy.
The KL divergence represents a measure of the divergence of a distribution from the other one.
On the other hand, the PPO introduced a simplified design to keep the ratio of the new policy to the old policy in a permissible range. Such an approach has been utilized in \cite{Pi.2020} to develop a trajectory tracking control for a quadrotor air vehicle.

Policy gradient theorem can also be extended to deterministic policies, which is called the Determinist Policy Gradient (DPG) \cite{Silver.2014}. To this end, consider again (\ref{Eq:131}) while substituting the probability distribution $\pi(u|x)$ with the Dirac delta function, i.e. $\pi(u|x)\sim\delta\left(u-\pi(x)\right)$, which is equivalent to a deterministic policy. Subsequently, knowing that $$\sum_{u} \left(\delta\left(u-\pi(x)\right)\right)Q_{\pi} (x,u)=Q_{\pi} (x,\pi(x)),$$ by estimating $\pi(x)$ as $\hat{W}_{\pi}^T \mu_{\pi}(x)$ and differentiating both sides of (\ref{Eq:131}), it is obtained that
\begin{align}
\begin{split}\label{Eq:137}
   \nabla V_{\pi}(x_0) =  \sum_{x}\rho_{\varpi} (x) \bigg(\nabla Q_{\pi} (x,\pi(x))+ \sum_{u}\left(\delta\left(u-\pi(x)\right)-\varpi(u|x) \right) \nabla Q_{\pi} (x,u) \bigg).
\end{split}
\end{align}
Thus, knowing that $$\nabla Q_{\pi} (x,\pi(x)) = \nabla \pi(x) \nabla_u \left. Q_{\pi} (x,u) \right|_{u=\pi(x)},$$ one can obtain the on-policy DPG algorithm by setting $\pi=\varpi$ in (\ref{Eq:137}) as follows:
\begin{equation}\label{Eq:138}
  \hat{W}_{\pi}(k+1)=\hat{W}_{\pi}(k)+\eta \nabla {\pi}(x) \nabla_u \left. \hat{Q}_{\pi} (x,u) \right|_{u=\pi(x)},
\end{equation}
where $\hat{Q}_{\pi} (x,u)$ is the estimated action-value function, which can be obtained using a critic network trained by the Sarsa algorithm through (\ref{Eq:126}).

Concerning the off-policy DPG, note that there is an additional term in (\ref{Eq:137}), while similar to the stochastic policy gradient theorem, it is neglected in \cite{Silver.2014}. Thus, the off-policy DPG equation is obtained again as (\ref{Eq:138}), whereas the estimated action-value function computed using the critic network should be trained by the Q-learning method.
Similar to the stochastic policy gradient, it is also possible to derive the DPG algorithm using the advantage function rather than the action-value function in (\ref{Eq:138}).

DPG would be more convenient in the control design process since a stochastic policy results in unpredictable behavior, which is not desirable in autonomous vehicles. However, the exploration strategy in DPG is of significant importance to avoid the convergence to local optima. A common choice to provide an acceptable exploration is adding white noise to the current optimized policy at each step to obtain an exploratory behavior policy.

A simplified version of the introduced (on-policy) actor-critic scheme has been employed in \cite{Enns.2002} and \cite{Enns.2003} to stabilize and control a nonlinear model of an Apache helicopter, respectively, while in \cite{Enns.2003}, three cascaded NNs have been employed in the action network (equivalently to conventional multi-loop control systems) to improve the training performance.
An on-policy DPG employing the Monte-Carlo method (rather than the TD method), which updates the actor and critic networks after the end of each episode, has been used in \cite{Hwangbo.2017} to control a quadrotor UAV, where a constrained optimization has been utilized in the design to avoid large policy updates at each step, similarly to the TRPO method. In addition, the \emph{natural gradient descent}, which attempts to include the effects of the performance function's curvature induced by higher-order derivatives into the updating rule \cite{Amari.1998}, has been employed in the training rule instead of the conventional gradient descent algorithm. The control scheme has been subsequently applied to a real quadrotor air vehicle, while it suffers from the huge computational cost of the (offline) training phase, which is performed in a simulation environment.

Deep DPG (DDPG) has been introduced in \cite{Lillicrap.2016} as a combination of the DPG and DRL to employ (deep) NN in a stable manner as the actor and critic estimators, where, here, \emph{target networks} (which are employed in DRL for the critic network) are defined for both the actor and critic networks.
An off-policy DDPG has been adopted in \cite{Wang.2019b} to control a quadrotor UAV considering external disturbances and actuator faults (only in the flight tests). Adopting the concept of DRL results in more efficient training with improved stability, while off-policy learning allows for utilizing an exploratory behavior policy, which is independent of the estimated target policy.
However, as discussed in the paper, the combination of the (neural network) function approximation, the bootstrapping scheme (due to the TD learning), and the off-policy learning can lead to significant bias and variance in estimations (while in some cases, it may result in the divergence of the algorithm \cite{Sutton.1998}). To deal with such an issue, an integrator has been placed at the input of the actor network, which significantly reduces the steady-state error. Besides, a hybrid offline-online training method has been employed to improve the target policy during the real flight, while no experimental results have been included in the paper.
DDPG has been utilized in \cite{RodriguezRamos.2019} to address the autonomous landing of a UAV on a moving platform, while the problem has been dealt with in a 2D environment. As mentioned in the paper, DDPG can be an optimal choice in control problems with low-dimensional continuous states and actions.
Further, the \emph{shaping} method has been utilized in the paper to design an appropriate reward function in which the \emph{progress} of the UAV in approaching the desired goal between two successive time steps has been considered as the reward function. It has been claimed that such a technique would results in a faster learning process \cite{Sampedro.2018}, though at the cost of significant design effort and possible change of the optimal solution. This is similar to the \emph{reward shaping} method introduced in \cite{Ng.1999}, wherein a potential-based function is summed with the basic reward function to speed up the learning process with no effect on the optimal policy.
To develop an intelligent UAV navigation system in large-scale complex environments (with no requirement for map reconstruction), authors in \cite{Wang.2019}, involved the concept of \emph{Partially Observable MDPs} (POMDP) within the framework of DRL. In a POMDP, at each step, we can observe only a part of the system state denoted by $o_t$, which does not satisfy the Markov property, and so, the current policy requires the entire previous trajectory $\tau_t=\left(u_0,o_0,\cdots,u_{t-1},o_{t-1}\right)$ to determine the control command. Such a framework provides the capability of capturing the complex features of the environment by storing the previous trajectory of the system.
Accordingly, a determinist policy gradient theorem, called the Fast-recurrent DPG, has been introduced in the paper to deal with POMDPs in which $\nabla V_{\pi}$ is computed similar to (\ref{Eq:137}) except that the current state $x$ is replaced by $\tau_t$.
In a similar manner, a combination of the POMDP and the deep Q-learning concepts has been utilized in \cite{Singla.2021} to address the obstacle avoidance problem in the case of a UAV with limited environment knowledge, where a recurrent NN has been employed as the estimator of the Q-function to better estimate the current system state using information from an arbitrarily long sequence of observations.

As a notable shortcoming, almost all of the RL-based control strategies require a remarkable time for offline training of NNs to be employed in a real application.
A preliminary study has been given in \cite{Lambert.2019} in which a quadrotor can learn to hover by a relatively small amount of training data using the model-based RL. The incoming data are first employed to build a dynamic model of the system followed by a policy updating algorithm, which uses an MPC-like cost function. However, the designed control system results in unstable behavior after about five seconds.

Besides, a well-known issue in utilizing RL in flight control systems arises from the fact that the learning process in RL relies on trial and error, which can simply make the air vehicle unstable.
Thus, the learning process (in the current form) should be performed in a simulation environment, and subsequently, the trained policy is employed in a real application. However, the employment of a policy, which is trained in a simulation environment, in a real experiment suffers from the well-known \emph{reality gap} problem. Different approaches have been proposed in the literature to overcome this issue \cite{Koos.2013}. The generalization of the policy through learning in different simulation environments with different flight conditions has been suggested in \cite{Sadeghi.2017}. Further, the utilization of \emph{abstracted} inputs and outputs in the learning process would be an effective approach to tackle the reality gap \cite{Scheper.2020}. In this regard, it may be a need for a mapping (or an intermediate controller) between the abstracted inputs-outputs and real signals in the control system.
Besides, one can employ a dynamic model, which involved probabilistic uncertainties in the model, in order to evaluate and bound the worst-case controller performance in real applications \cite{Bagnell.2001}.

A similar idea can also be beneficial to deal with the issue of the stability analysis in RL-based IFCSs. More specifically, a preliminary idea to analyze the closed-loop stability under the framework of RL could be maximizing the \emph{expected} rewards at the neighborhood of an action sequence rather than that of a specific action sequence \cite{Wang.2010d}. It can be a starting point to develop a \emph{probabilistic stability analysis} framework in contrast to well-known approaches to stability analysis (using the Lyapunov theorem or similar methods) to be employed in the case of dynamic systems controlled by an RL-based scheme. To develop such a framework, we should also provide appropriate answers to principal questions regarding the quality and quantity of data samples required in the learning process.

In addition to the above-mentioned RL scheme based on MDP, there are other types of policy optimization algorithms that directly search for the optimal policy as a black-box optimization without employing the estimated action-value function into the optimization algorithm.
Random search \cite{Mania.2018,Waslander.2005}, guided policy search \cite{Levine.2013}, and evolutionary algorithms \cite{Salimans.2017} are well-known approaches in this category, while due to the lack of a solid mathematical foundation, they are not widely employed in flight control systems yet.
A guided policy search based on MPC has been introduced in \cite{Zhang.2016c} in which a set of trajectories are first generated at each step using Linear Quadratic Gaussian (LQG) controllers, where their objective is to maximize a quadratic reward by penalizing the deviation from the current policy. Subsequently, a modified MPC was designed in the vicinity of obtained trajectories, where the sampled data from trajectories, which were generated by MPC, are then employed to train the policy network in a supervised learning framework.
However, there is a need for an approximate dynamic model in the proposed design. Such an approach has been utilized in \cite{Zhang.2016c} to control a quadrotor trajectory in the presence of obstacles. While the MPC in the training phase requires access to full state observation, the final NN policy employs the data gathered by only the onboard sensors.
Since in such a guided policy search, the control commands, in the training phase, are obtained using the MPC rather than a partially trained policy network, it is a beneficial approach to avoid a remarkable drawback of the RL, i.e the occurrence of catastrophic failures during the training. Accordingly, the training phase of RL can be performed safely in a real environment to avoid the reality gap.
Another approach to achieve this goal could be the design of a training scenario using \emph{gradually increasing control commands} to learn the optimal policy (in a safe environment) to avoid the systems' instability during the training. This is conceptually similar to teaching a child to walk by his/her parents (with no simulation environment!).
Such an idea could be a starting point on the way to a truly \emph{model-free} RL-based IFCS.

Finally, it is notable that the concept of adaptive optimal control can also be incorporated in the framework of the Stochastic Optimal Control (SOC) \cite{Kappen.2005}. Since few studies have addressed the application of such a design in flight control systems, the mathematical details are not given here. Briefly, considering an affine dynamic system and a quadratic cost with respect to system inputs, it can be proved that the HJB equation for a stochastic model can be transformed into a linear PDE by defining a \emph{desirability function} as an exponential value function. The solution of such a linear PDE, called the Cauchy problem, can be represented in a probabilistic manner by applying the Feynman-Kac formula, where the solution can be derived by an expectation over all possible system paths.
Accordingly, the Monte-Carlo method involving the importance sampling technique is utilized to approximate it \cite{Ha.2019}.
This problem can also be formulated within the framework of the \emph{information theory} by incorporating the concepts of the \emph{free energy} and the KL divergence, while there is no need for mentioned restrictions (such as an affine model) in such a formulation \cite{Williams.2017}.
To this end, the optimal probability distribution of the control command is first determined, where the control problem is then converted to the minimization of the KL divergence of the current probability distribution from the optimal distribution.
The solution is typically determined iteratively at each step considering a finite prediction horizon in the cost (value) function.
Such a method is also known as Model Predictive Path Integral (MPPI).
Indeed, MPPI is a variety of MPCs in which a set of trajectories are generated at each step by adding noises to system inputs, and then, future control commands are improved by utilizing a Monte-Carlo sampling and computing the corresponding cost of each trajectory.
The first control command in the computed sequence is then applied to the system and the remaining terms are used as the baseline in the next time step \cite{Lee.2020}.
Such a method, which is somewhat similar to the guided policy search, results in a more efficient exploration rather than MPCs based on random trajectories \cite{Liang.2019}. Consequently, it can be an efficient alternative to conventional RL-based control systems, thereby providing the significant potential to be employed in IFCSs in the future.
A vision-based MPPI has been given in \cite{Lee.2020} in which a deep NN was utilized to learn the optical flow of each pixel in the image, and then an MPC attempted to bring a target pixel to the center of the camera field of view while controlling the UAV path.
Note that, as the MPC requires a prediction model, these methods are expressible within the framework of the model-based RL.
An iterative learning control has been adopted within the introduced information-theoretic MPPI scheme in \cite{Williams.2017} and \cite{Liang.2019} for obstacle-avoidance control of a quadrotor trajectory and to provide a missile guidance law, respectively, where the system dynamics have been modeled by feedforward NNs.
In this regard, as a key requirement, there is a need for a large number of samples in sampling-based MPCs, while the mathematical foundation for the analysis of the algorithm convergence and the closed-loop stability in the above-mentioned method should still be strengthened.

Various challenges remain in the way of efficient RL-based model-free control systems yet, and in some cases, a simple PID or LQR controller may behave more effectively than existing RL-based control approaches \cite{Recht.2019}.
However, RL has provided a window into a new look at the control problem of complex systems in complex environments, and it is expected that such a framework can lead to a generic, fully autonomous, truly model-free, and safe control methodology in the near future such that it can be reliably employed in the case of more complex aerial vehicles (such as nonconventional aircraft) and more complex problems (such as the presence of severe external disturbances and actuator faults).

Five tables are given in the following. Principal characteristics of some key research addressing the NN-based control of VTOL aerial vehicles, HFVs (and NSVs), fixed-wing aircraft, and nonconventional air vehicles are listed in Tables \ref{tab:table1}-\ref{tab:table4}, respectively. Different specifications of each research, i.e., the control objective, the consideration of system constraints in the design, the use of an MLP technique, the employment of an OFB control scheme, and the type of uncertain dynamics considered in the model, as well as the main features and limitations arising from each control methodology are briefly reported. The provided data would be advantageous to identify considerable capabilities, complexities, and limitations of each control strategy for each type of aerial vehicle, to compare the importance and effectiveness of different control methods, and to figure out the challenging issues remaining unsolved. On the other hand, as a separate category, some novel \emph{high-level} control systems incorporating NNs in their design are given in Table \ref{tab:table5} in which the learning method, the control objective, key features, and considerable limitations of each research are listed. The combination of such high-level control strategies with existing low-level intelligent control systems would be a critical research area to develop an intelligent flight management unit.

%
\captionsetup{width=\textwidth}
\begin{scriptsize}
\begin{longtable}{>{\centering\arraybackslash}p{0.16in}|>{\centering\arraybackslash}p{0.62in}|>{\centering\arraybackslash}p{0.7in}|p{1.3in}|p{1.3in}}
\caption{Principal characteristics of some of the introduced intelligent control systems for VTOL aerial vehicles}\label{tab:table1} \\
Ref. & Controller  & Characteristics$^*$ & Main features & Limitations/ Complexities  \\ \hline
\cite{Zou.2015b} & Backstepping & TD & -Utilizing a switching function to integrate the NN and DO & -Neglecting the approximation error of differentiators \\ \hline
\cite{Lai.2016} & Neuroadaptive & TMD & -Considering aerodynamic frictions in the model \newline -Considering unknown inertia matrix & -Avoiding attitude singularity problem using a BLF rather than employing the well-known quaternion formulation \\ \hline
\cite{Kayacan.2016} & Neuroadaptive & TD & -SMC-like-based training algorithm for FNNs  & -The system should be decoupled into a set of SISO models \newline -Applicable in second-order systems \newline -Concerns with the stability analysis \\ \hline
\cite{Vamvoudakis.2017} & ADP ($H_2$ control) & T & -Event-triggered control  \newline -Using discounted cost & -On-policy learning \newline -The necessity for the PE condition \newline -Several conservative assumptions in the stability analysis \newline -Requires entire dynamic model \\ \hline
\cite{Hwangbo.2017} & On-policy DPG & WR & -Performing a wide range of maneuvers, stably & -No stability analysis \newline -Huge off-line computational burden \\ \hline
\cite{Li.2017c} & Backstepping & ADIO & -Adopting combined NN and DO \newline -Using Nussbaum function to deal with input saturation \newline -Using BLF to tackle output constraints & -Concerns with  the stability analysis \newline -Consdeiring a SISO model \\ \hline
\cite{Fu.2018} & Backstepping & AOMD & -DSC for multi-rotor UAV \newline -Adopting combined NN and DO \newline -Using a time-varying BLF & -Large control actions caused by BLF \\ \hline
\cite{Ferdaus.2019}  & Neuroadaptive & AR & -SMC-like-based training algorithm for FNNs \newline -Generality of the control scheme & -The system should be decoupled into a set of SISO models \newline -Chattering phenomenon \newline -The plant must be stabilizable by a PID controller \newline -Concerns with the stability analysis\\ \hline
\cite{Wang.2019b} & Off-policy DDPG & WFR & -Hybrid offline-online learning algorithm \newline -Adopting integrators to eliminate steady-state error & -No stability analysis \newline -Significant off-line computational burden \\ \hline
\cite{Camci.2018} & Neuroadaptive & TR & -SMC-like-based training algorithm for FNNs & -The system should be decoupled into a set of SISO models \newline -The plant must be stabilizable by a PD controller \newline -Concerns with the stability analysis \\ \hline
\multicolumn{5}{C{\textwidth}}{{\scriptsize * Control objective: A: Attitude control, W: waypoint tracking, T: Trajectory tracking, L: Longitudinal mode/ I: Consideration of input constraints, O: Consideration of output (state) constraints/ M: Minimal-learning parameter/ K: Output feedback control/ D: Disturbance or noise rejection, F: Fault-tolerant control, R: Model-free control}}
\end{longtable}

\begin{longtable}{>{\centering\arraybackslash}p{0.16in}|>{\centering\arraybackslash}p{0.62in}|>{\centering\arraybackslash}p{0.7in}|p{1.3in}|p{1.3in}}
\caption{Principal characteristics of some of the introduced intelligent control systems for HFVs and NSVs}\label{tab:table2} \\
Ref. & Controller  & Characteristics & \multicolumn{1}{c}{Main features} & \multicolumn{1}{|c}{Limitations/ Complexities}  \\ \hline
\cite{MouChen.2014} & Backstepping & ADI & -DSC with WNN-based DO \newline -Using Nussbaum function to deal with input saturation & -Concerns with satiability analysis \\ \hline
\cite{Chen.2015c} & Backstepping & ADI & -Adopting combined NN and DO \newline -Using a modified tracking error to deal with input saturation \newline -Control allocation using a convex optimization solved by an RNN & -Neglecting the control allocation error in the stability analysis \\ \hline
\cite{Xu.2015b} & Backstepping & LI & -DSC with direct neural approximation \newline -Using Nussbaum function to deal with dead-zone input nonlinearity & \\ \hline
\cite{Bu.2018b} & Backstepping & LMO & -Funnel control to guarantee the transient performance \newline -Consideration of flexible states & -Many design parameters \newline -Availability of the third derivative of the tracking error \\ \hline
\cite{Wu.2017}  & Backstepping & LMIO & -FOSMD in the backstepping design \newline -Morphing aircraft (with pure-feedback
model) \newline -Using Butterworth filter to avoid algebraic loop in the control design & -Consideration of only the cruise phase \newline -Assuming the bounded filtering error \\ \hline
\cite{Xu.2015c} & Backstepping & L & -Using a discrete-time model \newline -Utlizing a prediciton model & \\ \hline
\cite{Bu.2015} & Backstepping & L & -Direct neural-backstepping scheme \newline -Using the integral of tracking error to eliminate the steady tracking error &  \\ \hline
\cite{Xu.2015d} & Neuroadaptive & LMK & -Defining an OFB model and using HGOs to avoid backstepping & -Large control commands at early times \\ \hline
\cite{Xu.2016d} & Backstepping & LMF & -Avoiding singularity problem using direct DSC & -Considering only the bias actuator fault \newline -Unusual formulation of NNs \\ \hline
\cite{Bu.2018} & Pseudocontrol & LM & -Avoiding the backstepping design through transforming the model into a normal feedback form \newline -No requirement for the contraction assumption & -Time derivatives of FPA should be measurable \\ \hline
\cite{Xu.2019} & Backstepping & LFDO & -FOSMD in the backstepping design \newline -Consideration of AOA constraint \newline -Neural fault identification & -Using a SISO model \\ \hline
\cite{Li.2020} & Backstepping & LMFDO & -Control of the transient response \newline Asymptotic tracking control & -Parameter drift in the updating rules \newline -Excessive control effort at the vicinity of permissible output bounds \\ \hline
\end{longtable}

\begin{longtable}{>{\centering\arraybackslash}p{0.16in}|>{\centering\arraybackslash}p{0.62in}|>{\centering\arraybackslash}p{0.7in}|p{1.3in}|p{1.3in}}
\caption{Principal characteristics of some of the introduced intelligent control systems for fixed-wing aircraft}\label{tab:table3} \\
Ref. & Controller  & Characteristics & \multicolumn{1}{c}{Main features} & \multicolumn{1}{|c}{Limitations/ Complexities}  \\ \hline
\cite{Nguyen.2008} & Pseudocontrol & AF & -Hybrid direct-indirect adaptive control \newline -Considering (a specific) structural damage & -No actuator dynamics \newline -Slow convergence of the algorithm \\ \hline
\cite{Chowdhary.2013} & Pseudocontrol & WF & -Modification of guidance commands to adapt to current flight condition  & -No stability analysis \\ \hline
\cite{Luo.2015} & ADP ($H_\infty$ control) & WD & -Off-policy learning \newline -Partially model-free control \newline -Employing single NN & -No stability analysis \\ \hline
\cite{Wang.2017c} & ADP ($H_\infty$ control) & WD & -Event-triggered control  \newline -Employing single NN & -On-policy learning \newline -Requires entire dynamic model \newline -The necessity for the PE condition \newline -Several conservative assumptions in the stability analysis \\ \hline
\cite{Abbaspour.2018} & Dynamic inversion & AF & -Indirect EKF-based fault identification & -Concerns with the stability analysis \\ \hline
\cite{Emami.2019b} & MPC & TIFR & -Multimodel FTC scheme \newline -Indirect RLS optimization-based fault identification & -Concerns with the feasibility of the proposed control design \\ \hline
\cite{Yu.2020} & Backstepping & AF & -Fractional-order backstepping control \newline -Decentralized control of multi-UAVs \newline -Adopting combined NN and DO & -Concerns with employing fractional-order control in practice \newline -Conservative assumptions on estimation errors \\ \hline
\cite{Yu.2020b} & Backstepping & TF & -DSC-based distributed formation flight control \newline -Adopting combined NN and DO \newline -Consideration of wake vortices & -Some simplifications in dynamic modeling \\ \hline
\end{longtable}

\begin{longtable}{>{\centering\arraybackslash}p{0.16in}|>{\centering\arraybackslash}p{0.62in}|>{\centering\arraybackslash}p{0.7in}|p{1.3in}|p{1.3in}}
\caption{Principal characteristics of some of the introduced intelligent control systems for nonconventional air vehicles}\label{tab:table4} \\
Ref. & Controller  & Characteristics & \multicolumn{1}{c}{Main features} & \multicolumn{1}{|c}{Limitations/ Complexities}  \\ \hline
\cite{He.2017} & Neuroadaptive & TOD & -Flapping wing micro aerial vehicle control \newline -Adopting combined NN and DO & \\ \hline
\cite{Won.2017} & Deep Q-learning & TR & -Control of flapping-wing aerial vehicles using RL \newline -Utilizing an evolutionary exploration \newline -Maximizing expected reward near an action sequence to improve the robustness & -No stability analysis \\ \hline
\end{longtable}

\begin{longtable}{>{\centering\arraybackslash}p{0.16in}|>{\centering\arraybackslash}p{0.62in}|>{\centering\arraybackslash}p{0.7in}|p{1.3in}|p{1.3in}}
\caption{Principal characteristics of some of the introduced intelligent high-level control methods}\label{tab:table5} \\
Ref. & Method  & Objective$^*$ & \multicolumn{1}{c}{Main features} & \multicolumn{1}{|c}{Limitations/ Complexities}  \\ \hline
\cite{Giusti.2016} & Supervised learning & C & -Using simple images for training with no need for determining characteristic features of an object & -Lack of strong mathematical foundation \newline -No stability analysis \\ \hline
\cite{Sadeghi.2017} & Deep RL & C & -Real flight experiments \newline -Using only monocular images as input & -No stability analysis \\ \hline
\cite{Zhang.2018} & Deep RL & DE & -Training mobile charging stations to autonomously move to the charging point in an optimal manner & -Considering the 2D problem \newline -No stability analysis \\ \hline
\cite{Wang.2019} & Modified DPG & CW & -Using the POMDP scheme \newline -Navigating in lrage-scale complex environment & -No stability analysis \\ \hline
\cite{RodriguezRamos.2019} & DDPG & W & -Auto-landing on a moving platform \newline -Real flight experiments & -Considering 2D problem \newline -No stability analysis \\ \hline
\cite{Zhang.2016c} & Guided policy search & CW & -MPC-based guided policy search \newline -Providing a safer training phase using MPC in the training & -Requiring approximate dynamic model \newline -No stability analysis \\ \hline
\cite{Williams.2017} & MPPI & CW & -Control of nonaffine dynamics \newline -Utilizing information theoretic MPC with a generic cost function & -Requiring the dynamic model \newline -No stability analysis \\ \hline
\multicolumn{5}{C{\textwidth}}{{\scriptsize * Obstacle or Collision avoidance: C/ Data collection: D/ Waypoint tracking: W/ Consideration of total energy constraint: E}}
\end{longtable}
\end{scriptsize}

\section{Concluding remarks and future directions}\label{Sec:Conclusions}
Intelligent flight control systems have been significantly evolved, particularly during the last two decades. They have been able to satisfactorily deal with different practical issues in a real flight, e.g. atmospheric disturbances, operational faults, model uncertainties, unmodeled dynamics, etc. In addition, concerning model-free control methods, there has been remarkable progress in both indirect adaptive controllers, which employ NNs to provide a valid dynamic model of the system, and direct adaptive control systems using the optimal control or the RL frameworks.
Besides, recently, intelligent approaches, particularly those based on RL, have been effectively adopted in high-level control systems to provide intelligent path planning and guidance loops in flight control systems. Such remarkable progress of IFCSs results in introducing aerial robots with an outstandingly high level of autonomy. Despite all these advances, there still is a long way to go to introduce a \emph{generic intelligent flight control system}. In the following, we address a set of crucial bottlenecks along with some suggestions for the direction of future research in developing such an intelligent flight control system.
\begin{enumerate}
  \item \emph{Design parameters}: The determination of appropriate design parameters in proposed IFCSs is a challenging issue, which is typically carried out by trial and error. Although the training of the controller's parameters (using an additional learning loop \cite{Baydin.2017,Emami.2018b} or evolutionary algorithms \cite{Abaspour.2015}) or the reduction of the design parameters (by incorporating self-organizing \cite{Pratama.2014} or new data analysis approaches \cite{Ferdaus.2019b}) can deal with such a problem to a certain extent, the development of \emph{generic} intelligent control systems with no (or at least very low) design parameters is still an open problem in the field of intelligent control. Thinking about more flexible control structures organized by the incoming system data using machine learning approaches can be a gateway to efficient solutions.
  \item \emph{High-level control}: An intelligent guidance loop to ensure that a feasible trajectory is commanded to the aircraft is critical in developing a reliable flight control system in the presence of internal and external disturbances. However, this loop is typically remained unchanged after the fault occurrence in classical IFCSs \cite{Chowdhary.2013}. Adaptive estimation of the flight envelope in the presence of operational faults would be the first step to provide a feasible FTC system \cite{Tang.2009}. In a more general view, the problem of the intelligent trajectory generation (for different purposes such as collision avoidance) is a challenging problem, which has received less attention from academia. Such a problem considering different design criteria, such as obstacle/collision avoidance, optimal resource allocation, etc., have been addressed in \cite{Sadeghi.2017,Zhang.2018,Wang.2019} using RL, while there are concerns about the definition of an appropriate reward function. In this respect, there is a significant need for the improvement and unification of such high-level control systems with conventional low-level IFCSs, consistently, to develop a fully autonomous flight management system. Further, by developing novel machine learning algorithms along with the development of the computing power, there will be an opportunity to redefine an entire flight control problem (which, in the existing framework, includes various control loops) as a new framework with a more integrated and concise structure that can map high-level commands to low-level inputs with less human intervention using novel machine learning methods.
  \item \emph{Evolutionary algorithms}: Although evolutionary algorithms are not currently a mainstream topic in aerial robotics, they may be an appropriate candidate in the near future to be adopted in NN-based flight control systems to enhance the effectiveness and efficiency of training algorithms, while reducing the computational complexity, and to learn the networks' architecture and learning hyperparameters \cite{Silva.2016,Scheper.2020}. To this end, there would be a requirement for a new mathematical framework (maybe in a probabilistic representation) to analyze the convergence of such learning approaches in order to develop a reliable control design procedure.
  \item \emph{Controllability region}: The provided stability analysis in almost all of the existing literature introduces a region of stability, where the boundaries of that region are determined by upper limits of a set of parameters, which do not have a physical meaning or are not measurable \cite{Song.2019}. This is a serious problem in utilizing such adaptive controllers in a real application because we can not determine the controllability region of the system based on physical parameters. This problem becomes more challenging in model-free control systems. In this regard, there is a need for introducing a set of \emph{tangible criteria} for analyzing closed-loop stability. More specifically, the borrowing of concepts from the information theory to analyze the controllability of the system according to different characteristics of incoming data would be an attractive idea to provide a beneficial stability analysis framework even for model-free control systems \cite{Williams.2017}.
  \item \emph{Input-output constraints}: Simultaneous consideration of input and output constraints in the control design process is a challenging problem. This is due to the fact that the satisfaction of input constraints may lead to larger tracking errors, and conversely, the consideration of output constraints may necessitate impractical control commands. Integration of the funnel control with input saturation constraint has been addressed in \cite{Hopfe.2010} and \cite{Hopfe.2010b} for a linear minimum phase and a SISO nonlinear system without considering model uncertainty in the control problem, while the problem becomes more challenging in the presence of uncertain dynamics. Reinforcement learning would be an effective candidate in such control problems. More precisely, using the RL, it is possible to learn an optimal policy as a mapping from permissible system inputs to desired outputs. Further, concerning \emph{fault-tolerant} flight control systems, an iterative learning control scheme integrated by RL methods (such as MPPI \cite{Liang.2019}) would be an appropriate solution in future studies.
  \item \emph{Structural constraints}: Elastic modes of an air vehicle can result in a variety of undesirable phenomena such as flutter and control reversal, which may affect the closed-loop performance. This can become more challenging in the case of damaged aircraft due to the uncertainty in structural margins and elastic modes shifting (caused by changes in the structural stiffness and mass of the airframe) \cite{Nguyen.2008}. The consideration of such structural constraints in the design process of IFCSs in future studies is a vital issue.
  \item \emph{RL computational complexity}: The considerable computational cost of RL is still a challenging issue, which is more problematic in the case of high dimensional problems \cite{Hwangbo.2017}. In this regard, the development of integrated multi-loop control systems in which RL is employed in the outer control loop design, while existing classical intelligent controllers (discussed in Section \ref{Sec:Model_based_IFCS}) are utilized in the inner loop, would be an effective solution to reduce the dimension of the action space, thereby significantly reducing the learning complexity.
  \item \emph{Reward function in RL}: The definition of an appropriate reward function in RL-based control systems is of great importance, where there is still no well-known intelligent approach to define that. Although \emph{reward shaping} is a well-known approach to speed up the learning process, there are significant concerns about the optimality of the computed policy and the convergence of the algorithm \cite{Ng.1999}. Inverse RL could be another solution to identify appropriate reward function using the \emph{learning from demonstration} (i.e. the task of learning from an expert) \cite{Abbeel.2004}. Further, the concept of learning from demonstration (also known as \emph{apprenticeship learning}) would be a useful approach to train an intelligent trajectory generation scheme \cite{Coates.2009}.
  \item \emph{NNs' adaptation speed}: There are still considerable concerns about the adaptation speed of NNs in both the model-based and model-free control systems, particularly in the case of time-varying systems and environments with rapid changes \cite{Emami.2019b,Emami.2019c}. To deal with such an issue, there would be a requirement for more effective learning schemes with a faster convergence rate, while not violating the robustness of the closed-loop system. On the other hand, the computation of the \textit{best} adaptation rate in different flight conditions is another complicated problem with no clear answer \cite{NairaHovakimyan.2011}. Concerning the adaptive control of dynamic systems with parametric uncertainty, the above-mentioned problems have been addressed in the past two decades within the framework of $L_1$-adaptive control, which attempts to decouple the estimation loop from the control loop in order to decouple the adaptation from the robustness \cite{Hovakimyan.2010}. However, several issues have been reported in the literature regarding the claims made about the capabilities of $L_1$-adaptive control \cite{Ioannou.2014}. In this regard, there is a serious need to develop novel learning frameworks to be used in the case of dynamics systems subject to rapid changes in the dynamic model and the environment (considering both parametric and nonparametric uncertainties).
  \item \emph{NNs' structure in RL}: Typically, feedforward NNs are used as the actor and critic networks within the actor-critic framework (particularly, in RL \cite{Wang.2019b}). RNNs can be an efficient alternative to feedforward NNs within such a framework to improve the closed-loop stability while reducing the bias and variance of the learning process.
  \item \emph{Time-dependent NNs}: There are more complex NNs in the literature, which can be adopted in the control design procedure to enhance the modeling and training performance. For instance, continuous-time RNNs \cite{Beer.1995} can be used to incorporate the time-varying sampling times \cite{Scheper.2020}. Further, in spiking NNs, which are more close to biological NNs, each neuron has a \emph{membrane potential} affected by incoming signals, where the neuron emits a spike once its membrane potential exceeds a specific threshold, and subsequently, the membrane potential is reset to a rest value. Although, due to the complex training process of such NNs, currently, they are mostly employed in relatively simple control problems \cite{Clawson.2016,Hagenaars.2020}, spiking NNs can be a beneficial choice to imitate complex behavior of intelligent systems, thereby providing a significant potential to be used in complex intelligent controllers. Further, as these NNs encompass the concept of time in their models, they could be appropriate solutions to deal with explicitly time-dependent model uncertainties and external disturbances.
  \item \emph{Complex NNs}: Different types of deep NNs, wavelet NNs, and CNNs have demonstrated their superior performance in the identification of complex nonlinear systems \cite{Yu.2019,Alexandridis.2013}. Although several studies have addressed the development of direct (and rarely indirect) adaptive flight control systems consisting of such complicated NNs \cite{Lin.2014,Lin.2015,Bu.2018,Kang.2019b}, the effective employment of them in both the identification and control design steps, which may also require the development of more efficient training algorithms rather than existing ones, can impressively reduce the NNs' estimation error, which in turn, results in reducing the conservativeness of designed controllers, significantly.
  \item \emph{Aggressive maneuvers}: Most of the introduced trajectory tracking control schemes have been designed and validated under simple trajectories with no aggressive maneuvers \cite{Won.2017}. In recent years, some researchers have addressed the development of autonomous aircraft aerobatics employing the concept of learning from demonstrations \cite{Abbeel.2010}. Such an approach can be effectively employed in the framework of IFCSs, particularly those based on RL to provide the ability to perform a wide range of maneuvers, and in the near future, IFCSs are expected to be able to fulfill more complex tasks, such as take-off, landing, and different aggressive maneuvers, with guaranteed performance.
\end{enumerate}

\section*{Acknowledgements}
This work was supported by Iran National Science Foundation (INSF) and Iran's National Elites Foundation (INEF) grant 98027065.

\bibliographystyle{unsrt}
\bibliography{Ref}

\end{document}